\documentstyle[12pt]{article}

\newcommand{\nl}{\nonumber \\}

\def\beq{\begin{equation}}
\def\eeq{\end{equation}}
\def\beqa{\begin{eqnarray}}
\def\eeqa{\end{eqnarray}}

\newcommand{\sect}[1]{\setcounter{equation}{0}\section{#1}}

\newcommand{\eq}{\begin{equation}}
\newcommand{\en}{\end{equation}}
\newcommand{\bea}{\begin{eqnarray}}
\newcommand{\ena}{\end{eqnarray}}

\newcommand{\shalf}{\frac{1}{2}}

\def\one{{\hbox{ 1\kern-.8mm l}}}

\newcommand{\mat}[4]{\left( 
		     \begin{array}{cc}
		     {#1} & {#2} \\
		     {#3} & {#4} 
		     \end{array}
		     \right)
		    }

\newcommand{\vecc}[2]{\left(
		      \begin{array}{c}
		     {#1} \\
		     {#2} 
		     \end{array}
		     \right)
		    }

\def\part{\partial}
\def\aa{\alpha}
\def\bb{\beta}
\def\cc{\chi}
\def\dd{\delta}
\def\ee{\epsilon}
\def\gg{\gamma}
\def\lb{\bar\lambda}
\def\pp{\psi}
\def\th{\theta}
\def\tb{\bar\theta}

\begin{document}

\setcounter{page}{1}
\rightline{NORDITA 98/11-HE}
\rightline{hep-th/9803026}
\rightline{\hfill February, 1998}

\vskip .3cm

\centerline{\Large \bf Duality in ${\bf N=2,4}$ supersymmetric gauge theories
\footnote{Work partially supported by the European Commission TMR programme
ERBFMRX-CT96-0045.}} 

\vskip 1cm

\centerline{\bf Paolo Di Vecchia, 
		   \footnote{e-mail: DIVECCHIA@nbivms.nbi.dk} }
\centerline{\sl NORDITA}
\centerline{\sl Blegdamsvej 17, DK-2100 Copenhagen \O, Denmark}

\vspace{1cm}
\centerline{ {\it Lectures given at Les Houches Summer School (August 1997)}}
\vspace{1cm}







\begin{abstract}
In these lectures we present a detailed description of various aspects of 
gauge theories with extended
$N=2$ and $N=4$ supersymmetry that are at the basis of recently found exact 
results. These results include the exact calculation of the
low energy effective action for the light degrees of freedom in the $N=2$
super Yang-Mills theory and the conjecture, supported by some checks, that the
$N=4$ super Yang-Mills theory is dual in the sense of Montonen-Olive. 
\end{abstract}

\vspace{1cm}

\sect{Introduction}
\label{intro}

In the last few years a number of exciting results on the non-perturbative 
behaviour of four-dimensional gauge theories and string theories in various
dimensions have been obtained. They are all based on the fundamental idea of
duality. Duality is a symmetry that  already appears in free electromagnetism
and corresponds to the fact that the free Maxwell equations are invariant under 
the exchange of electric and magnetic fields. Such a symmetry is not discussed
in  elementary courses on electromagnetism because it is lost when an
interaction is introduced that, due to the absence of magnetic monopoles, has 
only terms with the electric current and with the electric charge density. 

If, however, we forget for a moment that no magnetic monopole has yet been  
detected in experiments, and we introduce in the Maxwell equations also
a  magnetic current and a magnetic charge density, we immediately
discover that the interacting Maxwell equations with these terms added preserve
the invariance under the exchange of electric and magnetic fields provided
that at the same time the electric and magnetic currents and charge 
densities are also exchanged. But, if both electric $q_i$ and magnetic 
$g_j$ charges are present, their quantum theory can only be consistent if 
they both are 
quantized in terms of an elementary electric and magnetic charge. This is a
direct consequence of the famous Dirac quantization condition~\cite{DIRAC}:
\beq
q_i g_j = 2 \pi \hbar n_{ij}  
\label{Dir45}
\eeq
where $n_{ij}$ is an integer. This relation, in fact, implies that both the 
electric
and magnetic charges are quantized in terms of an elementary electric $q_0$
and magnetic $g_0$ charges that satisfy themselves the Dirac quantization
condition with an integer $n_0$. Consequently a theory in which the 
fundamental electric charge $q_0$ is small, corresponding to a perturbative
electric theory, is necessarily a theory in which the magnetic charges are
large, corresponding to a strongly interacting magnetic theory and viceversa.
Thus we can only have a perturbative theory either in the electric
or in the magnetic charge, but not in both. 

Another apparently different kind of duality, the so-called 
Kramers-Wannier~\cite{ISING}  duality,
was discovered in  two-dimensional Ising model. This is a model for 
spins $\sigma_i$, taking the values $\pm 1$, living on a square
lattice and interacting according to a nearest-neighbour interaction with 
strenght $J$. We call $Z (K)$ its partition function, that is a function of the
temperature $T$ through the relation $K= J/(k_B T)$, where $k_B$ is the 
Boltzmann
constant and $J$ is the coupling between the spins. It is known that the 
Ising model can be formulated either on the 
original lattice with coupling constant $K$ and partition function $Z(K)$ or 
on the dual lattice, constructed from the original
lattice by selecting the central points of each square of the lattice, with 
coupling constant $K^{*}$ and partition function $Z^{*} ( K^{*} )$. 
It turns out that the two partition functions $Z$ and $Z^{*}$ are equal if the
two couplings constants are related by the relation
\beq
\sinh 2 K  = 1/(\sinh 2 K^{*})
\label{isingmo}
\eeq
The formulation on the original lattice provides a good description of the 
system at high temperature $T$ or weak coupling $J$ (small $K$), while the one
on the dual lattice gives a good description of the system at low temperature 
or strong coupling (small $K^{*}$).
The relation in eq.(\ref{isingmo}) is also used to show that, if the system
has a unique phase transition, it must occur at the self-dual point $K = K^{*}$.

The fact that a certain theory can be represented by two perturbatively 
completely different theories, as in the Ising model by expanding in $K$ or
in $K^*$, became evident in the middle of the 
seventies with the proof of the quantum  
equivalence~\cite{LUTHER,COLE,MANDESTA} 
between a purely bosonic 
field theory as the sine-Gordon one described by the Lagrangian:
\beq
L = \frac{1}{2} \partial_{\mu} \Phi \partial^{\mu} \Phi + \frac{M^2}{\beta^2}
\left( \cos \beta \Phi -1 \right)
\label{sineGo}
\eeq
and a purely fermionic one as the massive Thirring model described by the 
Lagrangian:
\beq
L= {\bar{\Psi}} \left( i \gamma_{\mu} \partial^{\mu} + m \right) \Psi -
\frac{g}{2} {\bar{\Psi}} \gamma^{\mu} \Psi {\bar{\Psi}} \gamma_{\mu} \Psi
\label{Thir}
\eeq
provided that the two coupling constants are related by
\beq
\frac{\beta^2}{4 \pi} = \frac{1}{1 + g/\pi}
\label{sinth}
\eeq

As in electromagnetism and in the Ising model discussed before also in
this case weak coupling in one theory (for instance in the sine-Gordon theory)
corresponds to strong coupling in the other one (the Thirring theory) as 
expressed by the relation in eq.(\ref{sinth}). It was also recognized that the
sine-Gordon theory contains together with a perturbative scalar particle, 
corresponding to the scalar field present in the sine-Gordon Lagrangian, also
a soliton solution that does not correspond to any field  in the sine-Gordon 
Lagrangian, but that
can be shown to correspond to the fermion field of the Thirring Lagrangian.
The soliton has also the important property that its mass is large in 
sine-Gordon perturbation theory (small $\beta$). All these considerations 
made it soon clear that the sine-Gordon-Thirring theory was a unique theory
that can be either formulated in terms of the sine-Gordon Lagrangian 
containing a fundamental scalar particle (we call it fundamental because it 
corresponds to the field $\Phi$ present in the sine-Gordon Lagrangian) or in 
terms of the Thirring Lagrangian containing a fermionic particle, that is
fundamental from the point of view of the Thirring Lagrangian, 
corresponding to the fermionic field $\Psi$, but that is
solitonic from the point of view of the sine-Gordon Lagrangian.

The understanding of these properties in two-dimensional theories, together
with the discovery of the 't Hooft-Polyakov monopole~\cite{thooft,Pol} 
solution as a soliton 
in the four-dimensional Georgi-Glashow model, opened the way to the beautiful 
suggestion, made by Montonen and Olive~\cite{MONTOLI}, of duality between the 
original theory in 
terms of the fundamental particles described by the fields of the original 
Georgi-Glashow Lagrangian, that in this specific case were the $W$-bosons,
and the dual theory in which the fundamental particles are replaced by the
monopoles that are  solitons of the original theory. The original 
formulation of Montonen and Olive implied that the two theories, the original 
and the dual called also electric and magnetic, were essentially described by 
the same Lagrangian with their gauge coupling constants related to each other
as the elementary electric and magnetic charges $q_0$ and $g_0$ are related
through the Dirac quantization condition (see eq.(\ref{Dir45}) and what 
follows). 
Although Montonen and Olive brought a number of arguments in support of their 
suggestion, as for instance the fact that
the masses of all particles were given by the same duality invariant formula,
it became soon clear that their beautiful idea could not be realized in the
Georgi-Glashow model. This was mainly due to two reasons. The first one
was that the mass formula was a classical formula and there was no
evidence that it will keep the same form in the full quantum theory. 
The additional problem was how to obtain solitons with spin $1$ as 
required by the fact that the fundamental particles of the original theory,
the $W$-bosons, had spin equal to $1$. 

It became soon clear~\cite{DADDA} that, in order to realize the beautiful 
duality idea of Montonen and Olive, one needed an additional 
ingredient, namely supersymmetry.
This was the reason to study~\cite{DADDA} the simplest supersymmetric theory 
with monopole and dyon solutions having the same structure as those in 
the Georgi-Glashow model, namely the $N=2$ super Yang-Mills theory. It was 
also soon recognized~\cite{WO} 
that this theory had a BPS mass formula that was a direct 
consequence of the quantum supersymmetry algebra and not just a formula valid 
in the classical theory as in the Georgi-Glashow model. This solved 
the first problem mentione above.
Quite soon, however, it became also clear~\cite{OSBORN} that the 
supersymmetry multiplet
to which the magnetic monopole belonged, did not contain a spin $1$ state and
consequently also the $N=2$ super Yang-Mills could not realize the 
Montonen-Olive
duality. This brought Osborn~\cite{OSBORN} to consider $N=4$ super Yang-Mills
in which the BPS classical mass formula is not changed by quantum corrections
and the $N=4$ supersymmetry multiplet contains a state with spin $1$. This is
a theory that passed all tests for realizing the Montonen-Olive duality and
in fact in the last section of these lectures we will present a reformulation
of it as discussed in a recent review written by David Olive~\cite{OLIVE}.  

The Montonen-Olive duality, that by now has been extended also to string 
theories for space-time dimensions $D \leq 10$ and has played a fundamental
role in the
understanding of the non-perturbative connections between various consistent
and perturbatively inequivalent string theories, seems to be working only for
theories that have enough supersymmetries to prevent classical formulas, as
the BPS mass formulas, to be modified by quantum corrections and also coupling 
constants to run. Those theories, having potentials with flat directions, are 
characterized by a manifold of inequivalent vacua. Such degeneracy is 
sometimes
modified by quantum corrections, but it is in general never wiped out.

On the other hand, these are not very interesting theories 
for hadron physics that at present energies  is well described by QCD in which
the strong fine structure constant runs and  the vacuum is uniquely fixed.
If we want to study theories that are closer to QCD, although still have flat 
directions, we must therefore go to 
$N=2$ supersymmetric theories or even better to $N=1$ supersymmetric theories.
 
In these lectures we will describe the Seiberg-Witten approach to the 
determination of the exact low-energy effective action for $N=2$ super 
Yang-Mills with gauge group $SU(2)$. We have unfortunately no time to also
discuss $N=1$ supersymmetric gauge theories. For reviews on them the reader
is adviced to consult Refs.~\cite{INTRI,SHIFMAN}. A number of reviews on 
duality in gauge theories and more specifically on the
Seiberg-Witten approach have also 
appeared~\cite{OLIVE,HARVEY,BILAL,LERCHE,KETOV,HASSAN,PESKIN,LINDSTROM,KLEMM}.

This is a much extended version of the lectures that I gave last year at the
ITEP Winter School~\cite{ITEP}. I have tried as much as possible to make
them self-contained. The plan of the lectures is as follows. 

In section (\ref{dua}) we discuss  duality in electromagnetism and in 
section (\ref{dirac}) the Dirac quantization condition.
In section (\ref{mono}) the 't Hooft-Polyakov magnetic monopole and 
the Julia-Zee dyon solutions in the Georgi-Glashow model are discussed in great
detail, while section (\ref{colleco}) is devoted to their semiclassical
quantization. In section (\ref{insta}) we discuss instanton solutions in
euclidean Yang-Mills theory, their difference with respect to monopoles and 
the introduction of the $\theta$ parameter in gauge theories. 
Section (\ref{MOdua}) is devoted to the Montonen-Olive duality conjecture. 
After the formulation of this beautiful idea it became
very soon clear that this duality property cannot be satisfied in the
Georgi-Glashow model where the quantum corrections invalidate the conclusions
based on  semiclassical considerations. The theories that have a chance to
realize it were those in which the semiclassical properties are not
destroyed by quantum corrections and those are the supersymmetric gauge 
theories. That is why in section (\ref{repre}) we 
discuss the representations of supersymmetry algebra with and without central 
charges, in sect. (\ref{supe}) we construct the supersymmetric Yang-Mills 
actions in four dimensions from dimensional reduction from $D=6,10$ and 
finally in section 
(\ref{susy}) we write the Lagrangians of  supersymmetric gauge theories
also with matter in the formalism of $N=1$ superfields. In sect. (\ref{semi}) 
we 
present the semiclassical analysis of $N=2$ super Yang-Mills theory, discuss 
the perturbative and instanton contributions to its low energy effective action
and show that this theory has monopole and dyon solutions as the Georgi-Glashow
model. Section (\ref{centra}) is devoted to the computation of the central
charges of the supersymmetric algebra and to the derivation of the mass formula
for the BPS states that is a direct consequence of the quantum supersymmetry 
algebra with central charges and not just a property of the classical theory 
as in the Georgi-Glashow model. In the second part of section (\ref{centra}) 
we show that the structure of the fermionic zero modes of $N=2$ super 
Yang-Mills theory is such that Montonen-Olive duality cannot be realized in it.
In sections (\ref{global}) and (\ref{singu}) 
we discuss respectively the global parametrization of the moduli space in
$N=2$ super Yang-Mills with gauge group $SU(2)$ and its singularity structure,
while in section  (\ref{exsolu}) we explicitly construct the Seiberg-Witten
solution. Section (\ref{Neq24}) is devoted to a quick review of the main
properties of $N=4$ super Yang-Mills theory. In sect. (\ref{RIMO}), 
following very closely Ref.~\cite{OLIVE}, we riformulate the Montonen-Olive 
duality conjecture adapted to the $N=4$ super Yang-Mills theory and we show 
that the various formulations are related to each others by the action of the 
modular group $SL(2,Z)$. 

Many details of the calculations are presented in three appendices. 
Appendix \ref{A} is devoted to many details concerning monopoles and dyons
in the Georgi-Glashow model, Appendix \ref{B} is a summary of the
$N=1$ superfield formalism and finally in Appendix \ref{C} we explicitly
construct the central charges of the supersymmetry algebra for $N=2$
super Yang-Mills.

\sect{Electromagnetic duality}
\label{dua}

The free Maxwell equations
\[
{\bf \nabla} \cdot {\bf E} = 0 \hspace{2cm} {\bf \nabla} \cdot {\bf B} = 0
\]
\beq
{\bf \nabla} \wedge {\bf E} + \frac{\partial {\bf B}}{\partial t}=0
\hspace{1cm}
{\bf \nabla} \wedge {\bf B} - \frac{\partial {\bf E}}{\partial t}=0
\label{Maxwell}
\eeq
are not only invariant under Lorentz and conformal transformations. They
are also invariant under a duality transformation :
\[
{\bf E} \rightarrow {\bf E} \cos \phi  - {\bf B} \sin \phi  
\]
\beq
\label{duality}
{\bf B} \rightarrow {\bf B} \cos \phi  + {\bf E} \sin \phi  
\eeq
In particular if we take $\phi = - \pi/2$ one obtains from eq. (\ref{duality})
a discrete duality transformation:
\beq
{\bf E} \rightarrow {\bf B} \hspace{2cm} {\bf B} \rightarrow - {\bf E}
\label{discdua}
\eeq
This transformation  is generated by the duality matrix  acting on the 
two-vector consisting of the electric and magnetic fields as
\beq
\left( \begin{array}{c} {\bf E} \\ 
			{\bf B} \end{array}  \right) 
\rightarrow 
\left( \begin{array}{cc} 0 & 1 \\ -1 & 0  \end{array}  \right)
\left( \begin{array}{c} {\bf E} \\ {\bf B} \end{array}  \right)
\label{discdua2}
\eeq

In terms of the  complex vector ${\bf E} + i {\bf B}$ the duality 
transformation in eq.(\ref{duality}) becomes
\beq
{\bf E} + i {\bf B} \rightarrow e^{i \phi} \left( {\bf E} + i {\bf B} \right)
\label{duacomple}
\eeq

Notice that the energy and momentum density of the electromagnetic field
given respectively by
\beq
\frac{1}{2} | {\bf E} + i {\bf B} |^2 = \frac{1}{2} \left({\bf E}^2 + {\bf B}^2 
\right)
\label{energyden}
\eeq
and
\beq
\frac{1}{2i} ({\bf E} + i {\bf B})^{*} \wedge ({\bf E} + i {\bf B}) =
{\bf E} \wedge {\bf B}
\label{momdens}
\eeq
are invariant under the duality transformation in eq.(\ref{duality}), while 
the real and imaginary part of 
\beq
\frac{1}{2} \left( {\bf E} + i {\bf B} \right)^2 = \frac{1}{2} \left( 
{\bf E}^2 - {\bf B}^2 \right) + i {\bf E} \cdot {\bf B}
\label{dou}
\eeq
that are respectively the Lagrangian of the electromagnetic field and the 
topological charge density, transform  under the duality group exactly as
the doublet $({\bf E}, {\bf B})$ in eq.(\ref{duality}), but with an angle 
equal to $2 \phi$.

If we perform a discrete duality transformations twice, 
we get
\beq
( {\bf E} , {\bf B} ) \rightarrow ( - {\bf E} , - {\bf B} ) 
\label{chconj}
\eeq
that corresponds to the charge conjugation operation.

The reason why this beautiful duality property of the free electromagnetic field
is not even mentioned in the courses on electromagnetism is due to the fact 
that it is lost when we introduce the interaction of the electromagnetic field
with matter by just adding in the right hand side of the Maxwell equations an
electric current ${\bf j}_e$ and an electric charge density $\rho_e$.
If we want to keep duality we must also introduce a magnetic current 
${\bf j}_{m}$ and a magnetic charge density $\rho_{m}$ together with their 
electric counterparts. If we do so the Maxwell equations given in 
eq. (\ref{Maxwell}) and written in complex notations become:
\beq
{\bf \nabla} \cdot ( {\bf E} + i {\bf B} ) = \rho_e + i \rho_m
\label{Maxcomple1}
\eeq
and
\beq
{\bf \nabla} \wedge ( {\bf E} + i {\bf B} ) = i \frac{\partial}{\partial t}
({\bf E} + i {\bf B} ) + i ({\bf j}_e + i {\bf j}_m)
\label{Maxcomple2}
\eeq

The previous equations are invariant under the duality transformation given in
eq.(\ref{duacomple}) if the electric and magnetic currents and densities 
transform as
\beq
\rho_e + i \rho_m \rightarrow e^{i \phi} ( \rho_e + i \rho_m )
\label{duacomple1}
\eeq
and
\beq
{\bf j}_e + i {\bf j}_m \rightarrow e^{i \phi} ( {\bf j}_e + i {\bf j}_m )
\label{duacomple2}
\eeq
 
In particular if we have only pointlike particles with both electric 
and magnetic charges $q$ and $g$ respectively, then duality implies the 
following transformation:
\beq
q+ ig \rightarrow e^{i \phi} ( q +i g )
\label{duacomple3}
\eeq

Particles with magnetic charge are not  introduced in usual electromagnetism
for the very simple reason that they are not observed in the experiments. If
we include them we must either think that their mass is higher than the 
presently available energy or find other reasons for their absence. However,
if we insist in preserving duality also in the presence of interaction, as 
shown by Dirac~\cite{DIRAC}, a theory with both electric and magnetic charges
$q_i$ and $g_j$ can be consistently quantized only if  the Dirac
quantization condition  is satisfied 
\beq
q_i g_j = 2 \pi \, \hbar \, n_{ij}
\label{dirqua}
\eeq
where $n_{ij}$ are arbitrary integers. We will derive the Dirac 
quantization condition in the next section.

The Dirac quantization condition is clearly not invariant under the duality 
transformation in eq.(\ref{duacomple3}). It is only invariant under the 
discrete transformation obtained from eq.(\ref{duacomple3}) for $\phi = - 
\frac{\pi}{2}$:
\beq
q \rightarrow g  \hspace{2cm} g \rightarrow - q
\label{disdua}
\eeq

\sect{The Dirac quantization condition}
\label{dirac}

In this section we show that in any consistent quantum theory containing both
electric and magnetic charges the Dirac quantization condition must be 
satisfied. 

A magnetic monopole located at the origin generates a magnetic field given
by:
\beq
{\bf B} = \frac{g }{4 \pi r^3}{\bf r}  \hspace{2cm} r = | {\bf r}|
\label{magne}
\eeq
The equation of motion of a particle with mass $m$ and charge $q$ moving in
the magnetic field ${\bf B}$ generated by the magnetic monopole is given by:
\beq
m {\ddot{{\bf r}}} =  q  {\dot{{\bf r}}} \wedge {\bf  B} 
\label{equamo}
\eeq
Using eqs.(\ref{magne}) and (\ref{equamo}) and the general relation valid
for three arbitrary vectors ${\bf A}$, ${\bf B}$ and ${\bf C}$
\beq
{\bf A} \wedge ( {\bf B} \wedge {\bf C} ) = {\bf B} ( {\bf A} \cdot {\bf C})
- {\bf C} ({\bf A} \cdot {\bf B})
\label{generel4}
\eeq
it is easy to see that
\beq
\frac{d}{dt} \left( {\bf r} \wedge m {\dot{\bf r}} \right) = 
{\bf r} \wedge m {\ddot{\bf r}} = \frac{qg}{4 \pi} \frac{d{\hat{\bf r}}}{dt}
\label{conse}
\eeq
where ${\hat{\bf r}} \equiv {\bf r}/r$.

The previous equation implies that the total angular momentum is conserved:
\beq
\frac{d}{dt} \left[  {\bf r} \wedge m {\dot{{\bf r}}} - \frac{qg}{4 \pi} 
{\hat{{\bf r}}} \right] =0
\label{totcon}
\eeq
The first term in the bracket is the angular momentum of the particle with 
mass $m$ and 
charge $q$, while the second term is the angular momentum of the 
electromagnetic field
generated by the electric and magnetic charges. In order to see this
let us compute the angular momentum of the electromagnetic field:
\[
{\bf J}^{(e.m.)} =  \int d^3 {\bf r}\, {\bf r} \wedge ( {\bf E} \wedge {\bf B} )
= \int d^3 {\bf r}\, \frac{g}{4 \pi r} \left({\bf E} - {\bf r} 
\frac{{\bf r} \cdot {\bf E}}{r^2} \right) =
\]
\beq
= \int d^3 {\bf r} \, {\bf E} \cdot {\bf \nabla}
\left( \frac{g {\hat{\bf r}}}{4 \pi} \right) = - \frac{qg}{4 \pi} 
{\hat{{\bf r}}}_p
\label{elect}
\eeq
where in the last step we have performed a partial integration and we have used
the equation:
\beq
{\bf \nabla} \cdot {\bf E} = q \delta^{(3)} ({\bf r} - {\bf r}_p)
\label{con}
\eeq
Eq.(\ref{elect}) shows that the second term in eq.(\ref{totcon}) is the 
angular momentum of the electromagnetic field.

In a quantum theory the projection of the angular momentum along a direction
is quantized. This implies that:
\beq
{\hat{\bf r}} \cdot {\bf J} = - \frac{qg}{4 \pi} = -\frac{1}{2} \hbar n
\label{quanti}
\eeq
that is the Dirac quantization condition in eq.(\ref{dirqua}).

Although the previous argument gives the correct result it is not very
convincing. In particular it does not explain why we should have half-integer
angular momentum without having fermions. Therefore in the following we give 
a more rigorous derivation of the Dirac quantization
condition. 

If we have a
magnetic monopole located at the origin, then the divergence of the
magnetic field is non zero and we cannot choose a vector potential that 
is regular everywhere and such that 
\beq
{\bf B} =  {\bf \nabla} \wedge {\bf A}
\label{vecpot}
\eeq

We can, however, introduce a vector potential ${\bf A}_N$ for the northern 
hemisphere and another one ${\bf A}_S$ for the southern hemisphere. In the
northern hemisphere we can compute the magnetic field using eq.(\ref{vecpot})
with ${\bf A} = {\bf A}_N$, while in the southern hemisphere the magnetic field
is computed again from eq.(\ref{vecpot}) but this time with 
${\bf A} = {\bf A}_S$. 
Along the equator the two vector potentials must match up to a gauge 
transformation:
\beq
{\bf A}_N = {\bf A}_S + {\bf \nabla} \chi (\theta)
\label{gautra}
\eeq
where $\theta$ is the angle around the equator.

Electrically charged particles are described by a wave function 
$\Psi_N (x)$ in the northern hemisphere and by another wave function 
$\Psi_S (x)$ in the
southern hemisphere. On the equator they must be equal up to a gauge 
transformation:
\beq
\Psi_N (x) = {\rm e}^{- i q \chi (\theta)/\hbar} \Psi_S (x)
\label{gautra2}
\eeq
where $q$ is the electric charge of the particle.

Since the wave function must be single valued when we go around the equator,
we must require that the parameter of the gauge transformation satisfies the
eq.
\beq
\chi ( \theta + 2 \pi)=  \chi (\theta ) + \frac{2 \pi \hbar}{q} n
\label{sin}
\eeq
where $n$ is an integer.

Let us now compute
\beq
\int_{eq.} d {\bf \ell} \cdot { \bf A}_N - \int_{eq.} d {\bf \ell} \cdot 
{\bf A}_S = \int_{0}^{2 \pi}
d \theta \frac{d  \chi}{d {\theta}} (\theta) = \chi(2 \pi) - \chi(0) = 
 \frac{2 \pi \hbar}{q} n
\label{linin}
\eeq
where the two integrals in the l.h.s. of the previous equation are performed 
along the equator and we have used eqs.(\ref{gautra}) and (\ref{sin}).

On the other hand the l.h.s. of eq.(\ref{linin}) can be rewritten by means
of the Stokes theorem as:
\beq
\int_{N} d{\bf S} \cdot {\bf B} + \int_S d {\bf S} \cdot {\bf B} = 
\int_{sphere} d{\bf S} \cdot {\bf B} = g
\label{stokes}
\eeq
where the integrals on the l.h.s. of the previous equation are performed
on the northern and southern hemisphere respectively and their sum is equal to
the integral over the entire sphere. In the last step in eq.(\ref{stokes})
we have used the fact that inside the sphere there is a magnetic monopole
with magnetic charge $g$. Finally comparing eqs.(\ref{linin}) and 
(\ref{stokes}) we get the Dirac quantization condition.

A consequence of the Dirac quantization condition is that both electric
and magnetic charges are quantized being multiples of the elementary magnetic
and electric charges $g_0$ and $q_0$~\cite{GO}:
\beq
q_i = n_i q_0 \hspace{2cm} g_j = n_j g_0
\label{quanti3}
\eeq
where $g_0$ and $q_0$ satisfy the relation
\beq
q_0 g_0 = 2 \pi \hbar n_0
\label{elequa4}
\eeq
The integer $n_0$ depends on the theory under consideration. Thus the Dirac
quantization condition provides an alternative mechanism, besides the one
of having the electric charge to be part of a non abelian group that is 
realized in grand unified theories, for the quantization of the electric
charge of the elementary particles that is a phenomenon largely observed in
experiments.

As noticed at the end of previous section the Dirac quantization condition 
is not duality invariant. In order to have duality invariance we must 
generalize it to particles having both electric and magnetic charges, called
dyons. If we have two dyons with electric and magnetic charges
equal respectively to $(q_i , g_i )$ and $( q_j , g_j )$   and if we go 
through
an argument as the one used in the beginning of this section we get the
following generalization of the Dirac quantization condition:
\beq
q_i g_j - q_j g_i = 2 \pi \hbar n
\label{dsz}
\eeq
that goes under the name of Dirac-Schwinger-Zwanziger~\cite{SCHWINGER,ZWANZI} 
(DSZ) quantization 
condition. In order to show its duality invariance we rewrite it as follows:
\beq
q_i g_j - q_j g_i = Q_{i}^{T} \Omega Q_j
\label{condi}
\eeq
where we have defined
\beq
Q_{i}^{T} = \left( \begin{array}{cc} q_i & g_i 
			 \end{array} \right)  \hspace{1cm}
\Omega = \left(
\begin{array}{cc} 0 & 1 \\
		 -1 & 0 \end{array} \right) \hspace{2cm}
Q_i = \left( \begin{array}{cc} q_i \\
			       g_i \end{array}  \right)
\label{mat}
\eeq
Under a duality transformation the vectors $Q$ transform as follows:
\beq
Q \rightarrow O Q \hspace{2cm} Q^T \rightarrow Q^T O^T \hspace{1cm}
O = \left( \begin{array}{cc} \cos \phi & - \sin \phi \\
			     \sin \phi & \cos \phi \end{array} \right)
\label{duatra}
\eeq
The invariance under duality transformation follows easily from the
identity:
\beq
O^T \Omega O = \Omega
\label{ooo}
\eeq

In these first few sections we have derived a number of consequences based 
on the existence of magnetic monopoles. But up to now there is no evidence
of their existence in nature. The Dirac quantization condition tells us,
because of relation (\ref{elequa4}) that in a theory in which the electric 
charge
$q_0$ is small , the magnetic charge $g_0$ is necessarily big. But those
considerations do not put any restriction on the mass of the monopoles. 
In the next section
we will see that magnetic monopoles naturally appear in gauge theories with
scalar Higgs particles transforming according to the adjoint representation
of the gauge group and that in these theories their mass is big when the
gauge coupling constant is small.

\sect{The 't Hooft-Polyakov monopole}
\label{mono}

In this section we will discuss the monopole solution found by {}'t 
Hooft~\cite{thooft} and Polyakov~\cite{Pol} in the Georgi-Glashow model. 
For more information about magnetic monopoles the reader is recommended 
to consult Ref.~\cite{GO}.
 
Let us consider the Georgi-Glashow model
\beq
{\cal{L}} = - \frac{1}{4} F_{a}^{\mu \nu} F_{a \,\, \mu \nu} + 
\frac{1}{2} (D_{\mu} \Phi )_a ( D^{\mu} \Phi )_a - V ( \Phi )
\label{ggmod}
\eeq
where the covariant derivative of the scalar Higgs field is given by
\beq
(D^{\mu} \Phi )_a  = \partial^{\mu} \Phi_a - e \epsilon_{abc} A^{\mu}_b \Phi_c
\label{covderi}
\eeq
and
\beq
F_{a}^{\mu \nu} = \partial^{\mu} A_{a}^{\nu} - \partial^{\nu} A_{a}^{\mu} -
e \epsilon_{abc} A_{b}^{\mu} A_{c}^{\nu}
\label{filstre}
\eeq
$\epsilon_{abc}$ is the Levi Civita tensor, because the gauge group
is $SU(2)$. For an arbitrary group one should substitute the Levi Civita
tensor with the structure constants $f_{abc}$ of the group.
In eqs.(\ref{covderi}) and (\ref{filstre}) we have used the generators of 
the gauge group in the adjoint representation that are given by:
\beq
(T_{a})_{bc} = - i \epsilon_{abc} \hspace{2cm} [ T_a , T_b ] = i 
\epsilon_{abc} T_c
\label{adgene}
\eeq
See Appendix A for the explicit expressions of the gauge transformations and
for a more detailed discussion of our notations.
  
The potential $V$ is equal to
\beq
V ( \Phi ) = \frac{\lambda}{4} \left( \Phi^2 - a^2 \right)^2
\label{pot}
\eeq
 
The classical equations of motion, that follow from ${\cal{L}}$, are 
\beq
\left( D_{\nu} F^{\mu \nu} \right)_a = - e \epsilon_{abc} \Phi_b \left( D^{\mu}
\Phi \right)_c
\label{equ1}
\eeq

\beq
\left( D_{\mu} D^{\mu} \Phi \right)_a = - \lambda \Phi_a ( \Phi^2 - a^2 )
\label{equ2}
\eeq
They must be considered together with the Bianchi identity

\beq
D_{\mu} {}^* F^{\mu \nu} =0  \hspace{2cm} {}^* F^{\mu \nu} \equiv \frac{1}{2}
\epsilon^{\mu \nu \rho \sigma} F_{\rho \sigma}
\label{BIA}
\eeq
$\epsilon^{\mu \nu \rho \sigma}$ is the antisymmetric Levi Civita tensor 
with $\epsilon^{0123} =1$. We use the metric $g_{\mu \nu} = (1, -1, -1, -1)$.

The energy is given by
\beq
E \equiv \int d^3 x \,\, \theta_{00} = \int d^3 x \left \{ \frac{1}{2} 
\left[ 
\left( B_{i}^{a} \right)^2  +  \left(  E_{i}^{a} \right)^2  + (\Pi^{a})^{2}
+  [\left( D_{i} \Phi \right)^{a}]^{2} \right] + V( \Phi ) 
\right\} 
\label{ene}
\eeq
where
\beq
\Pi_a = \left(  D^0 \Phi \right)_a \hspace{1cm} F_{a}^{i0} = E_{a}^{i}
\hspace{1cm} F_{a \,\, ij} = - \epsilon_{ijk} B_{a}^{k}
\label{em}
\eeq
with $\epsilon_{ijk} \equiv \epsilon^{0ijk}$ ($\epsilon_{123} =1$).

The energy is positive semi-definite. It vanishes if and only if
\beq
F^{\mu \nu}_a = \left( D^{\mu} \Phi \right)_a = 0 \hspace{1.5cm} V(\Phi ) =0
\label{vac}
\eeq
These conditions are satisfied by taking
\beq
\Phi_a = a \delta_{a3}  \hspace{2cm} A_{a}^{\mu} =0
\label{vac2}
\eeq
or equivalently any gauge rotated version of them.

This field configuration corresponds to the vacuum of our model and obviously
satisfies the equations of motions and the Bianchi identity in 
eqs.(\ref{equ1}),(\ref{equ2}) and (\ref{BIA}).

It is easy to see that, if $ a \neq 0$, the $SU(2)$ gauge group is broken to 
$U(1)$. With the v.e.v of
the Higgs field taken along the third direction ($\Phi_a = a \delta_{a3}$) 
the $U(1)$  gauge field $A^{\mu}_{3}$ remains massless, while the two charged
fields
\beq
W_{\pm} = \frac{1}{\sqrt{2}}\left( A^{\mu}_{1} \pm i A^{\mu}_{2}
\right)
\label{www}
\eeq
get a mass equal to
\beq
M_{W} = a |q| = a e \hbar
\label{mass}
\eeq
where $q$ is their electric charge.
They are charged with respect to the unbroken $U(1)$. Its generator $Q_e$ is
given by the generator of the $SU(2)$ gauge group that leaves invariant the 
v.e.v of the scalar field
\beq
Q_e = \frac{e}{a} T_a \Phi_a \hbar =  e T_3 \hbar 
\label{cha}
\eeq

Finally from the Higgs mechanism one gets also a neutral Higgs scalar particle
with mass equal to
\beq
M_H = \sqrt{2\lambda} \,\, a \,  \hbar
\label{Higmass}
\eeq
$\hbar$ has been explicitly written in some of the previous formulas,
while has been put equal to $1$ in most cases. We use also conventions
where the speed of light $c=1$.

In addition to the constant vacuum solution of eq.(\ref{vac2})
the equations of motion admit also static (time independent) solutions. The
simplest of them can be obtained starting with a radially symmetric ansatz:
\beq
\Phi_a = \frac{r^a}{e r^2} H(\xi) \hspace{1cm} A^{0}_{a} =0 \hspace{1cm}
A^{i}_{a} = - \epsilon_{aij}\frac{r^j}{e r^2} \left[1 - K (\xi) \right]  
\label{ans}
\eeq
where $\xi \equiv ae r$ is a dimensionless quantity.

Inserting this ansatz into the energy one gets:
\beqa
E  & = & \frac{4 \pi a}{e} \int_{0}^{\infty} \frac{d\xi}{\xi^2} \left[ \xi^2 
\left( \frac{dK}{d \xi} \right)^2  + K^2 H^2 + \right. \nl  
& + & \left. \frac{1}{2} \left( \xi \frac{dH}{d \xi}
-H \right)^2 + \frac{1}{2} \left( K^2 -1 \right)^2 +  
\frac{\lambda}{4 e^2} \left(H^2 - \xi^2 \right)^2  \right]
\label{ene2}
\eeqa
The insertion of the ansatz in the equations of 
motions (\ref{equ1}) and (\ref{equ2}) gives a system of coupled differential 
equations for the radial functions $H$ and $K$:
\beq
\xi^2 \frac{d^2 K}{d \xi^2} = KH^2 + K (K^2 -1)
\label{equ3}
\eeq
and
\beq
\xi^2 \frac{d^2 H}{d \xi^2} = 2 K^2 H + \frac{\lambda}{e^2} H ( H^2 - \xi^2 )
\label{equ4}
\eeq
They can also be obtained by minimizing the energy in eq.(\ref{ene2}).
In order to have a finite energy solution one must also impose
boundary conditions for both  $\xi=0$ and $\xi \rightarrow \infty$. They are
discussed in Appendix \ref{A}. It can be shown~\cite{GO} that the previous 
system of equations admits a finite energy solution.
However, in general, it is not possible to write it down explicitly 
unless one takes the parameter $\lambda$ of the potential of the Higgs field  
equal to $0$. This corresponds to the so called Bogomolny~\cite{BOG}, Prasad, 
Sommerfield~\cite{PS} (BPS) limit. In this limit one obtains:
\beq
K ( \xi ) = \frac{ \xi }{\sinh \xi} \hspace{2cm} 
H( \xi) = \frac{ \xi }{\tanh \xi } -1
\label{bpslim}
\eeq

In order to have a better understanding of this limiting case and to 
explicitly derive the solution in eq.(\ref{bpslim}) let us rewrite
the sum of the two terms appearing in the energy density that involve
the square of the non abelian magnetic field and the square of the space 
components of the covariant derivative of the Higgs field as follows
\beq
\left( B_{i}^{a} \right)^2  +  \left[ \left( D_{i} \Phi \right)^{a} 
\right]^{2} =
\left[   B_{i}^{a} \pm \left( D_{i} \Phi \right)^{a} \right]^2 \mp
2  B_{i}^{a} \left( D_{i} \Phi \right)^{a}
\label{ide}
\eeq

When we insert it in the energy (see eq.(\ref{ene})) we see that all terms
are positive except the last one in the r.h.s of eq.(\ref{ide}). We get 
therefore  a lower bound for the energy
\beq
E \geq \mp \int d^3 x   B_{i}^{a} \left( D_{i} \Phi \right)^{a}
\label{ine}
\eeq
that, after a partial integration and the use of the Bianchi identity in 
eq.(\ref{BIA}), becomes
\beq
E \geq \mp \int d^3 x \partial_i \left[ B_{i}^{a} \Phi^a \right]
\label{ine2}
\eeq

The equality sign is obtained if and only if the following equations are 
satisfied:
\beq
E_{a}^{i} = 0 \hspace{1cm}  \Pi_a = 0 \hspace{1cm}  \lambda =0
\hspace{1cm} B^{i}_{a} \pm \left( D^{i} \Phi \right)_{a} =0
\label{linequ}
\eeq
They are first order equations that imply  the second order
equations of motion (\ref{equ1}) and (\ref{equ2}). It may be confusing to 
allow for a non vanishing v.e.v. $a$ for the scalar field $\Phi$ in the BPS 
limit where we put $\lambda =0$. A better way of proceeding could be to start
with a small, but not vanishing value of $\lambda$, and only after we have 
obtained a non vanishing v.e.v. for $\Phi$ send $\lambda$ to zero. Another
possibility is to have a potential with flat directions that allows for a
non vanishing v.e.v. for $\Phi$, but it does not fix the value of $a$. This
last case is, in fact, what happens in supersymmetric theories as we will
see later on.

The ansatz in eq.(\ref{ans}) satisfies
the first two equations in  (\ref{linequ}). In order to find the 
functions $H$ and $K$, for the BPS limit ($\lambda =0$), one must impose the 
last equation in (\ref{linequ}) and show that one obtains the solution  in eq. 
(\ref{bpslim}). For the sake of simplicity we restrict ourselves to the case
of the monopole solution corresponding to the minus sign in the last eq. in 
(\ref{linequ}). In Appendix \ref{A} it is shown that the last equation in 
(\ref{linequ}) implies the following first order equations:
\beq
\xi K ' = - KH \hspace{2cm}  \xi H' = H +1 - K^2
\label{linequ2}
\eeq
whose solution with the boundary conditions for $\xi \rightarrow \infty$:
\beq
\lim_{\xi \to \infty} K( \xi) = 0 \hspace{2cm}
\lim_{\xi \to \infty} H ( \xi) = \xi
\label{bouncon}
\eeq
and with suitable boundary conditions for $\xi \rightarrow 0$, as discussed in
the Appendix \ref{A}, is given in eq.(\ref{bpslim}). The prime in 
eq.(\ref{linequ2}) means  derivative with respect to the argument. The boundary
conditions for $\xi \rightarrow 0$ and $ \infty$ are required in order to 
have a solution with finite energy (see eq.(\ref{ene2})). 

Inserting the static classical solution into the energy density given by the 
integrand in the r.h.s of eq.(\ref{ene}), one can see that it is concentrated
in a small region around the origin and goes to zero exponentially as $r$ goes
to infinity. 

We will see later on that in the quantum theory the classical solution 
corresponds to a new
particle of the spectrum that is an extended object (with size $\sim$ $1/a$) 
located in the region where the energy density is appreciably different from 
zero. 

We want to show now that the soliton solution given by the ansatz in 
eq.(\ref{ans}) is actually a magnetic monopole with respect to the unbroken
$U(1)$ group. 

It is easy to show that, in order to have a finite energy solution, for large 
enough values of $r$ the following equations must be satisfied:
\beq
D_{\mu} \Phi =0 \hspace{3cm}  \Phi^2 = a^2
\label{infi}
\eeq
apart from a small exponential correction. For instance from eq.(\ref{dfai}) 
it easy to see 
that the solution in eq.(\ref{bpslim}) satisfies eqs.(\ref{infi}).

Corrigan et al.~\cite{Corri} have shown that the most general solution of the 
previous equations corresponds to a vector field given by:
\beq
A^{\mu}_{a} = \frac{1}{a^2 e} \epsilon_{abc} \Phi_{b} \partial^{\mu} \Phi_{c}
+ \frac{1}{a} \Phi_{a} B^{\mu}
\label{solu}
\eeq
where $B^{\mu}$ is  arbitrary. The corresponding field strenght is entirely
in the direction of the Higgs field $\Phi^a$ and is equal to
\beq
F_{a}^{\mu \nu} = \frac{1}{a} \Phi_{a} F^{\mu \nu}
\label{fiestre}
\eeq
where
\beq
F^{\mu \nu} = \frac{1}{e a^3} \epsilon_{abc} \Phi_a \partial^{\mu} \Phi_b
\partial^{\nu} \Phi_c + \partial^{\mu} B^{\nu} - \partial^{\nu} B^{\mu}
\label{fiestre2}
\eeq

For large values of $r$ it satisfies the free Maxwell equations
\beq
\partial_{\mu} F^{\mu \nu} =0 \hspace{2cm}
\partial_{\mu} {}^{*}F^{\mu \nu} =0
\label{Maxeq}
\eeq
This follows from the fact that $F_{\mu \nu}^{a}$ in eq.(\ref{fiestre}) 
satisfies the eq. of motion (\ref{equ1}) and the Bianchi identity
in eq.(\ref{BIA}) together with the fact that eqs.(\ref{infi}) are valid for 
$r \rightarrow \infty$.

We see that outside the region where the extended particle is located
the non abelian field strenght is aligned along the direction of the Higgs
field $\Phi_a $ and is proportional to an abelian field strenght 
$F^{\mu \nu}$ that can be interpreted as the field strenght of the unbroken 
$U(1)$ electromagnetic.

From eq.(\ref{fiestre}) we can compute the non abelian magnetic field that 
is equal to
\beq
B_{i}^{a} =  \frac{1}{2} \epsilon_{ijk} F^{a}_{jk} =  \frac{1}{2 a^4 e } 
\Phi_{a} \epsilon_{ijk} \epsilon^{bcd} \Phi^b
\partial^{j} \Phi^{c} \partial^{k} \Phi^{d}
\label{nabemagn}
\eeq
where we have omitted the contribution of the last two terms in the r.h.s.
of eq.(\ref{fiestre2}), containing $B_{\mu}$, since they will drop out in
the calculation of the magnetic charge.
Inserting it in eq.(\ref{ine2}) one gets:
\beq
E \geq \pm g a   \hspace{2cm} g = - \frac{1}{a} \int d^3 x \partial_{i} 
\left[ B^{a}_{i} \Phi^a \right] = - \frac{4 \pi}{e} K 
\label{ine3}
\eeq
where the topological charge $K$ is given by   
\beq
K = \int d^3 x \;\; K_0 
\label{topcha}
\eeq
with
\beq
K_{\mu} = \frac{1}{8 \pi a^3} \epsilon_{\mu \nu \rho \sigma} \epsilon^{abc}
\partial^{\nu} \Phi^a \partial^{\rho} \Phi^b \partial^{\sigma} \Phi^c
\label{topcurr}
\eeq
We call it topological current because, unlike a Noether current, it is
conserved independently from the equations of motion as it can be trivially 
checked. It can also be seen (see Appendix \ref{A} and Ref.~\cite{GO}) that 
the topological charge
$K$ is an integer since it counts the number of times that the two-sphere, 
defined by the second equation in eq.(\ref{infi}), is covered when the 
two-sphere at infinity in space is covered once. The topological charge $K$,
that is an integer, should not be confused with the function $K (\xi)$
introduced in the ansatz (\ref{ans}).

To summarize we get
\beq
E \geq \pm a g 
\label{enemagn}
\eeq
where $g$ is the magnetic charge of the soliton solution that is 
obtained by integrating the equation:
\beq
\partial_i B_i =  \frac{4 \pi}{e} K_0
\label{divb}
\eeq
that follows from eq.(\ref{fiestre2}). One gets:
\beq
g =- \frac{1}{a} \int d^3 x \partial_i \left( \Phi_a B_{i}^{a} \right) =
- \int d^3 x \partial_i B_i = - \frac{4\pi}{e} K
\label{magncha}
\eeq

In the case of the static solution corresponding to the ansatz in 
eq.(\ref{ans}) it is easy to see that $K = \pm 1$ in such a way that $E =
a |g| $. This can be seen by observing that, using the Bogomolny eq. in 
(\ref{linequ}), one can rewrite $g$ as follows:
\beq
g = \pm \frac{1}{2a} \int d^3 x \partial_i \partial_i \left( \Phi^2 \right)
= \pm \frac{1}{2a} \int d^3 x \partial_i \partial_i 
\left(\frac{H^2}{e^2 r^2} \right) 
\label{g83}
\eeq

Rewriting the Laplacian in polar coordinates and inserting the explicit
expression for $H$ given in eq.(\ref{bpslim}) we get:
\beq
g = \pm \frac{4 \pi}{2 e} \int_{0}^{\infty} d \xi \frac{d}{d \xi} \left[
 \xi^2 \frac{d}{d \xi} \left(\frac{1}{\tanh \xi} - \frac{1}{\xi} \right)^2
\right] = \pm \frac{4 \pi}{e}
\label{g81}
\eeq
that is obtained from the contribution of the integrand at $\xi = \infty$.

The value obtained for the magnetic charge $g = \pm \frac{4 \pi}{e}$ is
consistent with the Dirac quantization condition (with n=2) given in eq.
(\ref{dirqua})
\beq
q g = 4 \pi \hbar
\label{dirac2}
\eeq
where $q$ is the charge of the $W$-boson given in eq. (\ref{cha}) for $T_3 
= \pm 1$. The fact 
that we obtain $n=2$ is a consequence of the fact that the gauge bosons 
transform according to the triplet representation of the gauge group $SU(2)$.
The value $n=1$ would have been obtained with matter fields transforming
according to the fundamental doublet representation being their electric
charge in this case quantized in terms of half-integers.

In conclusion we see that the Georgi-Glashow model does not contain only 
perturbative states as a photon, a massless Higgs field in the BPS limit
and a couple of charged bosons, all corresponding to the fields present in the 
Lagrangian in eq. (\ref{ggmod}) and having either a zero mass or a mass 
proportional to the gauge
coupling constant. It contains also additional particles that are soliton 
solutions of the classical equations of motion whose mass is instead 
proportional to the inverse of the gauge coupling constant as follows from 
eq.(\ref{ene2}) and therefore are very massive in the weak coupling limit 
(small $e$). In particular their mass in terms of the $W$ mass is given by
\beq
M_{sol} = \frac{4 \pi a}{e} =  4 \pi \frac{M_W}{e^2}
\label{solimass}
\eeq

The soliton solution following from the ansatz in eq.(\ref{ans}) has a 
non vanishing magnetic charge, but has zero electric charge. In order to also 
have a solution with a non vanishing electric charge~\cite{juzee} we must 
allow for a non vanishing electric potential of the type
\beq
A^{0}_{a} = \frac{r^a}{e r^2} J (\xi)
\label{dyonansa}
\eeq
instead of the vanishing ansatz given in eq.(\ref{ans}). In the case of the 
monopole solution the dimensionless parameter $\xi = ea r$ was introduced
using the dimensional parameter $a$ corresponding to the asymptotic value
for $r \rightarrow \infty$ of the Higgs field. For the dyon instead we 
introduce $ \xi = e{\hat{a}}r$ where ${\hat{a}}$ will be determined later 
on in terms of the asymptotic value $a$ of the Higgs field and the ratio
between the electric and magnetic charge of the dyon. With this ansatz 
the equations of 
motion in eqs. (\ref{equ3}) and (\ref{equ4}) are modified as follows:
\beq
\xi^2 \frac{d^2 K}{d \xi^2} = K \left[ K^2 + H^2 - J^2 -1 \right]
\label{equ3'}
\eeq
and
\beq
\xi^2 \frac{d^2 H}{d \xi^2} = 2 K^2 H + \frac{\lambda}{e^2} H ( H^2 - \xi^2 )
\hspace{1cm}  \xi^2 \frac{d^2 J}{d \xi^2} = 2 K^2 J
\label{equ4'}
\eeq
In the BPS limit where $\lambda=0$ one can obtain an analytical solution
given by:
\beq 
H ( \xi ) = \frac{1}{\cos \theta} \left[ \frac{\xi}{\tanh \xi} -1 \right]
\hspace{1cm}
K (\xi ) = \frac{\xi}{\sinh \xi} 
\label{dyonsol1}
\eeq
and
\beq
J ( \xi ) = \tan \theta \left[ \frac{\xi}{\tanh \xi} -1 \right]
\label{dyonsol2}
\eeq
where $\theta$ is an arbitrary constant.

Also in this case the explicit solution given in eqs.(\ref{dyonsol1}) and 
(\ref{dyonsol2}) can be found by minimizing the energy in eq.(\ref{ene}) as
we have done in the case of the monopole. In fact by using the identity:
\[
\left( B^{i}_{a} \right)^2  +  \left(  E^{i}_{a} \right)^2  + 
\left(  D^{i} \Phi \right)^{2}_{a} = 
\left[ B^{i}_{a} -  (D^{i} \Phi)_a \cos \theta \right]^2 +  
\]
\beq
+ \left[ E^{i}_{a} -  (D^{i} \Phi)_a \sin \theta \right]^2  +
2  E^{i}_{a}  (D^{i} \Phi)_a \sin \theta +
2  B^{i}_{a}  (D^{i} \Phi)_a \cos \theta
\label{ide2}
\eeq
from eq.(\ref{ene}) we get
\beq
E \geq   \sin \theta \int d^3 x E^{i}_{a}  (D^{i} \Phi)_a  +
\cos \theta \int d^3 x  B^{i}_{a}  (D^{i} \Phi)_a 
\label{ene4}
\eeq

The identity sign in the previous eq. holds if the following eqs. are 
satisfied:
\beq
\left( D^0 \Phi\right)_a =0 \hspace{2cm} V(\Phi) =0
\label{eqs2}
\eeq
together with
\beq
B^{i}_{a} -  (D^{i} \Phi)_a \cos \theta =0 \hspace{2cm}
E^{i}_{a} -  (D^{i} \Phi)_a \sin \theta =0 
\label{bpscon3}
\eeq

The Bianchi identity in eq.(\ref{BIA}) and the eq. of motion in (\ref{equ1})
imply
\beq
(D_i B_i )_a =0  \hspace{2cm} \Phi^a ( D_i E_i )_a =0
\label{con3}
\eeq
that allow to rewrite eq.(\ref{ene4}) as follows:

\beq
E \geq - a \left( \sin \theta q + \cos \theta g \right)
\label{ene6}
\eeq
where $q$ and $g$ are the electric and magnetic charges of the unbroken
$U(1)$ given by 
\beq
g = - \frac{1}{a} \int d^3 x \partial_i \left( \Phi^a B_{i}^{a} \right) 
\hspace{1cm}
q  = - \frac{1}{a}\int d^3 x \partial_i \left( \Phi^a E_{i}^{a} \right)
\label{elemacha}
\eeq
The dimensional parameter $a$ that we have introduced in eqs.(\ref{ene6})
and (\ref{elemacha}) is the v.e.v. of the Higgs field for 
$r \rightarrow \infty$, that is related to the parameter ${\hat{a}}$ used
for defining the dimensionless parameter $\xi$ by:
\beq
\Phi^a \rightarrow a \frac{r^a}{r}  \hspace{2cm} a = 
\frac{{\hat{a}}}{\cos \theta}
\label{vevphi}
\eeq
The relation between $a$ and ${\hat{a}}$ can be obtained by inserting the 
asymptotic behaviour for $r \rightarrow \infty$ of $H$ in eq.(\ref{dyonsol1}) 
in the ansatz for the Higgs field in eq.(\ref{ans}).

Let us now obtain directly the solution in eqs.(\ref{dyonsol1}) and 
(\ref{dyonsol2}) from the first order eqs.(\ref{eqs2}) and (\ref{bpscon3})
and show that the parameter $\theta$ is determined in terms of $q$ and
$g$.

The ansatz in eqs.(\ref{ans}) and (\ref{dyonansa}) satisfies the first
eq. in (\ref{eqs2}), while the second equation is satisfied by requiring
the vanishing of the coupling constant $\lambda =0$. Inserting then the
ansatz in the two eqs.(\ref{bpscon3}) one obtains 
(see eq.(\ref{linequa})) the two eqs.
\beq
\xi K' = - K {\hat{H}} \hspace{2cm} \xi {\hat{H}}' = 1 -K^2 +{\hat{H}}
\label{dieq2}
\eeq
where the new function ${\hat{H}}$ is related to the two functions $J(\xi)$
and $H(\xi)$ appearing in the ansatz through the following relations
\beq
{\hat{H}}(\xi) = \cos \theta H (\xi) \hspace{2cm} J(\xi) = \sin \theta H(\xi)
\label{rel8}
\eeq
From the solution of the eqs.(\ref{dieq2}) for $K$ and, in this case, for 
${\hat{H}}$ given in eq. (\ref{bpslim}), by means of eqs.(\ref{rel8}), one can 
immediately obtain $H$, $K$ and $J$ given in eqs. (\ref{dyonsol1})
and (\ref{dyonsol2}). Up to now $\theta$ is an arbitrary parameter. In the
following we show  that $\theta$ is 
determined by the ratio between the electric and magnetic charges of the 
dyon. In fact, the magnetic charge of the dyon can be computed from the 
first
eq.(\ref{elemacha}) starting from eqs.(\ref{infi}) and (\ref{solu}), 
proceeding as in the case of the monopole and therefore
obtaining the same result as before: $g = - \frac{4 \pi}{e}$. 
On the other hand the electric charge $q$ in the second eq.(\ref{elemacha}) 
can be 
computed in terms of the magnetic charge $g$ by using the eqs.(\ref{bpscon3}). One gets 
\beq
q = g \tan \theta 
\label{elecha}
\eeq
Inserting $\theta$ determined by the previous eq. in eq.(\ref{ene6}) one 
gets the mass of the dyon in terms of its electric and magnetic charges:
\beq
M = a \sqrt{q^2 + g^2}
\label{dyonmass2}
\eeq
This formula has been deduced for the dyon soliton solution in the BPS limit, 
but  it is actually valid for any 
particle of the spectrum. Notice also that it is invariant under the duality 
transformation in eq.(\ref{duacomple3}).

\sect{Collective coordinates}
\label{colleco}

In this section we proceed to the semiclassical quantization of the monopole
solution or more generally of any classical solution of the Bogomolny equation,
in the approximation in which we only allow for motions along the collective
coordinates of the classical solution. 

Let us consider a solution of the Bogomolny equation:
\beq
B_i = D_i \Phi \hspace{2cm}, \hspace{1cm} i=1,2,3.
\label{bogo}
\eeq
with a definite topological charge K and let us work in the temporal gauge 
$A_0 =0$. This is different from what we have done in the previous section 
when we 
constructed a classical solution corresponding to a dyon by allowing 
$A_0 \neq 0$. If we work in the temporal gauge $A_0 =0$ there is no static 
dyon solution. We will see that, in the temporal gauge, dyons will 
instead emerge  as time dependent solutions. 

Given a solution of eq.(\ref{bogo}) we can
usually find an entire family of solutions with the same energy.
The parameters labelling these different solutions are called collective 
coordinates or moduli and the space of the solutions with fixed energy or
with fixed topological charge is
called the moduli space of the solutions. For instance in the case of the
monopole solution we have always assumed that the monopole is located at
the origin, but, because of translational invariance, it could have been
located in any other point. This means that, starting from the monopole
solution found in Sect.\ref{mono}, we could make the substitution ${\bf r} 
\rightarrow
{\bf r} - {\bf R}$ obtaining another classical solution that depends on the 
arbitrary three-dimensional vector ${\bf R}$. The three coordinates of 
${\bf R}$ are the three translational collective coordinates.

Let us consider the Lagrangian
of the Georgi-Glashow model in the BPS limit (without the $\Phi^{4}$ 
potential). In the temporal gauge  it can be written in the form:
\beq
L = T - V
\label{geogla}
\eeq 
where the kinetic energy is given by
\beq
T = \frac{1}{2} \int d^3 x \left[ {\dot{A}}_{i}^{a} {\dot{A}}_{i}^{a} +
{\dot{\Phi}}^{a}{\dot{\Phi}}^{a} \right]
\label{kine}
\eeq
and the potential energy by
\beq
V= \frac{1}{2} \int d^3 x \left[ B_{i}^{a} B_{i}^{a} +
{D_{i}{\Phi}}^{a}{ D_{i} {\Phi}}^{a} \right]
\label{pote}
\eeq
The dot means derivative with respect to time.
For a solution of the Bogomolny eq. (\ref{bogo}) the potential term is
just proportional to the topological charge  $K$:
\beq
V= \frac{1}{2} \int d^3 x \left[ B_{i}^{a} -
{D_{i}{\Phi}}^{a} \right]^2  - ag  = \frac{4 \pi a}{e} K
\label{bogo2}
\eeq
The temporal gauge is preserved by the dynamics if we impose the Gauss's law 
(corresponding to the $A_0$ equation of motion) as a constraint between 
physical fields:
\beq
D_i {\dot{A}}_i + i e [\Phi, {\dot{\Phi}}] =0
\label{gauss}
\eeq
where for $\Phi$ and $A_{i}$ we use the matrix notation  discussed at the 
beginning of Appendix \ref{A}.

In the gauge $A_0 =0$ the configuration space of the fields is given by 
${\cal{C}}= {\cal{A}}/ {\cal{G}}$ where ${\cal{A}}= \{A_i (x) , \Phi (x) \}$
is the space of finite energy solutions and ${\cal{G}}$ is the set of all
"small" residual gauge transformations. 
Residual gauge transformations are those preserving the temporal gauge $A_0 =0$:
\beq
\delta A_0 = D_0 \epsilon =0 \Longrightarrow {\dot{\epsilon}}=0
\label{resi}
\eeq
They are time independent gauge transformations. A residual gauge 
transformation is 
said to be "small" when it goes to the identity at spatial infinity. Two 
finite energy solutions that are related by a "small" residual gauge 
transformation are equivalent and should not be counted twice in the 
configuration
space of all finite energy solutions. We distinguish "small" from "large"
gauge transformations because they have very different physical meaning.
A "small" gauge transformation connects two equivalent field configurations,
while a "large" gauge transformation corresponds to  a global symmetry, as 
the translational symmetry, and therefore two field configurations related by
a "large" gauge transformation are physically inequivalent. For each 
symmetry of the theory we have a collective coordinate. In the case of the
monopole three collective coordinates correspond to the possibility of
having a monopole located in any point of the three-dimensional space, while
a fourth collective coordinate is related to the possibility of transforming
the monopole solution by means of a "large" gauge transformation obtaining a
monopole with non zero electric charge, i.e. a dyon. Using semiclassical
quantization we will show that the electric charge of the dyon is quantized
as the one of the $W$-boson.

Defining the conjugate momenta:
\beq
(\Pi_{A})^{a}_{i} = {\dot{A}}_{i}^{a}  \hspace{2cm}
\Pi_{\Phi}^{a} = {\dot{\Phi}}^{a} 
\label{mom}
\eeq
the hamiltonian corresponding to the Lagrangian in eq.(\ref{geogla}) is 
given by
\beq
H = T + V
\label{hami}
\eeq
where now $T$ is given in terms of the conjugate momenta:
\beq
T = \frac{1}{2} \int d^3 x \left[ ({\Pi_{A}})_{i}^{a} ({\Pi_{A}})_{i}^{a} +
{\Pi_{\Phi}}^{a}{\Pi_{\Phi}}^{a} \right]
\label{kine2}
\eeq
It is convenient to group together the gauge field and the Higgs field
$A_{I} = ( A_i ,\Phi )$ by introducing a four-dimensional formalism where the 
index $I$ runs over the three-dimensional space index $i=1,2,3$ and over a 
fourth index corresponding to the Higgs field with the extra condition that
$A_{I}$ depends only on the first three coordinates: $\partial_4 A_{I} =0$.
In this notations both the kinetic energy in eq.(\ref{kine}):
\beq
T = \frac{1}{2} \int d^3 x  {\dot{A}}_{I}^{a} {\dot{A}}_{I}^{a} 
\label{kine3}
\eeq
and the Gauss's law in eq.(\ref{gauss})
\beq
D_I {\dot{A}}_I =0
\label{gauss2}
\eeq
can be rewritten in a more compact form.
 
The BPS monopoles are static solutions of the eqs. of motion that minimize 
the potential energy in eq.(\ref{pote}). Any motion, corresponding to a 
dependence on the time,
will give a contribution to the kinetic term increasing the energy of the
solution. On the other hand when we construct a quantum theory of monopoles
we must expand around the classical solutions by writing~\cite{OSBORN2}:
\beq
A_I (x) = A_{I}^{BPS} (x, z^{\alpha}) + \Delta A_{I} (x) 
\label{vari}
\eeq
where the variable $z^{\alpha}$ labels the moduli space of monopoles with a 
specific monopole charge and $\Delta A_I$  is the quantum field.
If, however, we restrict ourselves to motions with very small velocity and we
start the motion along the flat directions of the potential energy i.e. in the
moduli space of the static BPS monopoles, then the conservation of energy will
keep the monopole close to this space pretty much in the same way as for a 
point-particle slowly moving along the flat directions of a potential. For
very small velocity we can neglect oscillations along the transverse directions
as if we had an infinitely steep potential
and limit ourselves only to a motion in the moduli space. In this approximation
the expression in eq.(\ref{vari}) becomes:
\beq
A_{I} (x,t) = A_{I}^{BPS} ( x, z^{\alpha}(t) ) 
\label{modu}
\eeq
and the dependence on the time is only through the collective coordinates 
that we assume to vary with time. 

Since the quantity in eq.(\ref{modu}) is a solution of the BPS eq.(\ref{bogo})
for any value of $z^{\alpha}$, then the zero mode obtained by taking the
derivative of (\ref{modu}) with respect to the collective coordinate for small 
values of $z^{\alpha}$ satisfies the linearized Bogomolny equation. This
quantity is, however, in general not orthogonal to the gauge transformation,
i.e. it does not satisfy the Gauss's law in eq.(\ref{gauss2}). For this reason,
following Ref.~\cite{WEINBERG}, it is more convenient to start from a gauge 
rotated version of (\ref{modu}) given by
\beq
A_{I} (x,t) = g^{-1} (x,t) A_{I}^{BPS} (x, z^{\alpha} (t))  g (x,t) + 
\frac{1}{ie} g^{-1} (x,t) \partial_{I} g (x,t)
\label{modu2}
\eeq
and still keep $A_0 =0$. This means that the transformation in 
eq.(\ref{modu2}) is not a true gauge transformation because it acts only on
the space components of the vector potential and on the Higgs field, but it
leaves unchanged $A_0 =0$. 
We are in the temporal gauge and, as explained in Ref.~\cite{WEINBERG}, we are 
not going out of it by considering
time dependent gauge transformations. Eq.(\ref{modu2}) should rather be 
understood as if we have a family of static "gauge transformations" that are
different from a time to another. This means that the function in the l.h.s
of the previous equation does not only vary with the time through the time
dependence of $z^{\alpha}$, but also through the choice of a different "gauge
transformation". 

Let us now compute the electric field strenght in the gauge $A_0 =0$:
\beq
F_{0 I} = \partial_0 A_I = g^{-1} {\dot{z}}^{\alpha} 
\frac{\partial}{\partial z^{\alpha}} g + \frac{1}{ie} \partial_0 \left[ 
g^{-1} \partial_I g \right] + [g^{-1} A_{I}^{BPS} g , g^{-1} \partial_0 g ]
\label{ele}
\eeq
Using the equation
\beq
\partial_0 \left[g^{-1} \partial_I g  \right] = 
\partial_I \left[g^{-1} \partial_0 g  \right] +
\left[ g^{-1} \partial_I g , g^{-1} \partial_0 g \right] 
\label{equa}
\eeq
and remembering that the gauge functions are independent of $z^{\alpha}$
we get
\beq
F_{0 I} = {\dot{A}}_I = {\dot{z}}^{\alpha} \frac{\partial}{\partial 
z^{\alpha}} A_I (x,t) + \frac{1}{ie} D_I \left( g^{-1} \partial_0 g \right) 
\label{ele2}
\eeq
Introducing the quantity
\beq
\frac{1}{ie} g^{-1} \partial_0 g = - {\dot{z}}^{\alpha} \epsilon_{\alpha}
\label{gaugetra}
\eeq
and defining 
\beq
\delta_{\alpha} A_I =  \frac{\partial}{\partial z^{\alpha}} A_I (x,t)  - D_I  
\epsilon_{\alpha}
\label{defi}
\eeq
we can rewrite eq.(\ref{ele2}) as follows
\beq
{\dot{A}}_{I} =  {\dot{z}}^{\alpha} \delta_{\alpha} A_I 
\label{ele3}
\eeq
The parameter $\epsilon$ and correspondently the gauge transformation $g(x,t)$
is fixed by requiring that the 
quantity in eq.(\ref{ele3}) satisfies the Gauss law: 
\beq
D_I {\dot{A}}_I =   {\dot{z}}^{\alpha} D_I \left( \delta_{\alpha} A_I \right)
=0
\label{gaus2}
\eeq
An alternative but equivalent way of proceeding  is to start instead from the 
ansatz:
\beq
A_{I} (x, z^{\alpha} (t) ) = A_{I}^{BPS} (x, z^{\alpha} (t))  \hspace{2cm}
A_0 = {\dot{z}}^{\alpha}\epsilon_{\alpha}
\label{alte}
\eeq
and compute the electric field strenght obtaining
\beq
F_{0 I} = \partial_0 A_I - \partial_I A_0 + ie [A_0 , A_I ] =
{\dot{z}}^{\alpha} \delta_{\alpha} A_I 
\label{ele4}
\eeq
after the use of eq.(\ref{defi}). We want to stress that both eqs.(\ref{modu2})
and (\ref{alte}) do not perform a true gauge transformation because they never
transform simultaneoulsy both $A_I$ and $A_0$. In the first case only $A_I$ is
transformed, while in the second case only $A_0$.

Inserting the expression obtained in eq.(\ref{ele3}) in the kinetic term
(\ref{kine3}) we get
\beq
T = \frac{1}{2}  {\cal{G}}_{\alpha \beta} {\dot{z}}^{\alpha}
{\dot{z}}^{\beta}
\label{ene3}
\eeq
The total Lagrangian is then the one found by Manton~\cite{MANTON} and is 
given by:
\beq
L= \frac{1}{2}  {\cal{G}}_{\alpha \beta} {\dot{z}}^{\alpha}
{\dot{z}}^{\beta} - \frac{4 \pi a}{e} K
\label{ene5}
\eeq
that describes precisely the motion of a free particle propagating in the 
moduli space of the Bogomolny solutions with magnetic charge $K$ with metric:
\beq
{\cal{G}}_{\alpha \beta} (z) = \int d^3 x \left[ \delta_{\alpha} A_{i}^{a}
\delta_{\alpha} A_{i}^{a} + \delta_{\alpha} \Phi^{a} \delta_{\alpha} \Phi^{a}
\right]= \int d^3 x \delta_{\alpha} A_{I}^{a} \delta_{\alpha} A_{I}^{a}  
\label{met}
\eeq

Let us consider now the simplest case with $K=1$ corresponding to the
original 't Hooft-Polyakov solution. In this case there are four collective
coordinates. Three of them are due to translational invariance, as we have
already discussed, because the monopole can be centered around any point
of the three-dimensional space.
The time evolution of the position ${\bf R}$ is described by the Lagrangian
of a free particle with mass equal to $4\pi a/e$. Therefore the kinetic part 
of the Lagrangian 
corresponding to these three collective coordinates is equal to:
\beq
L = \frac{2 \pi a}{e} {\dot{\bf R}}^2
\label{lag}
\eeq
The fourth one is more subtle and requires some discussion. We have seen that
two field configurations that are related by a small gauge transformation 
(with gauge parameter $\epsilon$ that tent to zero at spatial infinity) are
equivalent. But let us consider now the possibility of a gauge transformation
whose parameter does not tent to zero at spatial infinity. As discussed above
it should not be considered as a gauge transformation that connects equivalent 
field configurations, but rather as a symmetry transformation as in the case 
of the
translational invariance considered above. In particular let us consider a 
time-dependent "gauge transformation" given by
\beq
A_{I} (t) = g^{-1} A_{I}^{BPS} g + \frac{1}{ie} g^{-1} \partial_I g
\hspace{2cm}
g = {\rm e}^{i \chi (t) \Phi/a}
\label{tdepgau}
\eeq
The dependence on time is in the parameter $\chi$ that satisfies the
condition $\chi (0) =0$. The previous transformation is a "large" gauge
transformation because it does not go to the identity at spatial infinity since
$\Phi^a \rightarrow a \frac{r^a}{r}$ as $ r \rightarrow \infty$.
From eq.(\ref{tdepgau}) we can easily compute
\beq
{\dot{A}}_{i} (t) = \frac{{\dot{\chi}}}{ea} D_{i} \Phi
\hspace{2cm}
{\dot{A}}_{4} (t) \equiv {\dot{\Phi}} (t) =0
\label{coderi}
\eeq
Inserting the previous eqs. in the kinetic term in eq.(\ref{kine3}) we get
\beq
T =   \frac{{\dot{\chi}}^2}{2 e^2 a^2}\int d^3 x 
\left(D_i \Phi \right)^a B_{i}^{a} =  
\frac{{\dot{\chi}}^2}{2 e^2 a^2} \frac{4 \pi a}{e} = \frac{2 \pi}{e^3 a}
 {\dot{\chi}}^2
\label{kine4}
\eeq
where in the first step we have used the Bogomolny eq. and in the second one
the fact that the three-dimensional integral gives the magnetic charge
of the monopole (see eq.(\ref{magncha})).

In conclusion we get the following action for the motion in the monopole 
moduli space:
\beq
S = \frac{1}{2} \int dt {\dot{z}}^{\alpha} {\dot{z}}^{\beta} 
{\cal{G}}_{\alpha \beta}
\label{actmodu}
\eeq
where
\beq
{\cal{G}}_{\alpha \beta} = \frac{4 \pi a}{e} \left( \begin{array}{cc}
			    \one_3 & 0 \\
			    0 &  1/(e^2 a^2) \end{array}    \right)
\label{metr4}
\eeq
In eq.(\ref{actmodu}) we have neglected the second term in the r.h.s. of 
eq.(\ref{ene5}) that is inessential being just a constant independent of 
$z^{\alpha}$.
However, while the three components of ${\bf R}$ are non compact variables, it turns out
that $\chi$ is a compact variables varying in the interval $(0, 2 \pi)$. In 
fact from the definition of the gauge transformation $g$ in eq.(\ref{tdepgau})
it is easy to get:
\beq
g( \chi + 2 \pi ) = g(\chi ) g (2 \pi )  \,\,\,;\,\,\, g(2 \pi ) 
= {\rm e}^{2 \pi i \Phi/a}  
\label{angu}
\eeq
Since for $ r \rightarrow \infty$ $g ( 2 \pi) \rightarrow 1$, $g (2 \pi)$ is 
a "small" gauge 
transformation. This means that $g( \chi + 2 \pi)$ and $ g (\chi)$ are related
by a "small" gauge transformation implying that $\chi$ is a compact variable:
\beq
\chi \sim \chi + 2 \pi
\label{angu4}
\eeq   
The moduli space action in eq.(\ref{actmodu}) is then equal to the action of 
a free particle moving in the manifold $R^3 \otimes S^1$ with flat metric.

Let us define the conjugate momenta to the variables ${\bf R }$ and $\chi$.
They are given by:
\beq
{\bf P} = \frac{4 \pi a}{e} {\dot{{\bf R}}} \hspace{2cm}
\Pi = \frac{4 \pi}{a e^3} {\dot{\chi}}
\label{conjmo}
\eeq
The hamiltonian can be easily computed to be:
\beq
H = \frac{e}{8 \pi a} {\bf P}^2 + \frac{a e^3}{8 \pi} \Pi^2 + \frac{4 \pi a}{e}
\label{hami7}
\eeq
Since $\chi$ is an angular variable the corresponding conjugate momentum $\Pi$
must be quantized: $ \Pi = n_e \hbar$, where $n_e$ is an integer.

The states of minimum energy are obtained by putting ${\bf P} =0$. In 
this case the hamiltonian becomes:
\beq
H = \frac{4 \pi a}{e} \left[1 + \frac{e^4}{32 \pi^2} ( n_e \hbar)^2 \right] 
\sim a \sqrt{\left(\frac{4 \pi}{e} \right)^2 + (n_e \hbar e )^2} = a 
\sqrt{g^2 + q^2} 
\label{hami5}
\eeq
where in the middle step we have used the fact that our calculations are 
valid for small $e^2$ and we have neglected higher orders in $e^2$ and
in the last step we have used the expressions for the electric and magnetic 
charges of a dyon. Actually there is no doubt about the expression of the 
magnetic charge given in eq.(\ref{magncha}). As far as the electric charge
is concerned we have still to show that the electric charge of a dyon is 
proportional to $\Pi$. This can be seen by
computing the electric field corresponding to the field configuration
given in eq.(\ref{coderi}). In the gauge $A_0 =0$ we get:
\beq
E_{i}^{a} = F^{a}_{i0} = - \frac{{\dot{\chi}}}{ea} (D_{i} \Phi)^a = - 
\frac{e^2}{4 \pi} \Pi B_{i}^{a}
\label{elefi}
\eeq
where we have used the Bogomolny eq.(\ref{bogo}) and the explicit expression 
for $\Pi$ given in eq.(\ref{conjmo}) in terms of ${\dot{\chi}}$. From 
eq.(\ref{elefi}) we can compute the electric charge:
\beq
- \int d^3 x \partial_{i} \left( E^{a}_{i} \Phi^{a}/a \right) =  
\frac{e^2}{4 \pi} \Pi \int d^3 x \partial_i \left( B^{a}_{i} \Phi^a /a 
\right) =  e \Pi
\label{elefi3}
\eeq
where in the last step we have used the explicit expression for the magnetic 
charge of a dyon. We have therefore shown that the electric charge of the dyon
is proportional to $\Pi$ that is the conjugate momentum corresponding to a
compact variable $\chi$. This means that, while at the classical level the
charge of the dyon can assume any real value, at the quantum level instead 
the electric charge of a dyon is quantized as also the charge of the elementary
$W^{\pm}$-bosons.

\sect{Instantons and Witten effect}
\label{insta}

We have seen that the monopole and more in general the dyon  classical 
solutions of the eqs. of motion of the Georgi-Glashow model correspond to
new particles of the spectrum that are not described by any of the fields
present in the Lagrangian. In this section we want to quickly describe
another kind of classical solution of the eqs. of motion called instantons.
While the monopole is a static classical solution of the Minkowski theory,
the instantons are classical solutions of the euclidean eqs. of motion and do not 
correspond to new particles of the spectrum. They should, instead, be
interpreted as field configurations contributing as saddle points to the 
euclidean functional integral. The simplest of these configuration was first
found~\cite{INSTA} in $SU(2)$ Yang-Mills theory using an argument that is the
generalization of the Bogomolny one used earlier for the
monopoles. One starts from the Yang-Mills euclidean action:
\beq
S_E = \frac{1}{4} \int d^4 x F_{\mu \nu}^{a} F^{\mu \nu}_{a}
\label{euclym}
\eeq
By rewriting it as 
\beq
S_E = \frac{1}{8} \int d^4 x \left\{ \left[ F^{a}_{\mu \nu} \pm {}^* 
F^{a}_{\mu \nu} \right]^2  \mp 2 F_{\mu \nu}^{a} {}^* F^{\mu \nu}_{a}
\right\}
\label{bogo4}
\eeq
one gets a lower bound for the action:
\beq
S_E \geq \mp \frac{1}{4} \int d^4 x F_{\mu \nu}^{a} {}^* F^{\mu \nu}_{a}
\label{lowbou}
\eeq
The equality sign holds if
\beq
F_{\mu \nu}^{a} \pm {}^* F^{a}_{\mu \nu} =0
\label{seldua}
\eeq
By defining the topological charge
\beq
Q =  \int d^4 x  q(x)  \hspace{2cm} 
q(x) = \frac{e^2}{32 \pi^2} F_{\mu \nu}^{a} {}^* F^{\mu \nu}_{a}
\label{topchaeu}
\eeq
that is an integer, since it is the topological number corresponding to the
mapping of the three-dimensional sphere into the gauge group $SU(2)$, one
can get the following expression for  the euclidean action if  eq. 
(\ref{seldua}) is satisfied 
\beq
S_E = \mp \frac{8 \pi^2}{e^2} Q
\label{eucliac3}
\eeq

The simplest solution to eq.(\ref{seldua}), corresponding to a value of the
topological charge $Q=1$, was found in Ref.~\cite{INSTA} and is given by:
\beq
i e A_{\mu} = \frac{x^2}{x^2 + \lambda^2} g^{-1} (x) \partial_{\mu} g (x)
\hspace{2cm} g (x) = \frac{ \one x_4 + i {\vec{\sigma}} \cdot 
{\vec{x}}}{\sqrt{x^2}}
\label{solu8}
\eeq 

It can be shown that the topological charge density $q(x)$ is a total
derivative, but, because of the existence of field configurations 
contributing to the  functional integral with integer non zero values of $Q$, 
one can 
introduce in $QCD$ or in any gauge theory a so called $\theta$-term by adding 
to the gauge action $S$ a term proportional to the topological charge
term:
\beq
S \rightarrow S + \theta Q 
\label{theta2}
\eeq
obtaining a theory that does not only depend on the gauge coupling constant $e$,
but also on the parameter $\theta$. If $\theta \neq 0$ $CP$ is not a symmetry.

Since $Q$ is an integer the physics is periodic
with period $2 \pi$. This means that every physical quantity depends on 
$\theta$ through a function of $\theta$ that is periodic with period $2 \pi$.
Actually, because of the axial $U(1)$ anomaly, in the case of $QCD$ in 
presence of at least one 
massless quark flavour one can always cancel the dependence on $\theta$. But,
if all quarks are massive, then there is an effective $\theta$ dependence.
It turns out that experiments require a very small value for the $\theta$-angle
in $QCD$. For a discussion of the physical consequences of the $\theta$-term
in $QCD$ see Ref.~\cite{DV}.

In the following we will introduce a $\theta$-term in the Georgi-Glashow
model studied in section (\ref{mono}) and, as a consequence of it, we will 
show that the electric charge of a
dyon gets an extra contribution. This effect of the
$\theta$-term is called Witten effect~\cite{WITTEN}. 

Let us consider a gauge rotation with a small angle $\varphi$ around the 
direction of the gauge field $\Phi^a$ with gauge parameter $\Lambda^a = 
\Phi^{a}/a$, where $\Phi^a$ is the Higgs field in the monopole background. 
At spatial infinity this is
a gauge transformation corresponding to the unbroken $U(1)$. Its generator
corresponds to the $U(1)$ electric charge defined in eq.(\ref{cha}). Under 
this transformation the 
Higgs field is, of course, left invariant while the vector potential gets
transformed as follows:
\beq
\delta A^{a}_{\mu} = -\frac{\varphi}{ea} (D_{\mu} \Phi )^a
\label{u1gau}
\eeq

The generator of this transformation is obtained from the Lagrangian of the 
Georgi-Glashow model in eq.(\ref{ggmod}) with the addition of the $\theta$ 
term:
\beq
{\cal{L}}_{GG} \rightarrow {\cal{L}}_{GG} + \theta Q
\label{ggtheta}
\eeq
and is given by:
\beq
N = \int d^3 x \left( \frac{\delta {\cal{L}}}{\delta ( \partial_0 A_{i}^{a})}
\frac{\delta A_{i}^{a}}{\varphi} \right)  
\label{N}
\eeq
Using the eqs.
\beq
\frac{\delta F^2}{\delta ( \partial_0 A_{i}^{a})} = - 4 E_{a}^{i}
\hspace{2cm}
\frac{\delta (F {}^* F) }{\delta ( \partial_0 A_{i}^{a})} = - 4 B_{a}^{i}
\label{e26}
\eeq
one gets:
\beq
N= - \frac{1}{ea} \int d^3 x \left\{ E^{i}_{a} (D_{i} \Phi )^a -
\frac{\theta e^2}{8 \pi^2} B^{i}_{a} (D_{i} \Phi )^a \right\}
\label{N0}
\eeq
Remembering the definition of the electric and magnetic charges in 
eq.(\ref{elemacha}) and using eq.(\ref{con3}) we get
\beq
N = - \frac{1}{e} \left(q - \frac{\theta e^2}{8 \pi^2} g \right)
\label{N1}
\eeq

Since $Q_{e}/e$ is an integer, as follows from eq.(\ref{cha}), the finite
$U(1)$ transformation generated by 
\beq
{\rm e}^{2 \pi i Q_{e}/e} =1
\label{111}
\eeq
must be equal to $1$. This implies that the same finite transformation
generated by $N$ must also be equal to $1$ and therefore also $N$ must be
an integer:
\beq
{\rm e}^{2 \pi i N} =1  \Longrightarrow N = - \frac{1}{e} 
\left(q - \frac{\theta e}{2 \pi} n_{m} \right) = - n_e
\label{inte3}
\eeq
where $n_e$ is an integer and we have used the Dirac quantization condition 
$e g =  4 \pi n_m$. We have not restricted ourselves to a monopole with
topological charge $K=1$, but we have allowed for any value of $K=n_m$. 

From eq.(\ref{inte3}) we get finally
\beq
q = n_e e + \frac{\theta e}{2 \pi} n_m
\label{gencha}
\eeq
showing that, in absence of the $\theta$-term the electric charge is
quantized in agreement with what obtained in eq.(\ref{hami5}), while, in 
presence of a  $\theta$-term, one gets an extra term
proportional to $\theta$ and to the magnetic charge of the dyon.

Since $\theta \sim \theta +2 \pi n$ eq.(\ref{gencha}) implies that, if a dyon
with a certain value of the electric charge $n_e$ exists, then dyons with any
integer value must exist. In conclusion the electric charge of the dyon is not
only quantized, but dyons with any integer value $n_e$ of the electric charge
must exist in the spectrum. This is an important consequence of $\theta$ 
periodicity. Notice also that the electric charges given by formula 
(\ref{gencha}) and the magnetic charges $g =  \frac{4 \pi}{e} n_m$ satisfy
the DSZ quantization condition (\ref{dsz}). Viceversa it can also be 
shown~\cite{OLIVE} that, if we require the DSZ quantization condition, the 
electric and magnetic charges of dyons must lie on a two-dimensional lattice
given by:
\beq
q +i g = q_0 \left( n_e + n_m \tau \right) \hspace{2cm} \tau = 
\frac{\theta}{2 \pi} + i \frac{2 \pi \hbar n_0}{q_{0}^{2}}
\label{chalatt}
\eeq
with lattice periods equal to $q_0$ and $q_0 \tau$. In the 
case of the 't Hooft-Polyakov monopole we have $n_0 =2$ as in any theory
containing no field  transforming according to the fundamental representation
of the gauge group.
  
In the last part of this section we rewrite the Georgi-Glashow model with
zero potential and with the addition of the $\theta$ term in a more compact 
way in terms
of gauge and Higgs fields normalized in such a way to include the gauge coupling
constant in them. This is the formalism that is currently used in the
Seiberg-Witten approach. In terms of the rescaled fields:
\beq
A_{\mu} \rightarrow e A_{\mu} \hspace{2cm} \Phi \rightarrow e \Phi
\label{nefi}
\eeq
and of the quantity
\beq
\tau = \frac{\theta}{2 \pi} + i \frac{4 \pi}{e^2}
\label{tau}
\eeq
one can rewrite the Lagrangian of the Georgi-Glashow model with the $\theta$
term as follows:
\beq
{\cal{L}} = - \frac{1}{16 \pi} Im \left[\tau \left( F^{a}_{\mu \nu} 
F_{a}^{\mu \nu} - i F^{a}_{\mu \nu} {}^* F_{a}^{\mu \nu}\right) \right] +
\frac{1}{2 e^2} (D_{\mu} \Phi )^2
\label{ggmod4}
\eeq
The topological charge in eq.(\ref{solu8}) becomes in this formalism:
\beq
q(x) = \frac{1}{32 \pi^2}F^{a}_{\mu \nu} {}^* F^{\mu \nu}_{a}
\label{solu9}
\eeq
If we call again $a$ the vacuum expectation value of the rescaled field $\Phi$
defined in eq. (\ref{nefi}) we have to rewrite the mass formula in 
eq.(\ref{dyonmass2}) as follows:
\beq
M = \frac{a}{e} | q + ig| = a | n_e + \tau n_m |
\label{nema}
\eeq

\sect{Montonen-Olive duality}
\label{MOdua}

Leaving aside for a moment the dyon solution discussed in the previous
sections we have 
found  that the semiclassical spectrum of the Georgi-Glashow model in the 
BPS limit consists of two neutral particles, a massless photon and a massless 
Higgs particle, of an electrically
charged $W$ boson with charge equal to $ q_0 = \pm e \hbar$ and of a 
magnetic monopole with magnetic charge equal to $g_0 = \pm \frac{4 \pi}{e} =
\frac{4 \pi \hbar}{q_0}$.

If there is duality invariance as suggested for instance by the formula in eq. 
(\ref{nema}) we can make a duality transformation
with angle $\phi = - \frac{\pi}{2}$ such that
\beq
q_0 \rightarrow g_0 \hspace{2cm} g_0 \rightarrow - q_0
\label{duality2}
\eeq
This transformation implies that
\beq
q_0 \rightarrow \frac{4 \pi \hbar }{q_0}
\label{duality3}
\eeq

Based on this observation Montonen and Olive~\cite{MONTOLI} suggested that 
there are
two equivalent formulations of the same theory dual to each other. In the 
first one, that we call electric, the $W$'s are elementary particles while
the magnetic monopoles are solitons. In the second one, that we call magnetic,
the elementary particles are instead the magnetic monopoles while the $W$ 
bosons are solitons. They
also suggested  that the two formulations had essentially the same Lagrangian.
The only important difference between them is that the
electric theory is weakly coupled when $q_0 \rightarrow 0$ ($e \rightarrow 0$)
while the magnetic theory is weakly coupled when $g_0 \rightarrow 0$ 
corresponding to $e \rightarrow \infty$. They brought the following arguments
in support of their duality conjecture.
 
\begin{enumerate}
\item{The mass formula in eq.(\ref{dyonmass2}), valid for all particles of the 
theory, is duality invariant.}
\item{Since there is no interaction between two monopoles, while there is a
non zero interaction between a monopole and an antimonopole, if duality is
correct, one must expect that the interaction between equal charge $W$-bosons
must be zero while that between  opposite charged $W$-bosons must be non 
vanishing.
This is actually verified in the BPS limit because in this limit the Higgs 
field is also
massless and  contributes with opposite sign with respect to the photon for 
equal charge 
$W$, while it contributes with the same  sign for opposite charge $W$.}
\end{enumerate}

The Montonen-Olive duality proposal, leaves, however some unanswered 
questions that we list:

\begin{enumerate}
\item{The elementary $W^{\pm}$ bosons have spin equal to $1$. If the 
magnetic monopoles are dual to them they must also have spin equal to $1$. 
But how can this happen?}

\item{The previous considerations are based on a mass formula that is only 
valid classically. How are the quantum corrections going to modify it?}

\item{In the previous considerations we have neglected the dyons. What is their 
role in the all picture?}
\end{enumerate}

The previous questions do not have an answer in the framework of the 
Georgi-Glashow model discussed in the previous section since the quantum 
Georgi-Glashow model is, actually, not duality invariant. But it was soon 
recognized~\cite{DADDA} that, in order to have a theory with Montonen-Olive 
duality, one
must include supersymmetry since in a supersymmetric theory the quantum 
corrections coming from the bosons and the fermions tend to cancel each others
preserving the structure of the classical mass formula and thus solving the 
second problem above. Actually the argument 
used  in Ref.~\cite{DADDA} for the $N=2$ super Yang-Mills theory is too naive 
and in fact wrong
as pointed out in Refs.~\cite{SCHONFELD,KAUL83,KAUL84,IMBIMBO84,IMBIMBO85} 
because this theory is not ultraviolet
finite. In order to have a classical mass formula that is not modified by 
quantum corrections one must consider the $N=4$ super Yang-Mills theory that 
is free from ultraviolet divergences~\cite{MANDE,BRINK,HOWE1,HOWE2,SOHNIUS}, 
as it was done by 
Osborn~\cite{OSBORN} who made also the important observation that in this case
magnetic monopoles and dyons have also supersymmetric partners with spin equal
to $1$. The introduction of the $N=4$ theory opens the way to the solution of
the first two puzzles discussed above. In the meantime Witten and 
Olive~\cite{WO} found
out that the structure of the duality invariant mass formula in 
eq.(\ref{nema}) for a BPS state in the $N=2$ theory is a direct consequence
of the supersymmetry algebra opening the way to the quantum exact determination
of the mass of the BPS states. This observation is  playing an essential
role also in recent developments in string theories.  

In conclusion it seems that the Montone-Olive duality can be realized in 
a four-dimensional gauge field theory provided that the theory is 
supersymmetric.
Therefore in the next section we discuss the representations of supersymmetry
algebra and in sections (\ref{supe}) and (\ref{susy}) we construct the 
Lagrangians of supersymmetric gauge theories in four dimensions.

\sect{Representations of supersymmetry algebra}
\label{repre}

As in the case of the Poincar{\'{e}} group the representations of the 
supersymmetry algebra for massive particles are different  from those
for massless particles. The supersymmetry algebra is given in both cases by:
\beq
\{ Q_{\alpha}^{i} , {\bar{Q}}_{\dot{\alpha}}^{j} \} = 2 
(\sigma_{\mu})_{\alpha {\dot{\alpha}}} P^{\mu} 
\delta^{ij}  \hspace{2cm} i,j=1 \dots N
\label{sualg}
\eeq
\beq
\{ Q_{\alpha}^{i} , Q_{\beta}^{j} \} =
\{ {\bar{Q}}_{\dot{\alpha}}^{i} , {\bar{Q}}_{\dot{\beta}}^{j} \} = 0
\label{sualg2}
\eeq
The difference between the two cases is due to the fact that in the massive 
case one can
always choose a center of mass frame where $P_{\mu} = ( M , \vec{0} )$, while 
this is not possible in the massless case. 

In the massive case in the center of mass frame one gets the following algebra:
\beq
\{ a^{i}_{\alpha} , (a^{j}_{\beta})^{\dagger} \} = \delta_{\alpha \beta} 
\delta^{ij}
\label{sualg3}
\eeq
and
\beq
\{ a^{i}_{\alpha} , a^{j}_{\beta} \} = 
\{ (a^{i}_{\alpha})^{\dagger}  , (a^{j}_{\beta})^{\dagger} \} = 0
\label{sualg4}
\eeq
where
\beq 
 a^{i}_{\alpha}  = \frac{1}{\sqrt{2 M}} Q_{\alpha}^{i} \hspace{1cm}
 (a^{j}_{\beta})^{\dagger}  = \frac{1}{\sqrt{2 M}} {\bar{Q}}^{j}_{\dot{\beta}}
\label{defi1}
\eeq

The representation of the fermionic harmonic oscillator algebra is constructed
starting from a vacuum state $|0>$ satisfying the equation:
\beq
a_{\alpha}^{i} | 0 > =0
\label{vac1}
\eeq
and acting on it with the creation operators:
\beq
\frac{1}{\sqrt{n!}} ( a_{\alpha_1}^{i_1} )^{\dagger} 
(a_{\alpha_2}^{i_2} )^{\dagger} 
\dots (a_{\alpha_n}^{i_n} )^{\dagger} |0 > \hspace{1cm} n = 0,1 \dots 2N 
\label{state1}
\eeq
The number of states  in eq. (\ref{state1}) is equal to
$ \left( \begin{array}{c} 2N \\
			   n  \end{array} \right)$.
Since $n$ runs from $0$ to $2N$
the total number of states in the representation of the massive supersymmetry
algebra is equal to:
\beq
\sum_{n=0}^{2N} \left( \begin{array}{c} 2N \\
			   n  \end{array} \right) = 2^{2N}
\label{totsta}
\eeq
The states in the representation have a maximum helicity gap $\Delta \lambda =
N$. Half of them are fermions and the other half are bosons.

In the massless case we can instead choose a frame where $P_{\mu} = (E,0,0,-E)$.
In this frame the supersymmetry algebra becomes:
\beq
\{ a^i , ( a^{j} )^{\dagger} \} = \delta^{ij}
\label{zerosal}
\eeq
\beq
\{ a^{i} , a^{j} \} = 
\{ (a^{i})^{\dagger}  , (a^{j})^{\dagger} \} = 0
\label{sualg5}
\eeq
where  
\beq 
 a^{i}  = \frac{1}{2 \sqrt{E}} Q_{1}^{i} \hspace{1cm}
 (a^{j})^{\dagger}  = \frac{1}{2 \sqrt{E}} {\bar{Q}}^{j}_{\dot{1}}
\label{defi2}
\eeq
The anticommutators involving the generators of the supersymmetry algebra with
indices $ \alpha =2$ and $ {\dot{\alpha}} = \dot{2}$ are all vanishing and
therefore they can be consistently put equal to zero:
\beq
Q^{i}_{2} = {\bar{Q}}^{i}_{\dot{2}} =0
\label{zecomm}
\eeq

Starting again from the vacuum state annihilated by the annihilation operators
$a^{i}$ we can construct the states of the representation acting on it
with the creation operators obtaining the state:
\beq
\frac{1}{\sqrt{n!}} ( a^{i_1} )^{\dagger} ( a^{i_2} )^{\dagger} 
\dots (a^{i_n} )^{\dagger} |0 > \hspace{2cm} n=0,1 \dots N 
\label{state2}
\eeq
that contains $ \left( \begin{array}{c} N \\
					n \end{array} \right)$ states. 
The total number of states in the massless representation is equal to:
\beq
\sum_{n=1}^{N}  \left( \begin{array}{c} N \\
					n \end{array} \right) = 2^{N}
\label{totsta2}
\eeq
that is smaller than in the case of a massive representation. The maximum 
helicity in this case is $ \Delta \lambda = N/2$. In the case $N=1$ one gets only 
one fermionic and one bosonic
state. In most cases, however, we must add another multiplet with opposite
helicity in order to have a $CPT$ invariant theory ($CPT$ reverses the sign of
helicity).

Let us finally consider the representation of the massive $N=2$ algebra with
non vanishing central charges~\cite{HAAG}. In this case in the center of mass 
frame the algebra is 
\beq
\{ Q_{\alpha}^{i} , {\bar{Q}}_{\dot{\alpha}}^{j} \} = 2 M 
\delta^{ij}  \delta_{\alpha {\dot{\alpha}}} 
\label{sualgcc}
\eeq
\beq
\{ Q_{\alpha}^{i} , Q_{\beta}^{j} \} = \epsilon^{ij} \epsilon_{\alpha \beta}
{\hat{Z}} \hspace{2cm}
\{ {\bar{Q}}_{\dot{\alpha}}^{i} , {\bar{Q}}_{\dot{\beta}}^{j} \} = 
\epsilon^{ij} \epsilon_{\dot{\alpha}  \dot{\beta}} \bar{{\hat{Z}}}
\label{sualgcc2}
\eeq
One can get rid of the phase in $\hat{Z}$ by a supercharge redefinition and
can rewrite the previous algebra in terms of the two quantities:
\beq
a_{\alpha} = \frac{1}{\sqrt{2}} \left[ Q^{1}_{\alpha} + \epsilon_{\alpha \beta}
{\bar{Q}}^{2}_{\beta} \right] \hspace{2cm}
b_{\alpha} = \frac{1}{\sqrt{2}} \left[ Q^{1}_{\alpha} - \epsilon_{\alpha \beta}
{\bar{Q}}^{2}_{\beta} \right] 
\label{defnew}
\eeq
obtaining
\beq
\{ a_{\alpha} , (a_{\beta} )^{\dagger} \} = ( 2 M + |{\hat{Z}}| ) 
\delta_{\alpha \beta}
\hspace{1cm}
\{ b_{\alpha} , (b_{\beta} )^{\dagger} \} = ( 2 M - |{\hat{Z}}| ) 
\delta_{\alpha \beta}
\label{defnew2}
\eeq
while all the other anticommutators are vanishing.

If $2M = |{\hat{Z}}|$ all anticommutators involving the oscillators $b$ are 
vanishing
and therefore we can put them equal to zero. We can then use only the 
oscillators $a$ for constructing the representation, obtaining the same number
of states as in the massless case. In the case $N=2$ here considered we get
the following four states:
\beq
|0>  \hspace{2cm} a^{\dagger}_{\alpha} | 0> \hspace{2cm} a^{\dagger}_{\alpha} 
a^{\dagger}_{\beta} | 0> 
\label{staneq2}
\eeq
instead of the $16$ states that we found in the case without central charge
(see eq. (\ref{totsta}) for $N=2$). 

Extending the previous procedure to the case $N=4$ we obtain a short 
representation 
with $16$ states instead of the one with $2^8 = 256$ states obtained without 
central charge (See eq. (\ref{totsta}) for $N=4$).

The fact that the representations of extended supersymmetry with non vanishing 
central charges are shorter and have the same dimension of those for the 
massless case makes it possible to have a consistent supersymmetric Higgs 
mechanism since in this case one has the
same number of degrees of freedom before and after the Higgs mechanism.
 
\sect{Supersymmetric Yang-Mills actions}
\label{supe}

In this section we construct the supersymmetric extension of Yang-Mills theory
in $D=4$ by dimensional reduction from higher dimensions~\cite{BSS}.

Let me start from the following action in $D$ dimensions
\beq
S = \int d^D x \left\{ - \frac{1}{4} F_{M N}^{a} F^{a \, M N} - 
\frac{i}{2} {\bar{\lambda}}^a \Gamma_{M} (D^{M} \lambda )^a \right\}
\label{susyd}
\eeq

If we perform the following supersymmetry transformation
\beq
\delta A_{M}^{a} =  \frac{i}{2} \left[ {\bar{\lambda}}^a \Gamma_{M} \alpha
- {\bar{\alpha}} \Gamma_{M} \lambda^{a} \right]
\label{sutra1}
\eeq
together with
\beq
\delta \lambda_a = \sigma_{R S} F_{a}^{RS} \alpha
\hspace{2cm}
\delta {\bar{\lambda}}_a = - {\bar{\alpha}} \sigma_{RS} F_{a}^{RS} 
\hspace{2cm} \sigma_{RS} = \frac{1}{4}[ \Gamma_{R}, \Gamma_S ]
\label{sutra2}
\eeq
it can be seen, by using the useful identities,
\beq
\Gamma^{M} \sigma^{R S} = \frac{1}{2} \left[ g^{M R} \Gamma^{S} - 
g^{M S} \Gamma^{R} - \frac{(-1)^{D/2}}{(D-3)!} \epsilon^{M R S N_1 \dots 
N_{D-3}} \Gamma_{D+1} \Gamma_{N_1} \dots \Gamma_{N_{D-3}}  
\right]
\label{ide1}
\eeq
and
\beq
\sigma^{R S} \Gamma^{M} = \frac{1}{2} \left[ - g^{M R} \Gamma^{S} + 
g^{M S} \Gamma^{R} - \frac{(-1)^{D/2}}{(D-3)!} \epsilon^{M R S N_1 \dots 
N_{D-3}} \Gamma_{D+1} \Gamma_{N_1} \dots \Gamma_{N_{D-3}}  
\right]
\label{ide3}
\eeq
that the term with the $\epsilon$ tensors cancel using the Bianchi identity
for $F_{\mu \nu}$, while the other terms are equal to
\[
\delta S = \int d^D x \left\{  \frac{i}{2}\partial_{M} \left(
{\bar{\alpha}}\sigma_{RS} F^{RS}_{a} \Gamma^{M} \lambda^{a} \right) 
+ \frac{i}{2}  F_{M N}^{a}\left[  {\bar{\alpha}} \gamma^{N} (D^{M} \lambda)^a
- (D^{M} {\bar{\lambda}})^a \Gamma^{N} \alpha \right] \right.
\]
\beq
\left. - \frac{i}{2} {\bar{\alpha}} (D^M F_{RM} )^a \Gamma^{R} \lambda^a 
- \frac{i}{2} {\bar{\lambda}}^a (D^M F_{MN} )^a \Gamma^{N} \alpha + 
i \frac{e}{2} {\bar{\lambda}}^a \Gamma^M f^{abc} \delta A_{M}^{b} \lambda^c
\right\} 
\label{equa39}
\eeq
From this eq. it follows that the action in eq. (\ref{susyd}) transforms as a 
total derivative 
\beq
\delta S = \frac{i}{2} \int d^{D} x \partial_{M} \left[  {\bar{\alpha}} 
\sigma_{R S}
F_{a}^{R S} \Gamma^{M} \lambda_{a} +  F_{a}^{M N} \left( {\bar{\alpha}} 
\Gamma_{N} \lambda^{a} - {\bar{\lambda}}^{a} \Gamma_{N} \alpha \right) 
\right]
\label{sustra}
\eeq
provided that the last term in eq.(\ref{equa39}) is vanishing
\beq
\left({\bar{\lambda}}^{a} \Gamma_{M} f^{abc} \lambda^{c}  \right)  
\left[ {\bar{\alpha}} 
\Gamma^{M} \lambda^{b} - {\bar{\lambda}}^{b} \Gamma^{M} \alpha \right] =0
\label{equasu}
\eeq
$\alpha$ and $\lambda$ are spinors in $D$-dimensions, $\Gamma_{D+1} =
\Gamma_0 \dots \Gamma_{D-1}$ and 
\beq
F^{a}_{M N} = \partial_{M} A_{N}^{a} - \partial_{N} A_{M}^{a} - e f^{abc} 
A_{M}^{b}
A_{N}^{c} \hspace{2cm} 
(D_{M} \lambda )^a = \partial_{M} \lambda^{a} - e f^{abc} A_{M}^{b}
\lambda^{c}
\label{defi3}
\eeq

Therefore the action in eq. (\ref{susyd}) is $N=1$ supersymmetric if the 
equation (\ref{equasu}) is satisfied. As shown in Ref.~\cite{BSS} this 
happens in the following cases:
\begin{enumerate}
\item{D=3, if $\lambda$ is a Majorana spinor }
\item{D=4, if $\lambda$ is a Majorana spinor }
\item{D=6, if $\lambda$ is a a Weyl spinor }
\item{D=10, if $\lambda$ is a Weyl-Majorana spinor}
\end{enumerate}
There is a simple way to understand this result by noticing that,
in all these cases, the number of on shell bosonic degrees of freedom, that is 
equal to $D-2$, is 
equal to the number of on shell fermionic degrees of freedom that is equal to 
$2^{[\frac{D}{2}]}$
multiplied with a factor $x =\frac{1}{2}$ if the spinor field $\lambda$ is a 
Majorana or Weyl spinor
and a factor $ x = \frac{1}{4}$ if the spinor field is a Weyl-Majorana spinor:
\beq
D-2 =  x \;\; 2^{[\frac{D}{2}]}
\label{degfre}
\eeq
where $[\frac{D}{2}]= \frac{D}{2}$ if $D$ is even and $[\frac{D}{2}] = 
\frac{D-1}{2}$ if $D$ is odd.

As a consequence of the invariance under supersymmetry one can construct
a supercurrent 
\beq
J^{M}_{A} = \sigma_{RS} F^{RS}_{a} \Gamma^{M} \lambda^{a}
\label{sucurre}
\eeq
that is conserved if the eqs. of motion are satisfied. 

In particular 
the action in eq. (\ref{susyd}) is $N=1$ supersymmetric if $D = 6 $ and $
10$. This fact can be used to write actions with extended $N=2,4$ 
supersymmetries in
four dimensions by the technique of dimensional reduction. Let us divide the
$D$ dimensional space-time component $x^{M} \equiv ( x^{\mu} , x^{i} )$ in a 
part $x^{\mu}$, where the index $\mu$ runs over the four-dimensional 
space-time, and in part $x^{i}$, where the index $i$ runs over the compactified
$D-4$ dimensions. We assume that the various fields are independent from
the compactified coordinates. 

Let us start to compactify the bosonic term in the action (\ref{susyd})
containing the non abelian field  strenght given in eq. (\ref{defi}). The
dimensional reduction of $F_{MN}$ gives respectively:
\beq
F^{a}_{\mu \nu } = \partial_{\mu} A_{\nu}^{a} - \partial_{\nu} A_{\mu}^{a} - 
e f^{abc} A_{\mu}^{b} A_{\nu}^{c}
\label{fmunu}
\eeq
\beq
F^{a}_{\mu i} = \partial_{\mu} A^{a}_{i} - e f^{abc} A_{\mu}^{b} A_{i}^{c}
\equiv \left( D_{\mu} A_{i} \right)^a
\label{fmui}
\eeq
and
\beq
F_{ij}^{a} = - e f^{abc} A_{i}^{b} A_{j}^{c}
\label{fij}
\eeq
Using the previous equations one obtains immediately the compactification
of the gauge kinetic term
\beq
- \frac{1}{4} F_{MN}^a F^{a \,MN} = - \frac{1}{4} F^{a}_{\mu \nu} F^{a \,
\mu \nu}
 + \frac{1}{2} \left( D_{\mu} A_{i} \right)^{a} \left( D^{\mu} A_{i} 
\right)^{a} - \frac{e^2}{4} f^{abc} A_{i}^{b} A_{j}^{c}
f^{ade} A_{i}^{d} A_{j}^{e}
\label{combos}
\eeq
where a sum over repeated indices is understood.

In order to perform the compactification of the fermionic term of the action
in eq. (\ref{susyd}) we have to distinguish the two cases $D=6$ and $D=10$.

A representation of the Dirac algebra for $D=6$ is given by:
\beq
\Gamma_{\mu} = \gamma_{\mu} \otimes 1  \hspace{1cm} \mu = 0,1,2,3 
\label{gamma6}
\eeq
\beq
\Gamma_{4} = \gamma_5 \otimes i \sigma_{1}  \hspace{1cm}
\Gamma_{5} = \gamma_5 \otimes i \sigma_{2} \hspace{1cm} \Gamma_7 \equiv
\Gamma_0 \Gamma_1 \Gamma_2 \Gamma_3 \Gamma_4 \Gamma_5 = \gamma_5
\otimes \sigma_3 
\label{gamma6b}
\eeq
where the $\sigma$-matrices are the Pauli matrices and $\gamma_5 = 
i \gamma^{0} \gamma^{1} \gamma^{2} \gamma^{3}$. 

A Weyl spinor in $D=6$ satisfies the condition:
\beq
( 1 + \Gamma_7 ) \lambda =0
\label{weylco}
\eeq
that is automatically satisfied if we take
\beq
\lambda = \left( \begin{array}{c} \frac{1 - \gamma_5 }{2} \chi \\ 
				  \frac{1 + \gamma_5 }{2} \chi
		    \end{array} \right)
\hspace{2cm}
{\bar{\lambda}} = \left( \begin{array}{cc} 
{\bar{\chi}} ( \frac{1+ \gamma_5}{2} ) & 
{\bar{\chi}} (\frac{1- \gamma_5}{2}) \end{array} \right)
\label{sol}
\eeq
where $\chi$ is a Dirac spinor in four dimensions.

Inserting it in the fermionic term in eq. (\ref{susyd}) one gets:
\beq
i \left({\bar{\lambda}}\right)^a \Gamma^{M} D_{M} \lambda^a = i 
\left( {\bar{\chi}} \right)^a \Gamma^{\mu} \left( D_{\mu}
\chi \right)^a - e f^{abc} {\bar{\chi}}^{a} A_{4}^{b} \gamma_5 \chi^{c} + ie 
f^{abc} {\bar{\chi}}^{a} A_{5}^{b} \chi^{c}   
\label{comfer}
\eeq

The Lagrangian of $N=2$ super Yang-Mills is obtained by summing the bosonic
contribution in eq. (\ref{combos}) with the indices $i,j=1,2$ to the
fermionic contribution in eq. (\ref{comfer}). One gets
\[
{\cal{L}}= - \frac{1}{4} F^{a}_{\mu \nu} F^{a \, \mu \nu}
 + \frac{1}{2} \sum_{i=1}^{2} \left( D_{\mu} A_{i} \right)^{a} 
\left( D^{\mu} A_{i} 
\right)^{a} - \frac{e^2}{2} f^{abc} A_{4}^{b} A_{5}^{c}
f^{ade} A_{4}^{d} A_{5}^{e} +
\] 
\beq
- i \left( {\bar{\chi}} \right)^a \gamma^{\mu} \left( D_{\mu}
\chi \right)^a  + e f^{abc} {\bar{\chi}}^{a} A_{4}^{b} \gamma_5 \chi^{c} - ie 
f^{abc} {\bar{\chi}}^{a} A_{5}^{b} \chi^{c} 
\label{neq2sYM}
\eeq
after a redefinition of the Dirac spinor $ \chi \rightarrow \sqrt{2} \chi$.

The $N=4$ super Yang-Mills is instead obtained starting with a Weyl-Majorana
spinor in $D=10$. In $D=10$ the Dirac algebra can be represented
as follows:
\beq
\Gamma^{\mu} = \gamma^{\mu} \otimes 1 \otimes \sigma_3 \hspace{1cm}
\mu=0,1,2,3
\label{gmu}
\eeq
\beq
\Gamma^{3+i} = 1 \otimes \alpha^i \otimes \sigma_1 \hspace{1cm}
\Gamma^{6+i} = \gamma_5 \otimes \beta^i \otimes \sigma_3 \hspace{1cm}
i=1,2,3
\label{gi}
\eeq
where  the fourdimensional internal 
matrices $\alpha$ and $\beta$ satisfy the following algebra:
\beq
\{\alpha^i , \alpha^j \} = \{\beta^i , \beta^j \} = -2 \delta^{ij} 
\hspace{1cm} [ \alpha^i , \beta^j ] =0
\label{con2}
\eeq
and
\beq
[ \alpha^i , \alpha^j ] = -2 \epsilon^{ijk} \alpha^{k} \hspace{2cm}
[ \beta^i , \beta^j ] = -2 \epsilon^{ijk} \beta^{k}
\label{excon}
\eeq
Finally the correspondent of $\gamma_5$ in ten dimensions is given by:
\beq
\Gamma_{11} = \Gamma_0 \Gamma_1 \Gamma_2 \Gamma_3 \Gamma_4 \Gamma_5 \Gamma_6
\Gamma_7 \Gamma_8 \Gamma_9 = 1 \otimes 1 \otimes \sigma_2
\label{chiope}
\eeq

A Weyl-Majorana spinor satisfying the condition:
\beq
( 1 + \Gamma_{11} ) \lambda =0
\label{Weyco}
\eeq
can always be written as 
\beq
\lambda = \psi \otimes \frac{1}{ \sqrt{2}} \left( \begin{array}{c}
							1 \\
							-i
				  \end{array} \right)
\hspace{1cm}
{\bar{\lambda}} = {\bar{\psi}} \frac{1}{\sqrt{2}} \left( \begin{array}{cc}
						    1 & -i 
						\end{array} \right)
\label{weyspi}
\eeq
where the Majorana spinor $\psi$ has a four dimensional space-time index 
on which the Dirac matrices act and another internal four dimensional index
on which instead the internal matrices $\alpha$ and $ \beta$ act. 

Proceeding as in the $N=2$ case we arrive at the $N=4$ super Yang-Mills
Lagrangian:
\[
{\cal{L}}=  - \frac{1}{4} F^{a}_{\mu \nu} F^{a \, \mu \nu}
 + \frac{1}{2} \sum_{i=1}^{3} \left( D_{\mu} A_{i} \right)^{a} 
\left( D^{\mu} A_{i} 
\right)^{a} + \frac{1}{2} \sum_{i=1}^{3} \left( D_{\mu} B_{i} \right)^{a} 
\left( D^{\mu} B_{i} \right)^{a} - V(A_i , B_j) + 
\]
\beq
- \frac{i}{2} \left( {\bar{\psi}} \right)^a \gamma^{\mu} \left( D_{\mu} \psi 
\right)^a - \frac{e}{2}  
f^{abc} {\bar{\psi}}^{a} \alpha^{i} A^{b \, i} \psi^{c} - i 
\frac{e}{2} 
f^{abc} {\bar{\psi}}^{a} \beta^{j} \gamma_5 B^{b \, j} \psi^{c} 
\label{neq4sYM}
\eeq
where the potential is equal to:
\beq
V( A_i , B_j ) = \frac{e^2}{4} f^{abc} A_{i}^{b} A_{j}^{c}
f^{afg} A_{i}^{f} A_{j}^{g} + \frac{e^2}{4} f^{abc} B_{i}^{b} B_{j}^{c}
f^{afg} B_{i}^{f} B_{j}^{g} + \frac{e^2}{2} f^{abc} A_{i}^{b} B_{j}^{c}
f^{afg} A_{i}^{f} B_{j}^{g}
\label{potn=4}
\eeq

\sect{Supersymmetric gauge theories for D=4}
\label{susy}

In this section we start rewriting the  Lagrangians for super Yang-Mills 
theories 
in four dimensions using a $N=1$ superfield formalism that we briefly 
review in Appendix B in order to fix the notations. We then extend them by
also including the matter superfields.

As one can see in Appendix B the most general $N=1$ renormalizable
supersymmetric gauge
theory involves two kinds of superfields: a number of the chiral superfields 
$\Phi_i$ that
describe the matter and the vector superfield $V$ that describes the gauge
part of the action. In terms of those superfields  the most
general renormalizable supersymmetric gauge theory contains essentially three
kinds of terms: a so-called $F$ term corresponding to the last component of a
chiral superfield describing the super Yang-Mills part of the Lagrangian, a
so-called $D$ term corresponding to the last component of a real superfield
describing the kinetic term of the matter together with its interaction with 
the gauge field and the gaugino and another $F$ term corresponding to the
superpotential that must be at most cubic in the matter fields. In 
conclusion the most general $N=1$ supersymmetric gauge theory is described
by the following Lagrangian
by:
\beq
{\cal{L}} =  \int d^2 \theta \, d^2 {\bar{\theta}} \, \sum_i 
{\bar{\Phi}}_{i} {\rm e}^{2V} \Phi_i +  
\int \left\{
d^2 \theta  \left[ - \frac{1}{4} W^{\alpha} W_{\alpha} + W(\Phi_i) \right] 
+ h.c. \right\}
\label{susyn=1}
\eeq

Let us rewrite now the Lagrangians of the various supersymmetric theories
with the $N=1$ superfield formalism. Let us start from those for the pure 
super Yang-Mills theories. They are the following:
\begin{enumerate}

\item{{\bf ${\bf N=1}$ super Yang-Mills}}

This theory involves only the superfield strenght $W_{\alpha}^{a}$ of a vector 
superfield $V^a$ and its Lagrangian is given by
\beq
{\cal{L}}= - \frac{i}{16 \pi} \int d^2 \theta \, \tau \, W^{\alpha} 
W_{\alpha} + h.c. = \frac{1}{8\pi} Im \left[\tau \int d^2 \theta W^{\alpha}
W_{\alpha} \right]
\label{susyn=15}
\eeq
In terms of component fields we get 
\beq
{\cal{L}} = - \frac{1}{4e^2} F^{a}_{\mu \nu} F^{\mu \nu}_{a} - \frac{i}{e^2} 
{\bar{\lambda}}^a {\bar{\sigma}}^{\mu} (D_{\mu} \lambda )^a + \frac{1}{2 e^2}
D^2 + \frac{\theta}{32 \pi^2} F^{a}_{\mu \nu} {}^* F^{\mu \nu}_{a}
\label{susycom}
\eeq

\item{{\bf ${\bf N=2}$ super Yang-Mills}}

The Lagrangian of $N=2$ super Yang-Mills contains together with the superfield
strenght $W_{\alpha}^{a}$ of a vector
superfield $V^{a}$, that is already present in the $N=1$ theory, also a chiral 
superfield 
$\Phi^a$ transforming according to the adjoint representation of the 
gauge group. The Lagrangian of this theory is given by:
\beq
{\cal{L}} = \frac{1}{e^2} \int d^2 \theta \, d^2 
{\bar{\theta}} \, {\bar{\Phi}}\, {\rm e}^{2V} \,
\Phi - \left[ \frac{i}{16 \pi} \int d^2 \theta \, \tau \, W^{\alpha} W_{\alpha}+
h.c. \right]
\label{n=2}
\eeq
Since the first term in eq.(\ref{n=2}) is real we can rewrite eq.(\ref{n=2})
as follows:
\beq
{\cal{L}}= \frac{1}{4 \pi} Im \left\{ \tau \left[ \shalf \int d^2 \theta \,
W^{\alpha} W_{\alpha}  + \int d^2 \theta d^2 {\bar{\theta}}\, {\bar{\Phi}} 
{\rm e}^{2V} \Phi  \right] \right\}
\label{n=23}
\eeq
where $\tau$ is given in eq.(\ref{tau}).
By expanding the superfield $\Phi$ in terms of the component fields:
\beq
\Phi = \phi + \sqrt{2} \theta \psi +  \theta^2 F
\label{exp4}
\eeq
and rewriting the Lagrangian in eq.(\ref{n=23}) in terms of Dirac spinors we get
the same Lagrangian as in eq.(\ref{neq2sYM}) that we rewrite with all fields 
rescaled by the gauge coupling constant $e$ and with the addition of the 
$\theta$ term 
\[
e^2 {\cal{L}}= - \frac{1}{4} F_{\mu \nu}^{a} F_{a}^{\mu \nu} + 
\left({\overline{D_{\mu} \phi}} \right)^a \left( D^{\mu} \phi \right)^a +
\shalf \left( f^{abc} {\bar{\phi}}_b \phi_c \right)^2 
+ \frac{\theta e^2}{32 \pi^2} F_{\mu \nu}^{a} {}^* F^{\mu \nu}_{a} +
\]
\beq
- i {\bar{\chi}}^a \gamma^{\mu} (D_{\mu} \chi)^a - i \sqrt{2} f^{abc}
\left[ {\bar{\chi}}^a \frac{1+\gamma_5}{2} \phi^b \chi^c + 
{\bar{\chi}}^a \frac{1-\gamma_5}{2} {\bar{\phi}}^b \chi^c \right]
\label{compola}
\eeq
provided that we make the following identifications:
\beq
\phi = \frac{A_5 + i A_4}{\sqrt{2}} \hspace{2cm} \chi = \left( 
    \begin{array}{c} \psi_{\alpha} \\
		   - i {\bar{\lambda}}^{{\dot{\alpha}}} \end{array}  \right)
\hspace{2cm}  {\bar{\chi}} = \left( 
    \begin{array}{cc} i \lambda^{\alpha} &
		     {\bar{\psi}}_{{\dot{\alpha}}} \end{array}  \right)
\label{identi5}
\eeq

\item{{\bf ${\bf N=4}$ super Yang-Mills}}

The Lagrangian of $N=4$ super Yang-Mills contains, in addition to the vector
superfield $V$ as before, also three chiral superfields $\Phi_i$ 
transforming according
to the adjoint representation of the gauge group. It is given by\footnote{
I thank Kim Splittorff for helping me in deriving eq. (\ref{n=4}).} :

\[
{\cal{L}} = \frac{1}{e^2} \int d^2 \theta d^2 {\bar{\theta}}
\sum_{i=1}^{3} {\bar{\Phi}}_{i} {\rm e}^{2V} \Phi_{i} + \frac{1}{8 \pi} Im 
\left[ \int d^2 \theta \, \tau W^{\alpha} W_{\alpha} \right]
\]
\beq
- \left[ \int d^2 \theta  \sqrt{2} \Phi_1 \Phi_2 \Phi_3  + h.c. \right]
\label{n=4}
\eeq
\end{enumerate}

If we introduce also matter superfields $Q$ and ${\tilde{Q}}$ we can write 
the  supersymmetric versions
of $QCD$ that are described by the following two Lagrangians.
\begin{enumerate}

\item{{\bf ${\bf N=1}$ super QCD}}

\[
{\cal{L}} = \int d^2 \theta d^2 {\bar{\theta}}\left[{\bar{Q}} {\rm e }^{2V}
Q + {\tilde{Q}} {\rm e}^{-2V} {\bar{\tilde{Q}}} \right] +
\]
\beq
+ \frac{1}{8 \pi} Im \left[ \int d^2 \theta \, \tau W^{\alpha} W_{\alpha}
\right] + \left[ \int d^2 \theta\, m_f {\tilde{Q}}_f Q_f +  h.c \right]
\label{sqcd1}
\eeq

\item{{\bf ${\bf N=2}$ super QCD}}

\[
{\cal{L}} = \int d^2 \theta d^2 {\bar{\theta}}\left[{\bar{Q}} {\rm e }^{2V}
Q + {\tilde{Q}} {\rm e}^{-2V} {\bar{\tilde{Q}}} \right] + 
\frac{1}{e^2} \int d^2 \theta d^2 {\bar{\theta}}{\bar{\Phi}}{\rm e}^{2V}
\Phi
\]
\beq
+ \frac{1}{8 \pi} Im \left[ \int d^2 \theta \tau W^{\alpha} W_{\alpha} \right]
 + \left\{ \int d^2 \theta
\left[ \sqrt{2} {\tilde{Q}} \Phi Q + m_f {\tilde{Q}}_f Q_f  \right] + h.c. 
\right\} 
\label{sqcd2}
\eeq
\end{enumerate}

\sect{Semiclassical analysis of super $N=2$ Yang-Mills theory}
\label{semi}

The structure of the bosonic part of the Lagrangian in eq.(\ref{compola}) is
pretty much the same as the one of the Georgi-Glashow model in 
eq.(\ref{ggmod}). There are, however, few important differences. The first is the
presence of the fermion fields. The second one is that, unlike the 
Georgi-Glashow model, here the 
potential given in the case of a $SU(2)$ gauge group
by
\beq
\label{pot8}
V ( \phi ) = -\frac{1}{2e^2} [ \epsilon^{abc} \phi^b {\bar{\phi}}^c ]^2 
\eeq
does not fix uniquely the vacuum. In fact any field configuration of the type
\beq
\label{pot9}
\phi^a = ( 0, 0 , a ) = a \delta^{a3}
\eeq
corresponds to a minimum of the potential with vanishing value (since 
supersymmetry is not broken) for any complex number $a$.
The set of all values of $a$ is called the classical moduli space of
the theory. Actually a better parametrization of the vacua is  given in terms 
of the gauge invariant variable $u = \frac{1}{2} a^2 = Tr (\phi^2) $. 
A third difference with respect to the Georgi-Glashow model is that, because 
of the particular structure of the potential, in the supersymmetric case the 
BPS limit is obtained without needing to send to zero any piece of the potential
as it was instead necessary in the Georgi-Glashow model.

If $a \neq 0$, as in the Georgi-Glashow model, the $SU(2)$ gauge symmetry is 
broken to $U(1)$ by the supersymmetric Higgs phenomenon and the charged (with 
respect to the unbroken $U(1)$) components of the gauge fields $W^{\pm}$ 
together with the charged gauginos get a non  vanishing
mass, while the gauge field of the unbroken $U(1)$ remains 
massless together with its supersymmetric partners, a Dirac photino and a 
complex Higgs field. The massless fields
belong to a massless $N=2$ chiral supermultiplet.

Let us now list the symmetries of the Lagrangian in eq.(\ref{n=2}). 
\begin{enumerate}
\item{There is a $U(1)_J$ symmetry corresponding to the following superfield
transformations:
\beq
\Phi (\theta) \rightarrow \Phi ( {\rm e}^{-i \alpha} \theta) \hspace{1cm}
W_{\alpha} (\theta) \rightarrow {\rm e}^{i\alpha} W_{\alpha} 
( {\rm e}^{-i \alpha} \theta)
\label{u1j}
\eeq
This symmetry is actually  part of an $SU(2)_J$ that is, however, not manifest 
because we are using an $N=1$ superfield formalism in which also the second 
supersymmetry is not manifest. In order to have both the second supersymmetry
and the entire $SU(2)_J$ manifest we must use $N=2$ superfields.}
\item{The Lagrangian in eq.(\ref{n=2}) is also invariant under the 
$U(1)_R$ transformations given by:
\beq
\Phi (\theta) \rightarrow {\rm e}^{2i \beta} \Phi ( {\rm e}^{-i \beta} \theta)
\hspace{1cm}
W_{\alpha} (\theta) \rightarrow {\rm e}^{i\beta} W_{\alpha} 
( {\rm e}^{-i \beta} \theta)
\label{u1r}
\eeq
}
\end{enumerate}
In the quantum theory the $U(1)_R$ symmetry is broken by an anomaly. The 
corresponding Noether current satisfies the anomaly equation:
\beq
\partial_{\mu} J_{R}^{\mu} = 4 N_c q(x) \hspace{2cm} q(x) = \frac{1}{32 \pi^2}
F^{a}_{\mu \nu}  {}^{*} F_{a}^{\mu \nu} 
\label{u1rano}
\eeq
where $q(x)$ is the topological charge density and $N_c$ is the number of 
colours. We will be mainly considering the case $N_c =2$, but in many cases 
we will be writing formulas valid for an $SU(N_c )$ colour group.

There is, however, a subgroup $Z_{4 N_c}$ of $U(1)_R$ that is not anomalous.
It acts on the scalar field $\phi$ of the chiral superfield $\Phi$ as
\beq
\phi \rightarrow {\rm e}^{\frac{i \pi n}{N_c}} \phi  
\hspace{2cm} n=1 \dots 4N_c
\label{nanosgo}
\eeq
In the case of an $SU(2)$ gauge theory it acts on $\phi$ as a $Z_4$ and on 
$u = Tr (\phi^2)$ as a $Z_2$ transformation:
\beq
\phi \rightarrow {\rm e}^{i \frac{\pi}{2} n} \phi \hspace{2cm} u \rightarrow - u
\label{su2ano}
\eeq
An anomalous $U(1)_R$ transformation has the 
effect of modifying the $\theta$ angle by:
\beq
\theta \rightarrow \theta - 4 \beta N_c
\label{modtheta}
\eeq
However if we perform an anomalous $U(1)_R$ transformation together with an 
opposite shift of the $\theta$ angle
\beq
\theta \rightarrow \theta + 4 \beta N_c
\label{resy}
\eeq
then this is a "symmetry" of the theory.
This is not what is usually called a symmetry in
field theory because we transform not only the fields but also the parameters
appearing in the Lagrangian. If, however, as it happens in string theory, we
consider the parameters as related to the v.e.v. of some additional field and 
we insert those fields instead of their v.e.v. in the Lagrangian, then the 
previously discussed "symmetry" becomes a true field theory symmetry. Such a
symmetry, that is a symmetry of the microscopic Lagrangian, can be used, for
instance, for putting constraints on the construction of a low energy effective 
Lagrangian for the light degrees of freedom of the theory. This is an 
important observation that we will use later on.

The $N=2$ super Yang-Mills theory is an asymptotic free theory whose 
$\beta$-function gets, in perturbation theory, only a non vanishing 
contribution from one-loop diagrams~\cite{HSW,GRISARU}. 
As it follows from the last part of Appendix \ref{B} the $\beta$-function is 
given by:
\beq
\mu \frac{\partial e}{\partial \mu} \equiv \beta (e) = - 
\frac{b_0}{(4 \pi)^2} e^3 \hspace{2cm} b_0 = 2 N_c
\label{betafu}
\eeq
Integrating it one can compute the dependence of the  coupling constant in 
going from the scale $M$ to a scale $Q$ 
\beq
\frac{4 \pi}{e^2 (M)} + \frac{ N_c}{ \pi} \log \frac{Q}{M} \equiv
\frac{4 \pi}{e^2 (Q)}
\label{runcou}
\eeq
The renormalization invariant parameter $\Lambda$ is determined from the 
previous equation as the value of $Q$ such that the coupling constant is
divergent $ \lim_{Q \to \Lambda} e (Q) = \infty$. One gets:
\beq
\label{lambda}
\Lambda = M \, e^{- \frac{4 \pi^2}{N_c e^2 (M)}}
\eeq
In terms of $\Lambda$ one can rewrite eq.(\ref{runcou}) as follows
\beq
\label{run}
\frac{e^2 (Q)}{ 4 \pi} = \frac{2 \pi}{N_c \log \frac{Q^2}{ \Lambda^2}}
\eeq
showing that, because of asymptotic freedom, perturbation theory is good when
$Q$ is large.

If we introduce in eq.(\ref{lambda}) also the $\theta$ angle and we restrict 
ourselves to the case $N_c =2$ we get
\beq
\Lambda = M {\rm e}^{i \pi \tau/2} \hspace{2cm} \tau = \frac{\theta}{2 \pi}
+i \frac{4 \pi}{e^2}
\label{lambda2}
\eeq
Therefore under the shift in eq.(\ref{resy}) for $N_c =2$ we get
\beq
\Lambda \rightarrow \Lambda {\rm e}^{2 i \beta}
\label{lambdatra}
\eeq
If we compare this equation with eq.(\ref{u1r}) we see that the scalar field 
$\phi$ transforms
under the anomalous $U(1)_R$ with the weight equal to $2$ precisely as 
$\Lambda$ under the transformation in eq.(\ref{resy}). This implies that 
under an anomalous $U(1)_R$ transformation combined with a shift of $\theta$
as given in eq.(\ref{resy}) the ratio $\Lambda^{2}/\phi^{2}$ will stay 
invariant. 

If we are interested in studying the 
low-energy dynamics of the $N=2$ super Yang-Mills we can restrict ourselves 
to the massless fields and we can limit ourselves to a
Lagrangian with at most two derivatives and with no more than four-fermion 
couplings. $N=2$ supersymmetry fixes completely its form in terms of a unique
function ${\cal{F}}$. The most compact form of this low energy Lagrangian can
be obtained by using the $N=2$ chiral superfield $\Psi$, that is a function of
four variables $\theta$ without depending on their complex conjugate 
variables ${\bar{\theta}}$.
In the manifest $N=2$ superfield formalism the most general low energy 
Lagrangian has the form:
\beq
{\cal{L}}_{eff} = \frac{1}{16 \pi} Im \int d^4 \theta {\cal{F}} [\Psi (\theta)]
\label{sun=2eff}
\eeq
where ${\cal{F}}$ is a function to be determined. The $N=2$ superfield 
formulation of eq.(\ref{sun=2eff}) is not discussed in these lectures. It 
can be found in Ref.~\cite{DIVE} and Refs. therein. Here we just add that,
under the
anomalous $U(1)_R$, the $N=2$ chiral superfield $\Psi$ is transformed as
\beq
\Psi ( \theta) \rightarrow e^{2i \beta} \Psi ( \theta {\rm e}^{-i \beta})
\label{n=2trans}
\eeq
In the $N=1$ superfield formalism  Lagrangian in eq.(\ref{sun=2eff}) becomes:
\beq
{\cal{L}}_{eff} = \frac{1}{4 \pi} Im \left\{ \int d^4 \theta 
\frac{\part {\cal{F}}}{\part \Phi} {\bar{\Phi}} + \int d^2 \theta \shalf
\frac{\part^2 {\cal{F}}}{\part \Phi^2} W^{\alpha} W_{\alpha} \right\}
\label{sun=1eff}
\eeq
In general, in a $N=1$ supersymmetric theory the coefficient of kinetic terms 
of the gauge fields and that of the matter fields, called K{\"{a}}hler
potential, are completely independent. We see, 
instead, that the $N=2$ invariance requires that they are related being both 
derived from the same function ${\cal{F}}$. In terms of the ${\cal{F}}$ they 
are 
given by:
\beq
K( \Phi , {\bar{\Phi}}) = \frac{1}{4 \pi} Im \left[ 
\frac{\part {\cal{F}}}{\part \Phi}
{\bar{\Phi}} \right] \hspace{2cm} \tau (\Phi)  = 
\frac{\partial^2 {\cal{F}} }{\partial \Phi^2}
\label{kaelerpo}
\eeq
When expressed in terms of the component fields Lagrangian in 
eq.(\ref{sun=1eff}) becomes
\beq
{\cal{L}} = \frac{1}{4 \pi} Im \left\{ \tau (\phi) \left[ \partial_{\mu} 
\bar{\phi} 
\partial^{\mu} \phi - \frac{1}{4}\left( F^2 - i
 F_{\mu \nu} {}^{*}F^{\mu \nu } \right) + \frac{1}{2} ( 
f^{abc} {\bar{\phi}}_{b} \phi_c)^2 +  Fermions \right] \right\}
\label{genela}
\eeq

The main task is to determine the explicit form of the function ${\cal{F}}$
that in general will receive both perturbative and non-perturbative 
contributions.

Comparing eq. (\ref{sun=1eff}) with eq. (\ref{n=23}) we see that at the tree 
level the function ${\cal{F}}$ is given by
\beq
\label{clala}
{\cal{F}}_{cl} = \frac{1}{2} \tau_{cl} \Phi^2 \hspace{1cm} \tau_{cl}=
\frac{\theta}{2 \pi} + i \frac{4 \pi}{e^2}
\eeq

At one loop ${\cal{F}}$ is completely fixed by the $U(1)_R$ 
anomaly~\cite{DIVE,SEIBE}.
In fact we have seen that under the $U(1)_R$ transformations the chiral
superfield $\Phi$ transforms as in eq.(\ref{u1r}), while the transformation 
of the parameter $\tau$ follows from the insertion in $\tau$ 
(see eq.(\ref{lambda2})) of the $U(1)_R$ transformation of $\theta$ given in
eq.(\ref{modtheta}). This implies for ${\cal{F}}$ the following transformations
\beq
{\cal{F}}'' (\Phi {\rm e}^{2 i \beta}) - {\cal{F}}'' (\Phi ) = - 
\frac{2 \beta N_c}{\pi}
\label{Fdo}
\eeq
where the double prime means double derivative with respect to the argument.
The solution of the previous eq. is
\beq
{\cal{F}}'' (\Phi) = i \frac{N_c}{\pi} \log \frac{\Phi}{M}
\label{sFdo}
\eeq
where $M$ is an arbitrary parameter that has been introduced in order to make 
the argument of the 
logarithm dimensionless. Integrating two times the previous equation we 
get~\cite{DIVE,SEIBE}:
\beq
{\cal{F}}_{1} = i \frac{N_c}{4 \pi} \Phi^2 \log \frac{\Phi^2}{\Lambda^2}
\label{1loobeta}
\eeq
that is consistent with the $U(1)_R$ and scale anomaly~\cite{DIVE,SEIBE}. 

It can also be shown that higher loops do not give any contribution to 
${\cal{F}}$.
Only non perturbative effects, as for instance instantons, can give 
an additional contribution to ${\cal{F}}$~\cite{SEIBE}. We will show now
that the contribution of the instantons to ${\cal{F}}$ must be of the
following form:
\beq
{\cal{F}}_{inst.} = \sum_{k=1}^{\infty} {\cal{F}}_{k} \left( 
\frac{\Lambda}{\Phi} \right)^{4k} \Phi^2
\label{inscon}
\eeq
In fact, from eq.(\ref{eucliac3}) one can see that the contribution of an
instanton with topological charge $k$ is proportional to 
${\rm e}^{- 8 \pi^2 k/e^2}$.  Using the explicit form of the $\Lambda$ 
parameter in eq.(\ref{lambda}) computed in terms of the vacuum expectation
value of the scalar field $\phi$ that we call $a$ we get:
\beq
{\rm e}^{- 8 \pi^2 k/e^2} = \left( \frac{\Lambda}{a} \right)^{4k}
\label{instacon}
\eeq 
The previous expression is invariant under a $U(1)_R$ anomalous
transformation acting on $ < \phi> =a$ as given in eq.(\ref{u1r}) supplemented
with the change in the $\Lambda$ parameter as given in eq.(\ref{lambdatra}).
Remembering that $\Phi$ transforms under the anomalous $U(1)_R$ as in 
eq.(\ref{u1r})   one gets the form of the instanton contribution as given in eq.
(\ref{inscon}). 

In the last part of this section we show that $N=2$ super Yang-Mills has 
solitons that are magnetic monopoles and dyons that have precisely the same
structure~\cite{DADDA} as those already discussed in the Georgi-Glashow model 
in section (\ref{mono}).

From the Lagrangian for the $N=2$ super Yang-Mills given in eq.(\ref{compola})
one can compute the Hamiltonian that is equal to:
\[
H = \frac{1}{2 e^2} \int d^3 x \left\{ (E^{a}_{i})^2 + (B_{i}^{a})^2 +
[(D_{i} A_{4})^a ]^2 + [(D_{i} A_{5})^a ]^2 + \right.
\]
\beq
+ \left. [(D_{0} A_{4})_a ]^2 +
[(D_{0} A_{5})_a ]^2 + ( f^{abc} A_{4}^{b} A_{5}^{c} )^2 + fermions \right\}
\label{hamin=2}
\eeq
We follow the approach of Bogomolny and  rewriting the first four terms of 
the integrand in the previous eq. as follows:
\[
\left[ E^{a}_{i} - \cos \theta (D_{i} A_{4})^a - \sin \theta (D_{i} A_5 )^a
\right]^2 + \left[ B^{a}_{i} + \sin \theta (D_{i} A_{4})^a - \cos \theta 
(D_{i} A_5 )^a \right]^2 +
\]
\beq
+ 2 E_{i}^{a} \left[ \cos \theta (D_i A_4 )^a + \sin \theta (D_i A_5)^a \right]
+ 2 B_{i}^{a} \left[ - \sin \theta (D_i A_4 )^a + \cos \theta (D_i A_5)^a 
\right]
\label{rewri5}
\eeq
we get a lower bound for the Hamiltonian:
\[
H \geq \frac{1}{e^2} \int d^3 x \left\{
E_{i}^{a} \left[ \cos \theta (D_i A_4 )^a + \sin \theta (D_i A_5)^a \right] 
\right.
+
\]
\beq
\left. +B_{i}^{a} \left[ - \sin \theta (D_i A_4 )^a + \cos \theta (D_i A_5)^a 
\right] \right\}
\label{inequa5}
\eeq
The lower bound becomes an equality when the following equations are
satisfied:
\beq
(D_0 A_4 )_a = (D_0 A_5 )_a = ( f^{abc} A_{4}^{b} A_{5}^{c} )^2 =0
\label{equa78}
\eeq
\beq
E^{a}_{i} =  \cos \theta (D_{i} A_{4})^a + \sin \theta (D_{i} A_5 )^a
\label{bps69}
\eeq
\beq
B^{a}_{i} = - \sin \theta (D_{i} A_{4})^a + \cos \theta  (D_{i} A_5 )^a 
\label{bps70}
\eeq
and at the same time we put all fermionic fields equal to zero as it is
allowed from the classical eqs. of motion.

Introducing the ansatz in eqs.(\ref{ans}) and (\ref{dyonansa}) respectively
for the space and time component of the vector potential and the following 
ansatz for the Higgs fields:
\beq
A_{4}^{a}= \frac{r^a}{r^2} J_4 (\xi) \hspace{2cm}
A_{5}^{a}= \frac{r^a}{r^2} J_5 (\xi)
\label{ansa5}
\eeq
we see that eqs.(\ref{equa78}) are automatically satisfied. If we impose
eqs.(\ref{bps69}) and (\ref{bps70}), as shown in Appendix \ref{A}, we get 
that the function $K( \xi)$ in the ansatz in eq.(\ref{ans}) is again given 
by the expression in eq.(\ref{bpslim}), while the other functions are given
in terms of a unique function $R (\xi)$  
\beq
J_4 (\xi )= \alpha R (\xi) \hspace{1cm} J_5 (\xi) = \beta R ( \xi)
\hspace{1cm} J (\xi) = \gamma R( \xi) \hspace{1cm} R (\xi) = \xi \coth \xi -1
\label{solu98}
\eeq
where the dimensionless parameter $ \xi = {\hat{a}} r$ and  the constants
$\alpha$, $\beta$ and $\gamma$ are determined in terms of $\theta$ 
through the relations:
\beq
\alpha \sin \theta + \beta \cos \theta =1 \hspace{1cm}
\gamma = \alpha \cos \theta  - \beta   \sin \theta
\label{rela98}
\eeq
that imply
\beq
\alpha^2 + \beta^2 - \gamma^2 =1
\label{impl5}
\eeq

Let us see now what determines the constant $\theta$. Let us 
introduce the electric and magnetic charges:
\beq
q = - \frac{1}{a' e} \int d^3 x \partial_i \left[ E_{i}^{a} A_{5}^{a} -
B_{i}^{a} A_{4}^{a} \right]
\label{e23}
\eeq
\beq
g= - \frac{1}{a' e} \int d^3 x \partial_i \left[ B_{i}^{a} A_{5}^{a} +
E_{i}^{a} A_{4}^{a} \right]
\label{m23}
\eeq
where $a' = \sqrt{a_{4}^{2} + a_{5}^{2}}$. $a_4 \equiv \alpha {\hat{a}}$ and 
$a_5 \equiv \beta {\hat{a}}$ are respectively the asymptotic values for $r 
\rightarrow \infty$ of $A_4$ and $A_5$ and $a' = {\hat{a}} 
\sqrt{1+ \gamma^2}$. Using the previous formulas we get
\beq
E \geq - \frac{a'}{e} \left[ q \sin \theta + g \cos \theta \right]
\label{ene99}
\eeq
Inserting eqs.(\ref{bps69}) and (\ref{bps70}) in eqs.(\ref{e23}) and (\ref{m23})
we get
\beq
q = g  \tan \theta
\label{rel90}
\eeq
that implies for eq.(\ref{ene99})
\beq
E = \frac{a'}{e} \sqrt{q^2 + g^2 } = \frac{a'}{e} | q + ig |
\label{ebe67}
\eeq
In terms of the complex scalar field $\phi$ defined in eq.(\ref{identi5}),
calling $a$ the complex asymptotic value of $\phi$, in the BPS limit we can 
write the mass of the dyons in eq.(\ref{ene99}) as
\beq
M= \sqrt{2} \frac{|a|}{e} |q+ ig|
\label{mass99}
\eeq
where the magnetic charge is the same as in the Georgi-Glashow model:
\beq
g= + \frac{4 \pi}{e}n_m  \hspace{1cm} n_m = - 1
\label{elema}
\eeq
and the electric charge, after semiclassical quantization, is equal to
\beq 
q = n_e e
\label{q79}
\eeq
Using the two previous eqs. and introducing also a $\theta$ angle we can
rewrite eq.(\ref{mass99}) as follows:
\beq
M = \sqrt{2} |a| | n_e +\tau n_m|
\label{massbps5}
\eeq 

We have been considering a $N=2$ supersymmetric theory, but up to now the
fermionic part of the Lagrangian has not played any role. All our
considerations are based only on the structure of the bosonic part of the
Lagrangian since all fermionic fields have been put equal to zero consistently
with their classical eqs. of motion. In the following we will show that the 
$N=2$ supersymmetry is
essential for two reasons. The first is that the previous considerations,
for instance for the calculation of the mass of the dyons, are purely classical 
or at most 
semiclassical results. We will see that they will be valid in the full
quantum theory as a consequence of the supersymmetry  algebra that gets
modified by the presence of central charges. The second one is  the presence
of fermionic zero modes that in a supersymmetric theory implies that dyons 
and monopoles belong to supersymmetric short multiplets. As a consequence
the dyons and monopoles carry a non vanishing spin. We will discuss these
two aspects in the next section.

\section{Susy algebra and fermionic zero modes}
\label{centra}

In eqs.(\ref{sualgcc}) and (\ref{sualgcc2}) we have written the $N=2$
supersymmetry algebra in four dimensions including a complex central charge
${\hat{Z}}$. The
value of the central charge depends on the particular theory we are
considering. In the case of $N=2$ super Yang-Mills the central charge $Z$
has been computed by Olive and Witten in Ref.~\cite{WO}. In Appendix C we give
all necessary details of the calculations that bring to the result:
\beq
{\hat{Z}} = - 2 \frac{a'}{e} ( q - ig )
\label{cecha3}
\eeq
where $q$ and $g$ are given in eqs.(\ref{e23}) and (\ref{m23}). Introducing
the Majorana spinors:
\beq
Q^{i}_{A} = \left( \begin{array}{c} Q^{i}_{\alpha} \\
			     {\bar{Q}}^{i {\dot{\alpha}}} \end{array} \right) 
\hspace{2cm} 
{\bar{Q}}^{i}_{A} = \left( \begin{array}{cc} Q^{i \alpha} & 
			     {\bar{Q}}^{i}_{{\dot{\alpha}}} \end{array} \right) 
\label{MAspi}
\eeq
the $N=2$ algebra given in eqs.(\ref{sualgcc}) and
(\ref{sualgcc2}) can be rewritten in four-dimensional notations obtaining:
\beq
\{ Q^{i}_{A} , {\bar{Q}}^{j}_{B} \} = 2 \gamma^{\mu}_{AB} P_{\mu} \delta^{ij}
- 2 (\gamma_{5})_{AB}  \epsilon^{ij} V + 2i \epsilon^{ij} \delta_{AB} U
\label{dicech}
\eeq
where
\beq
2 U = - Im {\hat{Z}} = - \frac{2 a' g}{e} \hspace{3cm} 2 V = - Re {\hat{Z}} 
=  \frac{2 a' q}{e}
\label{condi2}
\eeq

From the algebra in eq. (\ref{dicech}) applied to a state in the center of 
mass frame where $P_{\mu} = ( M, \vec{0} )$ one gets:
\beq
\{ Q_{A}^{i}, ( Q_{B}^{j})^{\dagger}  \} = 2 M \delta^{ij} \delta_{AB} -
\frac{2a'}{e} L_{AB}^{ij}
\label{alge56}
\eeq
where $L$ satisfies the eq.
\beq
L^2 = (q^2 + g^2 ) \delta^{ij}\delta_{AB} \hspace{2cm}
L^{ij}_{AB} = \epsilon^{ij} \left[q (\gamma_5 \gamma^0 )_{AB} + ig 
(\gamma^0)_{AB} \right]
\label{lqua}
\eeq
This implies that the eigenvalues of the matrix $L$ are equal 
to $ \pm \sqrt{q^2 + g^2}$. Then, since the l.h.s. of eq.(\ref{alge56}) is
a positive definite operator, one gets a lower bound for the mass
\beq
M \geq \frac{a'}{e} \sqrt{q^2 + g^2 } = \frac{a'}{e} | q+ig| = \sqrt{2} 
\frac{|a|}{e} |q +ig |
\label{bpscondi}
\eeq
where we have introduced the complex parameter $a$ (as in eq.(\ref{mass99})) 
that is the asymptotic value of the complex field $\phi$.

We have obtained again the BPS condition, but now, unlike the case of
eq.(\ref{mass99}) that was derived in the classical theory, it is a direct 
consequence of the
supersymmetry algebra that is supposed to be valid in the full quantum theory.
For the BPS states, for which the equality sign holds, it is an exact mass 
formula. For this reason the introduction of an extended
supersymmetry allows one to overcome the difficulty mentioned in the second 
point toward the end of section (\ref{MOdua}). Actually, as we will see in 
section (13), this is not quite true in $N=2$ super Yang-Mills because the 
parameter $a$, that describes the moduli space of the theory, provides a 
good description of it only in the semiclassical region where $a$ is large.
In the strong coupling region the semiclassical formula (\ref{bpscondi})
must be modified.  

In the second part of this section we will discuss the fermionic zero modes.
Up to now, in order to get the monopole and dyon solutions, we have put all 
fermion fields equal to zero. On the other hand, if a theory is supersymmetric,
by means of a supersymmetry transformation, from a classical solution in which 
the fermionic fields are zero we get another classical solution in which 
they are non zero. If we perform a supersymmetry transformation, that can
be obtained by dimensional reduction from the expression in eq.(\ref{sutra2})
one gets
\beq
\delta \chi^a  = \left[ \sigma_{\mu \nu} F^{\mu \nu}_{a} - i \gamma_{\mu} 
\gamma_5 (D^{\mu} A_{4})^a  - \gamma_{\mu} (D^{\mu} A_{5})^a 
- i f^{abc} A_{4}^{b} A_{5}^{c} \gamma_5 \right] \alpha
\label{sutra7}
\eeq
Assuming that the fields in the previous eq. satisfy the BPS eq.(\ref{bps70})
satisfied by the monopole (taking for simplicity $\theta=0$ and $A_4 =0$) 
we can rewrite the previous supersymmetry transformation as
follows:
\beq
\delta \chi^{a} = - (D^{k} A_5 )^a \left[\gamma_{k} + \shalf 
\epsilon_{ijk} \gamma^i \gamma^j \right]\alpha
\label{sutra45}
\eeq
This implies that, if we choose the supersymmetry parameter $\alpha$
satisfying the equation:
\beq
\left[ 1 + \frac{1}{6} \epsilon_{ijk} \gamma^i \gamma^j \gamma^k \right]
\alpha = \left[ 1 + \gamma^1 \gamma^2 \gamma^3 \right] \alpha  = 0
\label{alpha43}
\eeq
then the supersymmetry transformation does not move the fermionic
field from zero. This means that the monopole solution preserves half of
the supersymmetry. On the other hand, if $\alpha$ satisfies instead the
condition with the opposite sign:
\beq
\left(1 - \gamma^1 \gamma^2 \gamma^3 \right) \alpha =0
\label{opposi}
\eeq 
then, by means of a supersymmetry transformation we go from $\chi^a =0$ to
\beq
\chi^a = -2 \gamma_{k} (D^k A_5 )^a \alpha
\label{zeromo5}
\eeq
that corresponds to  two fermionic zero modes. In fact, since $\alpha$ is a 
Dirac
spinor, it has four independent degrees of freedom. Half of them are killed by
the condition (\ref{opposi}) and therefore we are left with only two degrees
of freedom that is also the expected number of zero modes for fermions
transforming in the adjoint representation according to the Callias index 
theorem~\cite{CALLIAS}. 

The ansatz for the monopole in eq.(\ref{ans}) is transformed if we 
perform a space rotation, but is left invariant if we supplement the space 
rotation with a global gauge transformation whose parameter is related to 
the one of the space rotation. In other word the generator that leaves the
ansatz invariant is generated by the total angular momentum:
\beq
{\vec{J}} = {\vec{L}}+ {\vec{S}}+ {\vec{T}}
\label{totan}
\eeq
where ${\vec{L}}$ and ${\vec{S}}$ are the orbital and the spin angular
momenta, while ${\vec{T}}$ is the generator of the $SO(3)$ gauge 
transformations. If we have isovector spinors as in the $N=2$ super Yang-Mills
then the total angular momentum is half-integer, since the spin is equal to
$1/2$ and the orbital angular momentum is integer.  In particular since
$\shalf \otimes 1 = \frac{3}{2} + \shalf$ and since we have two zero modes 
the simplest possibility is that
the two fermionic zero modes, given in eq.(\ref{zeromo5}), transform
according to the $\shalf$ representation of ${\vec{J}}$. In the quantum theory
we can expand the field $\chi$ as follows:
\beq
\chi = a_{1/2} \chi_{0}^{1/2} + a_{-1/2} \chi_{0}^{-1/2} + \Delta \chi
\label{quafi}
\eeq
where $\Delta \chi$ is the quantum field and $\chi_{0}^{\pm 1/2}$ are the two
zero modes. The anticommutation relations satisfied by $\chi$ imply that the
coefficients of the zero mode expansion in eq.(\ref{quafi}) satisfy the algebra of the
fermionic harmonic oscillator:
\beq
\{a_{1/2} , {a}^{\dagger}_{1/2} \} = \{ a_{-1/2} , a^{\dagger}_{- 1/2} \} = 0
\label{harmosci}
\eeq
and all other anticommutators are zero.

The states of the supersymmetric multiplet of the monopole can be constructed
starting from the vacuum $|\Omega>$ that is annihilated by 
\beq
a_{1/2} |\Omega> =  a_{-1/2} |\Omega> =0
\label{anniva}
\eeq
and then acting with the creation operators on it. In this way we can
construct the states given in Table 1.

\begin{table}     
\caption{$N=2$ BPS multiplet.}
\begin{center}
\begin{tabular}{|c|c|c|}
\hline 
STATE              &$S_z$     & $\#$ OF STATES \\\hline
$|~ 0 >$           &0         & 1 \\ \hline
$a^{\dagger}_{\pm 1/2} |~ 0 >$    &$\pm 1/2$ & 2 \\ \hline
$a^{\dagger}_{1/2} a^{\dagger}_{-1/2} |~ 0 > $  
                   & 0         & 1 \\\hline
\end{tabular}
\end{center}
\end{table}
We see that the monopole multiplet contains spin $1/2$, but not spin $1$.
This means that $N=2$ super Yang-Mills cannot be duality invariant {\`{a}}
la Montonen-Olive. Notice that the hypermultiplet, that contains $4$ real
bosonic and $4$ real fermionic states, consists of two $N=2$ BPS multiplets.
 
\sect{Global parametrization of moduli space}
\label{global}
In the next three sections we will be shortly describing the beautiful 
paper of Seiberg and Witten~\cite{SEIWI} where an exact expression for 
$\tau (\Phi)$ (see eq.(\ref{kaelerpo})) in the low energy effective action 
of the $N=2$ super Yang-Mills theory has been constructed.

Unlike the $N=4$ theory which, as we will see, can be equivalently formulated 
either in 
terms of the original fundamental fields or in terms of the monopoles or
more in general of the dyons of the single particle spectrum with essentially 
the same Lagrangian, the $N=2$ theory cannot satisfy the Montonen-Olive duality
because the fundamental fields and the magnetic monopoles belong to two 
different $N=2$ superfields. The fundamental fields are in the chiral $N=2$ 
vector multiplet
while the magnetic monopoles and dyons are in the hypermultiplet~\cite{OSBORN}.

Nevertheless the $N=2$ theory can be formulated either in terms of the
variables $\phi, A_{\mu}$ and $\tau(\phi)$, as we have done in 
eq.(\ref{genela}),
or in terms of the dual variables $\phi_D , A_{D \,\mu}$ and 
$\tau_{D} (\phi_D )$ 
in pretty much the same way that free electromagnetism can be formulated
either in terms of the vector potential $A_{\mu}$ related to the field strenght
by $F_{\mu \nu} \equiv \partial_{\mu} A_{\nu} - \partial_{\nu} A_{\mu}$  or in
terms of the dual vector potential $A_{D \, \mu}$ related to the dual field 
strenght by ${}^* F_{\mu \nu} = \partial_{\mu} A_{D\, \nu} - \partial 
A_{D\, \mu}$.

In order to explain this let us first summarize some general property of the
$N=2$ super Yang-Mills theory. 
 
In section (\ref{semi}) we have seen that in this theory the
low energy effective action is completely fixed by giving a holomorphic
function ${\cal{F}}(\Phi)$. In terms of ${\cal{F}}$ we can construct 
the K{\"{a}}hler potential, as given in eq.(\ref{kaelerpo}), and the metric
\beq
(ds)^2 = \frac{\partial}{\partial \phi} \frac{\partial}{ \partial 
{\bar{\phi}}}
K(\phi, {\bar{\phi}}) d\phi d{\bar{\phi}}= Im ( \tau (\phi) ) d\phi 
d{\bar{\phi}}
\hspace{1cm} \tau (\phi) = \frac{\partial^2 {\cal{F}} (\phi)}{\partial \phi^2} 
\label{metri5}
\eeq
 
We have also seen that the moduli space of the $N=2$ theory is in the 
semiclassical
theory parametrized by the vacuum expectation value of the scalar field 
$\phi$ that
we have denoted by the complex number $a$. However $a$ cannot provide a
global description of the moduli space. In fact the metric $Im (\tau (a) )$, 
that is a positive definite harmonic function divergent for $|a| \rightarrow
\infty$, must have a minimum. But a globally defined harmonic function
cannot have a minimum and consequently the
variable $a$ cannot provide a global parametrization of the moduli space.

Therefore in Ref.~\cite{SEIWI} it was proposed to choose the gauge invariant
quantity $u = \frac{1}{2} Tr ( \phi^2 )$ as the one that provides a global
parametrization of the moduli space and to regard both $a (u)$  and the dual 
variable $a_D (u) \equiv \frac{\partial {\cal{F}}}{\partial a}$ as functions 
of $u$. In terms of both $a$ and $a_D$ the metric in eq.(\ref{metri5}) assumes
the form
\beq
(ds)^2 = Im \left( \frac{d a_D}{ d a} da d{\bar{a}} \right)=
Im \left( da_D d {\bar{a}} \right) = - \frac{i}{2} \left[ da_D d{\bar{a}}
- da d{\bar{a}}_D \right]
\label{newme}
\eeq
that is symmetric under the exchange $a \leftrightarrow a_D$.

Introducing the vector $v^{\alpha} = \left( \begin{array}{cc} a_D \\ a  
\end{array}\right)$ we can rewrite the metric in the more compact form:
\beq  
(ds)^2 = - \frac{i}{2} \epsilon_{\alpha \beta} \frac{d v^{\alpha}}{du}
\frac{d {\bar{v}}^{\beta}}{d{\bar{u}}} du d{\bar{u}}
\label{nmet}
\eeq 
that clearly show its invariance under the transformation:
\beq
v \rightarrow M v + c
\label{sl2r}
\eeq
where $M$ is a matrix of $SL(2,R)$ and $c$ is a constant vector.

An arbitrary matrix of $SL(2,R)$ is generated by the action of two independent
matrices $T_b$ and $S$. The first one 
\beq
T_b = \left(  \begin{array}{cc} 1 & b \\
				0 & 1  \end{array} \right)
\label{tb}
\eeq
leaves $a$ invariant and transforms $a_D$ according to
\beq
a_D \rightarrow a_D + b a
\label{tbtra}
\eeq
This implies that $\tau (a)$ is just translated
\beq
\tau (a) \rightarrow \tau (a) + b
\label{tautra}
\eeq
resulting in a translation for the vacuum angle $\theta$
\beq
\theta \rightarrow \theta + 2 \pi b
\label{thetra}
\eeq
Since physical quantities are invariant when
\beq
\theta \rightarrow \theta + 2 \pi n
\label{thetra1}
\eeq
for any integer $n$, comparing eqs.(\ref{thetra}) and (\ref{thetra1}) we
deduce that $b=1$ and consequently that the transformation associated to the
matrix $T_{b=1}$ is a symmetry of the theory. By selecting $b=1$ we have
reduced the original $SL(2,R)$ symmetry group to $SL(2,Z)$.

The other independent generator 
\beq
S = \left( \begin{array}{cc} 0 & 1 \\
			     -1 & 0 \end{array} \right)
\label{stra}
\eeq
does not correspond to a symmetry of the theory, but provides a transformation
between two different parametrizations of the theory. In fact the low energy 
effective Lagrangian can be represented either in terms of the variables
$( A^{\mu}, \lambda, \phi; \tau (\phi)) $ or in terms of the dual ones 
$( A^{\mu}_D, \lambda_D, \phi_D; \tau_D (\phi_D ) =
-1/ \tau (\phi) ) $. In order to more clearly see the relation between the two 
formulations it is convenient to set the vacuum angle $\theta=0$. Then we see 
that, if
$Im \tau (\phi)= \frac{4 \pi}{e^2}$, then $Im \tau_D (\phi_D )= 
\frac{e^2}{4 \pi}$.
Therefore one description may be more suitable for weak coupling, while the
other for strong coupling.

In the final part of this section we discuss the exact mass formula proposed
in Ref.~\cite{SEIWI} for the BPS saturated states in the $N=2$ theory. At the 
semiclassical level the mass of the BPS saturated states is given 
by~\cite{KAUL84,IMBIMBO85}
\beq
\label{mass4}
M= \sqrt{2}|Z| \hspace{2cm}
Z= a \left[ n_e + \left( \frac{\theta}{2 \pi} + i \frac{4 \pi}{e^{2}(\mu) } 
+ \frac{i}{\pi} \log \frac{a^2}{\mu^2 C} \right) 
n_m \right]
\eeq
where $\mu$ is the renormalization scale and $C$ is a scheme dependent constant.
The extra logarithmic term present in this formula with respect to the classical
mass formula is the effect of the renormalization of the gauge coupling constant
$e$.
Noticing that the coefficient of $n_m$ in eq.(\ref{mass4}), with a suitable
choice of $C$, is equal to $a_D \equiv \frac{\partial {\cal{F}}}{\partial a}$
with ${\cal{F}} = {\cal{F}}_{cl} + {\cal{F}}_1$ given in eqs.(\ref{clala}) and
(\ref{1loobeta}), eq.(\ref{mass4}) can be rewritten as follows
\beq
Z = a n_e + a_D n_m = \left( \begin{array}{cc} n_m & n_e \end{array} \right)
\left( \begin{array}{cc} a_D \\
			  a \end{array} \right) \hspace{2cm} M= \sqrt{2} |Z|
\label{mass4e}
\eeq

Seiberg and Witten~\cite{SEIWI} proposed eq.(\ref{mass4e}) as an exact
formula for the BPS states and made several checks for confirming its 
validity.

In particular $Z$ is invariant under the transformation
\beq
\left( \begin{array}{cc} a_D \\
			  a \end{array} \right)
\rightarrow M \left( \begin{array}{cc} a_D \\
			  a \end{array} \right)
\hspace{2cm}
\left( \begin{array}{cc} n_m & n_e \end{array} \right) \rightarrow
\left( \begin{array}{cc} n_m & n_e \end{array} \right) M^{-1}
\label{modinv7}
\eeq
where $M$ is a matrix of $SL(2,Z)$ because the vector
$\left( \begin{array}{cc} n_m & n_e \end{array} \right)$ has integer
entries and its transformed must also have integer entries. This is an 
independent way to derive the reduction of $SL(2,R)$ to $SL(2,Z)$.
Actually this procedure forces also the extra parameter $c$ in 
eq.(\ref{sl2r}) to be equal to zero.

\sect{Singularity structure of moduli space}
\label{singu}
In this section we study the singularity structure of $a$ and $a_D$ as 
functions of the variable $u$, that provides a global parametrization of the 
moduli space. 

In the semiclassical region corresponding to a large value of $u$ we get 
\beq
\label{5.1}
a = \sqrt{2u} \hspace{2cm} a_D \equiv \frac{\partial {\cal{F}}_1}{ \partial a}
\sim \frac{2 i a}{\pi}\log a = i \frac{2 \sqrt{2u}}{\pi}  \log 
\sqrt{2u}
\eeq
where at one loop ${\cal{F}}$ is given by (see eq.(\ref{1loobeta})):
\beq
{\cal{F}}_1 = \frac{i}{2 \pi} a^2 \log \frac{a^2}{\Lambda^2}
\label{F123}
\eeq
and 
\beq
\tau (u) \equiv \frac{\partial a_D }{\partial a} \sim \frac{2i}{\pi} 
\log a = \frac{2i}{\pi} \log \sqrt{2u}
\label{tau56}
\eeq
Under a rotation around $u = \infty$ given by
\beq
\label{5.2}
\log u \rightarrow \log u + 2i \pi
\eeq
$a (u)$, $a_D (u)$ and $\tau (u)$ are not monodromic functions. They 
transform according to
\beq
\label{5.3}
a \rightarrow - a \hspace{2cm} a_D \rightarrow - a_D + 2 a \hspace{2cm}
\tau (u)  \rightarrow \tau (u) -2
\eeq
The monodromy properties given in eq.(\ref{5.3}) are entirely determined by the
coefficient in front of the logarithm in eq.(\ref{tau56}) that is related to
the perturbative $\beta$-function in eq.(\ref{betafu}). More precisely this
coefficient is equal to a factor $(-i)$ times the value of the coefficient of
the $\beta$-function in eq.(\ref{betafu}), that is equal to $\frac{1}{4 \pi^2}$
for $N_c =2$, multiplied by a factor $4 \pi$ that is already present in front
of the effective action in eq.(\ref{sun=1eff}). The monodromy 
transformations in eq.(\ref{5.3}) are generated by acting on the vector 
$\left( \begin{array}{c} a_D \\
			  a  \end{array} \right)$ with the following monodromy 
matrix:
\beq
M_{\infty} = \left( \begin{array}{cc}  -1 & 2 \\
				       0 & -1 \end{array} \right)
\label{Minfty}
\eeq

The existence of a singularity requires the existence of at least another
singularity. But, if we had only one additional singularity, it is easy to 
see that $a$ would have been a good global parameter being the monodromy 
group an abelian group. Since this is not possible we must require 
the existence of at least two additional singularities. 

Following the example of what is happening in some $N=1$ supersymmetric
theories Seiberg and Witten assume that the singularities occur at those
points of the moduli space where additional massless particles appear in the
spectrum. In the classical theory this occurs for $a=0$ where the $SU(2)$
symmetry is restored and $W^{\pm}$ become massless. They bring strong 
indications against this possibility in the quantum theory  and instead
choose the singularities at the points $m^2$  where the monopole 
with  $(n_m ,n_e )= (1,0 )$ and $(m')^2$  where the dyon with $(n_m , n_e ) 
= (1, -1)$ become massless.

Using the exact formula in eq.(\ref{mass4e}) it is easy 
to see that this occurs at a certain value $u$ that we call  $ m^2$ for 
which $a_D ( m^2 ) =0 $ with $a ( m^2 ) \neq 0$  and when $a_D ( (m')^2) - 
a ((m')^2 ) =0 $ with $a ((m')^2 ) , a_D ( (m')^2) \neq 0$ respectively.
The existence of a $Z(2)$ symmetry that transforms $u$ in $- u$ suggests to
choose $(m ')^2 = - m^2$. 

The monodromy around the singularity at $u= m^2$ can be easily computed by 
observing that the low energy theory at the point $u= m^2$ consists of a
"magnetic" $N=2$ super QED (the matter has magnetic and non electric 
charge). This theory is not asymptotically free and the coefficient of the
$\beta$-function, besides a sign, has a factor $1/2$ of difference with  
respect to the $\beta$-function previously used for studying the singularity
around $u = \infty$. This means that in this case instead of the 
eq.(\ref{tau56}) we get the following expression:
\beq
\tau_{D} \equiv -\frac{\partial a}{\partial a_D} \sim  - \frac{i}{\pi} \log a_D
\label{taud45}
\eeq
Integrating the previous eq. one gets:
\beq
a(u) = k + \frac{i a_D }{\pi} \log a_D 
\label{au67}
\eeq
where $k \neq 0$ because we have assumed that at $u = m^2$ only the 
monopole and no other electrically charged particle become massless.
Assuming that $a_D$ is a good coordinate near $a_D =0$ we can write $a$ and 
$a_D$ as follows
\beq
a_D (u) \sim c_0 ( u - m^2 ) \hspace{2cm}
a(u) = k + \frac{i }{\pi} c_0 ( u - m^2 ) \log \left[ c_0 ( u - m^2 )\right]
\label{m256}
\eeq 
Under the monodromy transformation:
\beq
\log ( u - m^2 ) \rightarrow \log ( u - m^2 ) + 2 \pi i
\label{mono79}
\eeq
we get
\beq
\label{5.4}
a_D \rightarrow a_D \hspace{2cm} a \rightarrow a - 2 a_D \hspace{2cm}
\tau_{D} \rightarrow \tau_{D} +2 
\eeq
From the last equation we get the following transformation on $\tau$:
\beq
\tau (u) = - \frac{1}{\tau_D } \rightarrow 
\frac{\tau (u)}{1 - 2 \tau (u)}
\label{tau34}
\eeq
The monodromy matrix that, acting on the vector 
$\left( \begin{array}{c} a_D \\
			  a  \end{array} \right)$, generates the 
transformations in eqs.(\ref{5.4})
and (\ref{tau34}) is given by:
\beq
M_{m^2} = \left( \begin{array}{cc}  1 & 0 \\
				       -2 & 1 \end{array} \right)
\label{Mlam2}
\eeq

The singularity at $u = - m^2$ must be the mirror symmetric of the one
at $u = m^2$ because the two points are related by the $Z_2$ symmetry
that acts on $u$ as $u \rightarrow  {\rm e }^{i \pi} u$. Starting from the
expression valid in the semiclassical approximation for large and positive $u$
given in eq.(\ref{5.1}) and performing the $Z_2$ transformation we get
\beq
a \rightarrow {\tilde{a}} = i a \hspace{2cm} a_D \rightarrow {\tilde{a}}_D =
i ( a_D -a ) 
\label{tra456}
\eeq
${\tilde{a}}_D$ must have a single zero, that is the image of the zero of $a_D$
for $u$ near $m^2$: ${\tilde{a}}_D = {\tilde{c}}_0 (u + m^2 )$.
Then, proceeding as in the previous case, from 
\beq
{\tilde{\tau}}_D  = - \frac{\partial {\tilde{a}}}{\partial 
{\tilde{a}}_D} = - \frac{i}{\pi} \log {\tilde{a}}_D = - \frac{i}{\pi} \log
\left[{\tilde{c}}_0 ( u + m^2 ) \right]
\label{tau89}
\eeq
one gets
\beq
{\tilde{a}} = {\tilde{k}} + \frac{i}{\pi} {\tilde{a}}_D \log {\tilde{a}}_D
\label{tilda}
\eeq
In conclusion we get:
\beq
{\tilde{a}}_D = {\tilde{c}}_0 ( u + m^2 ) \hspace{2cm}
{\tilde{a}} = {\tilde{k}} + \frac{i}{\pi} {\tilde{c}}_0 ( u + m^2)  
\log \left[ {\tilde{c}}_0 (u + m^2 ) \right]
\label{vari97}
\eeq
Under a monodromy transformation 
\beq
\log ( u + m^2 ) \rightarrow \log ( u + m^2 ) + 2 \pi i
\label{mono78}
\eeq
we obtain
\beq
{\tilde{a}}_D \rightarrow {\tilde{a}}_D \hspace{2cm} {\tilde{a}} \rightarrow
{\tilde{a}} - 2 {\tilde{a}}_D
\label{mon79}
\eeq
Going back to the original variables $a$ and $a_D$ they become:
\beq
a \rightarrow 3 a - 2 a_D \hspace{2cm} a_D \rightarrow 2 a - a_D
\hspace{2cm} \tau (u)  \rightarrow \frac{2 - \tau (u) }{3 - 2 \tau (u)}
\label{mon67}
\eeq
They are generated by the monodromy matrix:
\beq
M_{- m^2} = \left( \begin{array}{cc}  -1 & 2 \\
				       -2 & 3 \end{array} \right)
\label{M-lam2}
\eeq
as it can be easily checked.

If there are only three singularities the three monodromy matrices must be
consistent. They in fact satisfy the consistency condition:
\beq
M_{\infty} = M_{m^2} M_{- m^2}
\label{consi7}
\eeq
It can also be seen that the three monodromy matrices span a subgroup of the
modular group, called $\Gamma (2)$, consisting of matrices of $SL(2,Z)$ 
congruent to the identity matrix modulo $2$.

\sect{Explicit solution}
\label{exsolu}
Having established the singularities and the monodromy transformations of 
$a$ and $a_D$ one could determine the solution algebraically apart from a
regular function that can then be fixed from the semiclassical behaviour. On 
the other 
hand the following  two points suggest to find the solution geometrically:

\begin{enumerate}
\item{$Im \tau (u) > 0$ as the parameter $\tau $ of a torus.}
\item{Under the various monodromies $\tau (u)$ transforms pretty much in the 
same way as the parameter $\tau$ of a torus transforms under a change of 
homology basis.}
\end{enumerate}

This suggests that we can define a torus for each value of $u$ and that its 
$\tau$ parameter is the desired $ \tau (u)$. A torus is usually represented as 
a parallelogram in the plane $z$ with opposite sides identified. The two 
sides of the parallelogram go from the origin to the points $1$ and $\tau$
respectively. Functions defined on a torus are doubly periodic functions.
On the torus there exists a unique holomorphic differential that in the 
previously described parametrization of the torus is equal to $\omega = d z$.
The parameter $\tau$ of the torus can be obtained as the ratio of the integrals
along the two cycles $a$ and $b$ of the torus. In the previous 
parametrization of the torus $\tau$ is given by:
\beq
\tau = \frac{\oint_b \omega}{\oint_a \omega} = \frac{\int_{0}^{\tau} dz}{
\int_{0}^{1} dz }
\label{tauto}
\eeq
A torus can also be described by a cubic equation:
\beq
y^2 = x^3 + B x^2 + C x + D
\label{cubeq}
\eeq
where $B,C$ and $D$ are arbitrary constants. The cubic defines a two sheet
function. There are four branch points that we call $x_1$, $x_2$ and $x_3$
and $\infty$. The first three branch points are located at the zeroes of
the cubic in eq.(\ref{cubeq}). The  cycle $a$ corresponds to a closed path that
encircles the cut between the first two branch points, while the cycle $b$ 
corresponds to a path that starts from the upper side of the other cut, goes 
to the upper side of the first cut, continues across the cut in the second sheet
and finally comes out again in the first sheet across the original cut.

In the parametrization of the torus in terms of the cubic equation the unique
holomorphic differential is equal to:
\beq
\omega = \frac{dx}{y (x)}
\label{holodi}
\eeq
and the $\tau$ parameter of the torus is given by:
\beq
\tau = \frac{\oint_b \frac{dx}{y(x)}}{\oint_a \frac{dx}{y(x)}}
\label{tauto6}
\eeq

This form is very similar to the expression for the parameter $\tau (u)$ 
whose monodromy properties we have described in the previous section:
\beq
\tau (u) = \frac{\frac{d a_D}{du}}{\frac{da}{du}}
\label{6.2}
\eeq

Assuming that for each value of $u$ we can define a torus whose parameter
is equal to $\tau (u)$ we get
\beq
\frac{d a_D}{du} = \alpha \oint_b \frac{dx }{y(x)} \hspace{2cm}
\frac{d a}{du} = \alpha \oint_a \frac{dx }{y(x)}
\label{dadad}
\eeq
where $\alpha$ is a dimensionless constant to be determined. In order to 
determine the previous
expressions we need to fix the coefficients $B,C$ and $D$ of the cubic in 
eq.(\ref{cubeq})  as functions of $u$ and $m^2$, that are the only dimensional 
parameters at our disposal. The form of the cubic can be
fixed with the procedure that we now describe.

Since $a$ and $a_D$ have dimension of a $mass$ and $u$ has dimension of a 
$[mass]^2$ eqs.(\ref{dadad}) implies that $x$ has dimension of a $[mass]^2$
as $u$ and therefore that the parameters $B,C$ and $D$ have respectively 
dimension of a $[mass]^2$, a $[mass]^4$ and a $[mass]^6$. They can only be
functions of the only two dimensional parameters at our disposal,
namely $u$ and $m^2$. As a consequence the most general expression for $B,C$
and $D$ can be written in terms of $u$ and $m^2$ with arbitrary dimensionless
coefficients:
\[
B = R u + S m^2 \hspace{2cm} C = T u^2 + V m^4 + W u m^2
\]
\beq
D = M m^6 + N m^4 u + P u^2 m^2 + Q u^3
\label{coefi89}
\eeq
The first requirement is that the cubic must be invariant if we perform a 
transformation of the $Z_2$ 
invariant group discussed around eq.(\ref{su2ano}). It acts on $x,y$ and $u$
as follows:
\beq
y \rightarrow \pm iy \hspace{2cm} x \rightarrow - x 
\hspace{2cm} u \rightarrow - u
\label{z2768}
\eeq
without transforming $m^2$. The invariance of the cubic under this 
transformation implies that a number of coefficients must be vanishing:
\beq
S= W = M = P= 0
\label{vancoe7}
\eeq
If these coefficients are vanishing it is easy to check that the cubic is also
invariant under a more general $U(1)_R$ transformation, under which $x, u$ and
$y$ transform respectively with weight $4, 4$ and $ 6$ provided
that in addition we also transform $\theta$ as in eq.(\ref{resy}) and 
correspondingly also $m$ as $\Lambda$ in eq.(\ref{lambdatra}). This is a 
consequence of the fact that this is a "symmetry" of the theory as discussed
after eq.(\ref{resy}).

The second requirement that restricts further the form of the cubic is the fact
that the cubic in the limit of large $u$ must reproduce the semiclassical
behaviour $\frac{da}{du} \sim 1/\sqrt{u}$. This implies
\beq
T = Q = 0
\label{con908}
\eeq

In conclusion the two previous requirements imply the following form for
the cubic:
\beq
y^2 = x^3 + Bux^2 + Cx m^4 + D m^4 u
\label{cub97}
\eeq
where now $B,C$ and $D$ are dimensionless parameters to be determined.

A torus that is described by the cubic in eq.(\ref{cub97}) degenerates when 
the discriminant $\Delta$ of the cubic is vanishing:
\beq
\Delta \equiv 4 m^4 B^3 D u^4 - \left[B^2 C^2 + 18 BCD - 27 D^2 \right]m^8 u^2
+ 4 m^{12} C^3 =0
\label{discri}
\eeq
Our third assumption is that the points of degeneracy of the torus precisely 
coincide with
the two points $u = \pm m^2$ where $\tau (u)$ is singular with monodromy 
properties given respectively in eqs.(\ref{tau34}) and (\ref{mon67}). In order
to impose this condition we need to distinguish two cases. The first one 
corresponds to a value of $D \neq 0$. In this case the solutions of 
eq.(\ref{discri}) are given by:
\beq
u^2 = m^4 \frac{R \pm \sqrt{R^2 - 64 B^3 C^3 D}}{8 B^3 D}
\hspace{2cm} R \equiv B^2 C^2 + 18 BCD - 27 D^2
\label{equaDneq0}
\eeq

The requirement that the discriminant in eq.(\ref{discri}) is vanishing only
for $u^2 = m^4$ gives the two eqs.
\beq
R^2 = 64 B^3 C^3 D \hspace{2cm} R = 8 B^3 D
\label{con768}
\eeq
They imply 
\beq
D = \frac{C^3}{B^3}
\label{detD}
\eeq
and the equation:
\beq
t^4 - 8 t^3 + 18 t^2 - 27 = (t+1) (t-3)^3 =0
\hspace{2cm} t = \frac{B^2}{C}
\label{equa658}
\eeq
In conclusion, if $D \neq 0$, the curve is given by:
\beq
y^2 = x^3 + B u x^2 + \frac{B^2 m^4}{t} x + \frac{B^3 m^4 u}{t^3}
\label{cuDneq0}
\eeq 
where $t$ can only assume the two values $t= -1$ or $t=3$.

If instead $D=0$ then the vanishing of the discriminant in eq.(\ref{discri})
implies:
\beq
u^2 = 4 \frac{C}{B^2} m^4  \Longrightarrow B^2 = 4 C
\label{D=0}
\eeq
and the cubic has the form:
\beq
y^2 = x^3 + Bu x^2 + \frac{B^2}{4} m^4 x
\label{D=0cu}
\eeq

The form of the curves in eqs.(\ref{cuDneq0}) and (\ref{D=0cu}) can be further
simplified by a rescaling of $x$ and $y$
\beq
y \rightarrow \lambda^3 y \hspace{2cm} x \rightarrow \lambda^2 x
\label{lamtra}
\eeq
with an arbitrary parameter $\lambda$, without changing the curve. In this way,
by choosing $\lambda^2 = -B$, we can eliminate $B$ from the two curves 
arriving at
\beq
y^2 = x^3 - u x^2 + \frac{m^4}{t} x - \frac{m^4 u}{t^3}
\label{fincu1}
\eeq
if $D \neq 0$ and 
\beq
y^2 = x^3 - u x^2 + \frac{m^4}{4} x
\label{fincu2}
\eeq
if $D=0$. This last curve is the one chosen by Seiberg and Witten in 
Ref.~\cite{SEIWI2}. 

By choosing $t =-1$ in eq.(\ref{fincu1}) we get the 
curve found by Seiberg and Witten in their original paper~\cite{SEIWI}:
\beq
y^2 = ( x - m^2 )( x + m^2 )(x -u)
\label{fincu3}
\eeq
Inserting the previous curve in eq.(\ref{dadad}) and
integrating over u we get:
\beq
a(u) = - 2 \alpha \oint_{a} dx \frac{\sqrt{x-u}}{\sqrt{x^2 -m^4 }}=
- 4 \alpha \int_{-m^2}^{m^2} dx \frac{\sqrt{x-u}}{\sqrt{x^2 -m^4 }}
\label{au5}
\eeq
and
\beq
a_D (u) = - 2 \alpha \oint_{b} dx 
\frac{\sqrt{x-u}}{\sqrt{x^2 -m^4 }} = - 4 \alpha \int_{m^2}^{u} dx 
\frac{\sqrt{x-u}}{\sqrt{x^2 -m^4 }}
\label{6.11}
\eeq
The factor $2$ follows from the fact that the integral under the cut gives
the same  contribution of the one over the cut.
Remembering that $a(u) \rightarrow \sqrt{2u}$ for $u \rightarrow \infty$
we get after some calculation:
\beq
\alpha = - \frac{1}{2 \sqrt{2} \pi}
\label{nor89}
\eeq
arriving at the explicit solution constructed by Seiberg-Witten~\cite{SEIWI}:
\beq
a(u) = \frac{\sqrt{2}}{\pi} \int_{-1}^{1} dx 
\frac{\sqrt{x-u}}{\sqrt{x^2 -m^4}}
\hspace{2cm}
a_D (u) = \frac{\sqrt{2}}{\pi} \int_{1}^{u} dx 
\frac{\sqrt{x-u}}{\sqrt{x^2 -m^4 }}
\label{6.1}
\eeq
It can be shown that both $a(u)$ and $a_{D} (u)$ are singular at the points 
$u= \pm m^2, \infty$ with exactly the monodromies discussed in the previous 
section. In terms of the previous functions one can construct the coefficient 
$\tau (u)$
of the kinetic term of the gauge field that satisfies by construction the 
important property: $Im \tau > 0$ for any $u$.
By a shift $x \rightarrow x + u/3$ we can rewrite the curve in 
eq.(\ref{fincu3}) in the form:
\beq
y^2 = x^3 + a x + b
\label{curve56}
\eeq
where
\beq
a= - m^4 - u^2 /3 \hspace{1cm};\hspace{1cm} b= - \frac{2}{3} u \left( m^4
- u^2 /9 \right) 
\label{ab678}
\eeq
Since a torus is uniquely identified by the complex parameter $\tau$ varying 
in the fundamental region ($|\tau| \geq 1$, $- \frac{1}{2} \leq Re \tau \leq 
\frac{1}{2}$ or equivalently by the ratio
\beq
r \equiv \frac{a^3}{b^2} = - \frac{9}{4} \frac{(m^4 + u^2 /3)^3}{u^2 
(m^4 - u^2 /9)^2}
\label{r345}
\eeq
we see that, for each value of $r$ corresponding to $\tau$ in the fundamental 
region, we get six values of $u$ by solving eq.(\ref{r345}). This means that $u$
varies in a region that corresponds to $6$ images of the fundamental region of
$SL(2,Z)$, that is, in fact, the fundamental region of $\Gamma_2$.

Finally if we choose instead $t =3$ in eq.(\ref{fincu1}) we get the curve
\beq
y^2 = x^3 - u x^2 + \frac{m^4}{3} x - \frac{m^4 u}{27}
\label{fincu4}
\eeq
A study of the monodromies around the points $\pm m^2$ shows that their degrees
is not infinite, as in the case of the curve in eq. (\ref{fincu3}) where one 
found logarithmic branch points\footnote{F. Gliozzi, private communication}. 
They do not seem to correspond to values of $u$
where additional particles become massless. We do not 
discuss the curve in eq.(\ref{fincu4}) further in these lectures.

\sect{$N=4$ super Yang-Mills}
\label{Neq24}
In the previous sections we have seen that supersymmetry is an essential
ingredient for having a dual theory in the sense of Montonen-Olive. We have
also seen that $N=2$ super Yang-Mills is the simplest supersymmetric theory 
containing monopole and dyon solutions whose mass is fixed in the full 
quantum theory by the supersymmetry algebra and not just given by a BPS 
formula valid in the classical theory as in the case of the Georgi-Glashow 
model. $N=2$ super Yang-Mills has also the attractive feature of having  a 
supersymmetry algebra that contains only two central charges, the electric 
and magnetic charges. 
On the other hand it is known from the work of Ref.~\cite{OSBORN} that the
the monopole solution of the $N=2$ theory belongs to the hypermultiplet that
does not contain a spin $1$, while the $W$-bosons, that have spin $1$, belong
to the $N=2$ chiral multiplet. This shows immediately that the monopoles and
the $W$-bosons of $N=2$ super Yang-Mills cannot be dual in the sense of 
Montonen-Olive. Therefore in the search of a theory in which the Montonen-Olive
duality is realized one is brought to consider a theory with more 
supersymmetry and one is naturally led to $N=4$ super Yang-Mills.

We start by rewriting the Lagrangian of this theory, 
given in eq.(\ref{neq4sYM}), in the same notation used for the $N=2$
theory. We get:
\[
L=  \frac{1}{4 \pi} Im \left\{ \tau \left[ - \frac{1}{4} \left( 
F^{a}_{\mu \nu} F^{a \, \mu \nu} - i F_{\mu \nu}^{a} {}^{*} F_{a}^{\mu \nu}
\right) + \frac{1}{2} \sum_{i=1}^{3} \left( D_{\mu} A_{i} \right)^{a} 
\left( D^{\mu} A_{i} \right)^{a} + \right. \right.
\]
\[
\frac{1}{2} \sum_{i=1}^{3} \left( D_{\mu} B_{i} \right)^{a} 
\left( D^{\mu} B_{i} \right)^{a} - V(A_i , B_j) +     
\]
\beq
\left. \left.  + 
- \frac{i}{2}  {\bar{\psi}}^a \gamma^{\mu} \left( D_{\mu} \psi 
\right)^a - \frac{1}{2}  
f^{abc} {\bar{\psi}}^{a} \alpha^{i} A^{b \, i} \psi^{c} - i 
\frac{1}{2} 
f^{abc} {\bar{\psi}}^{a} \beta^{j} \gamma_5 B^{b \, j} \psi^{c} \right] 
\right\} 
\label{neq4sYM2}
\eeq
where the potential is equal to:
\beq
V( A_i , B_j ) = \frac{1}{4} f^{abc} A_{i}^{b} A_{j}^{c}
f^{afg} A_{i}^{f} A_{j}^{g} + \frac{1}{4} f^{abc} B_{i}^{b} B_{j}^{c}
f^{afg} B_{i}^{f} B_{j}^{g} + \frac{1}{2} f^{abc} A_{i}^{b} B_{j}^{c}
f^{afg} A_{i}^{f} B_{j}^{g}
\label{potn=42}
\eeq
and $\tau$ is given in eq.(\ref{tau}).

It is known since long time that this theory, being free from ultraviolet 
divergences~\cite{MANDE,BRINK,HOWE1,HOWE2}, has a vanishing 
$\beta$-function~\cite{SOHNIUS} 
and no chiral anomaly. The vanishing of the $\beta$-function
and the absence of the chiral anomaly follow respectively from 
eqs.(\ref{bfunct}) and (\ref{chiano}) as explained at the end of Appendix 
\ref{B}. 
$N=4$ super Yang-Mills is  a conformal invariant theory at the
full quantum level. Conformal invariance is spontaneously broken if some
scalar field gets a non vanishing vacuum expectation value.

Many of the properties found in $N=2$ super Yang-Mills, as the existence of 
a manifold of inequivalent 
vacua and  of monopole and dyon solutions, are also valid for  
$N=4$ super Yang-Mills. The supersymmetry algebra contains also a number
of central charges~\cite{OSBORN} ($12$ and not $2$ as in the $N=2$ theory). 
As in the case of the Georgi-Glashow model all particles of the spectrum have
electric and magnetic charges that lie on a two-dimensional lattice
\beq
q +ig = q_0 ( n_e + \tau n_m )  \hspace{2cm} \tau = \frac{\theta}{2 \pi}
+i \frac{4 \pi}{q_{0}^{2}}
\label{chalatt7}
\eeq
with periods $q_0$ and $q_0 \tau$ ($q_0 =e$ is the electric charge of the 
$W$-boson). In this case, however, the charges are not modified by quantum 
corrections.
Again as a consequence of the supersymmetry algebra the BPS states of the 
theory have a mass given by the classical formula:
\beq
M = \sqrt{2} | a n_e + \tau a n_m | = \sqrt{2} \frac{|a|}{e}| q +ig|
\label{examass}
\eeq
that, because of $N=4$ supersymmetry, is not modified by quantum corrections.
The question now is how to select those states that are single particle 
states. This can be easily done if we restrict ourselves to BPS saturated 
states  having a mass  given in eq.(\ref{examass}).
A single particle BPS-saturated state with mass $M$ must be stable and this 
is the case if it cannot decay into a couple of BPS saturated states with
mass $M_1$ and $M_2$, i.e.
\beq
M < M_1 +M_2
\label{stabi}
\eeq
Using for the mass the expression in eq.(\ref{examass}) together with
the exact expression for the charge given in eq.(\ref{chalatt7}) one can easily
see, by means of the Schwarz inequality, that eq.(\ref{stabi}) is satisfied 
if and only if the integers $(n_e , n_m )$ in eq.(\ref{chalatt7}) are 
coprimes. This implies that
the stable states with zero magnetic charge $(n_e ,0)$ are only the three 
states with $n_e=0, \pm 1$; the states with magnetic charge corresponding to 
$n_m = \pm 1$ are all stable states; the states with magnetic charge 
corresponding  to $n_m = \pm 2$ are only stable if their electric charge 
corresponds to odd values of $n_e$; the states with magnetic charge 
$n_m = \pm 3$ are stable if $n_e$ is different from $0$ and is not a multiple 
of $3$ and so on. 

An explicit analysis of the fermionic zero modes, as we have done in the
case of the $N=2$ theory, shows that the number of zero modes is twice larger 
than that of the $N=2$ theory. This means that, instead of the expansion 
given in eq.(\ref{quafi}), we have in this case:
\beq
\chi = \sum_{i=1}^{2} \left[ a_{1/2}^{(i)} \chi_{0}^{1/2,i} + a_{-1/2}^{(i)} 
\chi_{0}^{-1/2,i} \right] + \Delta \chi
\label{quafi4}
\eeq
Therefore we get twice the number of creation and annihilation operators 
that we had in  $N=2$ super Yang-Mills and we can 
construct a bigger number of states given in Table 2.
\begin{table}     
\caption{$N=4$ monopole multiplet.}
\begin{center}
\begin{tabular}{|@{}c|c|c|l@{}|}\hline
STATE                                           &$S_z$&$\#$ OF STATES \\\hline
$|~ 0 >$                                             & 0         & 1 \\ \hline
$a^{(i)\dagger}_{\pm 1/2} |~0>$                    & $\pm 1/2$ & 4 \\ \hline
$a^{(i)\dagger}_{1/2}  a^{(j)\dagger}_{1/2}  |~0>$           & 1   &1\\ \hline
$a^{(i)\dagger}_{-1/2} a^{(j)\dagger}_{-1/2} |~0>$    & -1         & 1 \\ \hline
$a^{(i)\dagger}_{1/2}  a^{(j)\dagger}_{-1/2} |~0>$  &  0         & 4 \\ \hline
$a^{(i)\dagger}_{1/2}  a^{(j)\dagger}_{-1/2}a^{(k)\dagger}_{1/2} |~0>$ &  
1/2   & 2 \\ \hline
$a^{(i)\dagger}_{-1/2}  a^{(j)\dagger}_{1/2}a^{(k)\dagger}_{-1/2} |~0>$ 
&  -1/2   & 2 \\ \hline
$a^{(i)\dagger}_{1/2}  a^{(j)\dagger}_{1/2}a^{(h)\dagger}_{-1/2} 
a^{(k)\dagger}_{-1/2}|~0>$&0&1
\\ \hline
\end{tabular}
\end{center}
\end{table}
Those states fill a unique short representation
of $N=4$ supersymmetry containing one state of spin $1$, four states of spin 
$1/2$ and five states with spin $0$. Both the $W$-bosons and the monopoles 
belong to this unique multiplet and this fact makes  the realization 
of the Montonen-Olive duality in this theory~\cite{OSBORN} possible.

\sect{Riformulation of Montonen-Olive duality}
\label{RIMO}

We are now in a position to riformulate the Montonen-Olive duality for 
$N=4$ super Yang-Mills in a way in
which the $W$-bosons, the magnetic monopoles and more in general 
all the dyons of the spectrum are treated in a completely democratic 
way~\cite{OLIVE}. We will see that we will not just have
an electric and magnetic description, but we will have an infinite number
of descriptions depending on which states of the charge lattice we are 
choosing as fundamental particles. 

The usual formulation of the $N=4$ super Yang-Mills is obtained by considering
the states with zero magnetic charge and with electric charge equal to 
$\pm q_0$ corresponding to the $W$-bosons that get a mass 
through the Higgs mechanism, together with the massless 
states at the origin of the charge lattice having vanishing electric and 
magnetic charges and corresponding to the photon and Higgs particle. Selecting
these states we have
determined one of the periods of the lattice. The other period is also fixed
when we specify the value of the angle $\theta$. 
We then ascribe a short $N=4$ supermultiplet to each of the three states with 
charge equal to $0$, $q_0$ and $- q_0$ and, having fixed the value of $\theta$,
we can explicitly write the full Lagrangian of $N=4$ super Yang-Mills 
containing only the states of the lattice that we have chosen. If the theory
is dual in the sense of Montonen-Olive the other stable states of the charge
lattice must appear as solitons or bound states of solitons.

On the other hand if the theory is dual in the sense of Montonen-Olive one
could also start from another couple of stable states of the charge lattice
corresponding to a certain dyon of the theory with a complex charge given by 
$ \pm q_{0} '$ and with mass equal to $M = \sqrt{2} |a| | q_{0} '|/e$, 
together with the 
massless photon and Higgs states located at the origin of the charge lattice
and specify the vacuum angle $\theta$ by giving another vector 
$ q_{0} ' \tau ' $ of the lattice that is not aligned with $q_0 '$. We can 
again ascribe a $N=4$ short multiplet
to any of the states previously chosen and write, as before, a $N=4$ super
Yang-Mills Lagrangian containing the states with charges equal to $0$ and 
$ \pm q_0 '$ and with a specified vacuum angle $\theta$.  Also in this case
the remaining stable states of the charge lattice will show up as solitons
or bound states of solitons of the new Lagrangian. Duality 
in the sense of Montonen and Olive means that all the theories
based on any pair of independent vectors of the  charge lattice are 
equivalent.

Since the vectors $q_0 '$ and $ q_0 ' \tau '$ form an alternative basis of the
charge lattice it must be possible to express them  in terms of the original 
vectors $q_0$ and $ q_0 \tau$ through the relation:
\beq
q_0 ' \tau ' = a q_0 \tau + b q_0 \hspace{2cm} 
q_0 '  = c q_0 \tau + d q_0 
\label{modtra}
\eeq
with $a,b,c$ and $d$ integer numbers. 

Since it must also be possible to express $q_0 $ and $q_0 \tau$ in terms
of $q_0 '  $ and $q_0 '  \tau '$ the integer parameters of the transformation 
must satisfy the equation:
\beq
ad - bc =1
\label{deteq1}
\eeq

Therefore the transformations from a basis to another basis form the modular
group $SL(2, Z)$.

Eqs. (\ref{modtra}) imply a relation between $\tau$ and $\tau '$ given by
\beq
\tau ' = \frac{a \tau + b}{c \tau + d}
\label{modtra1}
\eeq
that provides a connection between the values of the parameters 
$( \theta, q_0)$ in the two choices of basis vectors and actions.

The modular group is generated by the two transformations:
\beq
T \hspace{1cm}: \hspace{1cm} \tau \rightarrow \tau +1 \hspace{1cm}
\rightarrow \hspace{1cm} \theta \rightarrow \theta + 2 \pi
\label{ti}
\eeq
that is a symmetry of the theory because the physics is periodic when we
translate $\theta$ by $2 \pi$, and
\beq
S \hspace{1cm}: \hspace{1cm} \tau \rightarrow - \frac{1}{\tau}  \hspace{1cm}
\rightarrow \hspace{1cm} q_0  \rightarrow \frac{4 \pi \hbar }{q_0} 
\hspace{1cm}(\,\, if \,\, \theta=0)
\label{si}
\eeq
that relates weak coupling with strong coupling. 

The mass of the BPS-saturated states of the theory is proportional to the 
absolute value of the charge
\beq
M \sim | q- ig| = | q_0 ( n_m \tau + n_e )|
\label{massbps}
\eeq
and is left invariant if we transform $\tau$ as in eq.(\ref{modtra1})
and $q_0$ and the charge vector $\left( \begin{array}{c} n_m \\
			n_e  \end{array} \right)$ as follows
\beq
q_0 \rightarrow q_0 ' = q_0 ( c \tau + d) \hspace{1cm}
\left( \begin{array}{c} n_m \\
			n_e  \end{array} \right) \rightarrow
		       \left( \begin{array}{c} n_m ' \\
					       n_e '  \end{array} \right) =
\left( \begin{array}{cc} d & -c \\
		 -b &  a   \end{array} \right)
  \left( \begin{array}{c} n_m \\
			n_e  \end{array} \right)
\label{modtrasf}
\eeq
with $ad-bc=1$.

The modular group does not only perform a transformation from a system of 
basis vectors to another one, but acts also on the integer charge vector
$\left( \begin{array}{c} n_m \\
			n_e  \end{array} \right)$ rotating it into a new integer
charge vector $\left( \begin{array}{c} n_m ' \\
			n_e '  \end{array} \right)$.
In other words a modular transformation transforms $q- i g$ expressed in terms 
of the basis vectors $q_0$ and $q_0 \tau$ and of the integers $n_e$ and $n_m$ 
into
an expression having the same form in terms of the new basis vectors $q_0 '$ 
and $q_0 ' \tau '$ and of the new integers $n_e '$ and $n_m '$ related to the 
old 
ones by eqs.(\ref{modtra1}) and (\ref{modtrasf}). The invariance under the 
modular group requires that the presence in the spectrum of a state with a 
certain pair of integers implies also the presence  of the
state with other integers obtained from the first ones by the action of a
modular transformation as in the second equation of  (\ref{modtrasf}).
 
In particular from  eq.(\ref{modtrasf}) it follows that, given the
existence in the spectrum of the $W^{+}$-boson corresponding to $n_m =0$ and 
$n_e =1$,
the invariance under the modular group implies also the existence of the
transformed state:
\beq
\left( \begin{array}{c} 0 \\
			1  \end{array} \right) \rightarrow
		       \left( \begin{array}{c} -c \\
					       a  \end{array} \right) =
\left( \begin{array}{cc} d & -c \\
		 -b &  a   \end{array} \right)
  \left( \begin{array}{c} 0 \\
			1  \end{array} \right)
\label{modtrasf1}
\eeq
Since the condition $ad-bc=1$ is equivalent to require that $c$ and $a$ are 
coprimes, the existence of the $W^+$-boson implies the existence in the
spectrum of all stable states of the charge lattice as discussed at the end of
the previous section. This is a direct consequence of the Montonen-Olive 
duality.

Let us consider the states with $c=-1$. They are of the type $ \left( 
			     \begin{array}{c}  1 \\
					       a  \end{array} \right) $
where $a$ is an arbitrary integer. These are the dyons of $N=4$ super 
Yang-Mills required, as discussed after eq.(\ref{gencha}), by the $\theta$
periodicity corresponding to the generator $T$ of the modular group.
The next case is $c =-2$. In this case we expect the existence of the states
$ \left( \begin{array}{c}                      2 \\
					       a  \end{array} \right) $
where $a$ is odd. The existence of such states was shown by Sen~\cite{SEN}.
Evidence for the existence of stable states with higher values of $c$ can be 
found in Ref.~\cite{PORRATI,SEGAL,LOWE1,LOWE2,FERRARI}.
 
\vskip 1.0cm

{\large {\bf {Acknowledgements}}}
\vskip 0.5cm
I thank many of those attending these lectures at Les Houches Summer School
and at the Universities of Milan and Naples for many questions that helped 
to write them hopefully in a clearer way. I  thank  M. Bill{\`{o}}, F. 
Gliozzi, D. Olive and I. Pesando for many enlightening discussions regarding 
the mathematical aspects that are at the basis of the Seiberg-Witten 
construction, Kim Splittorff for his critical reading of various drafts of 
the manuscript and A. Liccardo and F. Pezzella for helping me to correct many
misprints. I am also grateful to M. Bianchi and F. Fucito for many 
useful discussions on the Seiberg-Witten approach during my many visits in 
Rome at the University of Tor Vergata.

\vskip 1.0cm

\appendix
\sect{ Appendix A}
\label{A}

The action of the Georgi-Glashow model is invariant under the gauge
transformations:
\beq
\Phi \rightarrow U \Phi U^{-1} \hspace{2cm}
A_{\mu} \rightarrow U A_{\mu} U^{-1} + \frac{1}{ie} U \partial_{\mu} U^{-1}
\label{gt3}
\eeq
where
\beq
\Phi = \Phi_a T_a  \hspace{2cm} A^{\mu} = A^{\mu}_{a} T_a
\label{matrino}
\eeq
and $T^a$ are the generator of the gauge group in the adjoint representation:
\beq
[T^a ,T^b ] = i f^{abc} T^c \hspace{2cm} T^{a}_{AB} = i f^{AaB}
\label{ad4}
\eeq
The covariant derivative and the Yang-Mills field strenght are given 
respectively by:
\beq
D_{\mu} \Phi = \partial_{\mu} \Phi + ie [A_{\mu}, \Phi] \hspace{1cm}
F_{\mu \nu} = \partial_{\mu} A_{\nu} - \partial_{\nu} A_{\mu} + i e
[ A_{\mu} , A_{\nu} ]
\label{coma}
\eeq
Under a gauge transformation they transform as
\beq
D_{\mu} \Phi \rightarrow U D_{\mu} \Phi U^{-1} \hspace{2cm}
F_{\mu \nu} \rightarrow U F_{\mu \nu} U^{-1}
\label{gt5}
\eeq
If we write $U = {\rm e}^{ i \omega^a T^a }$ we can write the action of the 
gauge transformation on the fields $\Phi^a$ and $A_{\mu}^{a}$. We get:
\beq
\delta \Phi_a = - \epsilon_{abc} \omega_b \Phi_c
\hspace{2cm}
\delta A_{\mu}^{a} = - \frac{1}{e} (D_{\mu} \omega )^a
\label{gtfa}
\eeq

In the second part of this appendix we show that the quantity $K$ defined
in eq.(\ref{topcha}) is an integer. In fact making use of the Stoke's theorem
we can rewrite eq.(\ref{topcha}) in the following form:
\beq
K = \frac{1}{8 \pi a^3} \int d \Omega \epsilon^{abc} \Phi^a {\hat{r}}_i
\epsilon_{ijk} \frac{\partial \Phi^b}{\partial {\hat{r}}^{j}}
\frac{\partial \Phi^c}{\partial {\hat{r}}^{k}}
\label{top4}
\eeq
where ${\hat{r}}^i = r^i /r$.
Parametrizing the three components of $n^a \equiv \Phi^{a}/a$ at spatial
infinity, where $n^2 =1$, in terms of the angles $\omega$ and $\chi$:
\beq
n^1 = \sin \omega \cos \chi \hspace{2cm}
n^2 = \sin \omega \sin \chi \hspace{2cm}
n^3 = \cos \omega
\label{para6}
\eeq
and the unit sphere in three-dimensional space by:
\beq
{\hat{r}}^1 = \sin \theta \cos \varphi \hspace{2cm}
{\hat{r}}^2 = \sin \theta \sin \varphi \hspace{2cm}
{\hat{r}}^3 = \cos \theta 
\label{unitsph}
\eeq
the following relation can be shown 
\beq
\shalf \epsilon^{abc} n^a {\hat{r}}_{i} \epsilon_{ijk} 
\frac{\partial n^b}{\partial {\hat{r}}^{j}}
\frac{\partial n^c}{\partial {\hat{r}}^{k}} =
{\hat{r}}_i \epsilon_{ijk} \frac{\partial \cos \omega}{\partial {\hat{r}}^{k}}
\frac{\partial \chi}{\partial {\hat{r}}^j}= -\frac{1}{\sin \theta}
\frac{\partial ( \cos \omega, \chi)}{\partial ( \theta, \varphi)}
\label{identi8}
\eeq
where the last expression means the jacobian of the transformation from the
variables $\cos \omega$ and $\chi$ to the variables $\theta$ and $\varphi$.
Introducing the previous identity in eq.(\ref{top4}) one gets:
\beq
K = -\frac{1}{4 \pi} \int d \theta d \varphi 
\frac{\partial ( \cos \omega, \chi)}{\partial ( \theta, \varphi)}
\label{jaco6}
\eeq
showing that $K$ just counts the number of times that one covers the 
two-dimensional sphere described by the variable $n^a$ when the sphere 
at infinity in space is covered once.

In the last part of this appendix we will explicitly solve the Bogomolny
equation for the monopole and dyon.

Starting from the ansatz
\beq
\Phi_a = \frac{r^a}{e r^2} H(\xi) \hspace{1cm} 
A^{0}_{a} = \frac{r^a}{e r^2} J(\xi) \hspace{1cm}
A^{i}_{a} = - \epsilon_{aij}\frac{r^j}{e r^2} \left[1 - K (\xi) \right]  
\label{ans2}
\eeq
it is easy to compute
\beq
B_{i}^{a}= - \frac{\delta^{ai}}{e r^2} \xi K' + \frac{r^{i} r^{a}}{e r^4}
\left[ \xi K ' +1 - K^2 \right]
\label{bai}
\eeq

\beq
E_{i}^{a}= \frac{\delta^{ai}}{e r^2} J K + \frac{r^{i} r^{a}}{e r^4}
\left[ \xi J ' - J(1 + K) \right]
\label{eai}
\eeq
and
\beq
(D_i \Phi)_{a}= \frac{\delta^{ai}}{e r^2} H K + \frac{r^{i} r^{a}}{e r^4}
\left[ \xi H ' - H(1 + K) \right]
\label{dfai}
\eeq  
The ansatz in eqs.(\ref{ans2}) automatically satisfies the first eq. in
(\ref{eqs2}). The second eq. is satisfied by requiring $\lambda =0$.
Inserting the expressions in eqs.(\ref{bai}) and (\ref{dfai}) in 
eqs.(\ref{bpscon3}) we get:
\beq
\xi K' = - K{\hat{H}} \hspace{2cm} \xi {\hat{H}}' = {\hat{H}} +1 - K^2
\label{linequa}
\eeq
where ${\hat{H}} (\xi) = H(\xi) \cos \theta  = J (\xi) \coth \theta $.

If we insert instead eqs.(\ref{bai}), (\ref{eai}) and (\ref{dfai}) in 
eqs.(\ref{bps69}) and (\ref{bps70}) and we write $J (\xi)$, $J_4 (\xi)$ and 
$J_5 (\xi)$ in terms of $R (\xi)$ as in eqs.(\ref{solu98}) we get the 
following equations 
\beq
\xi K ' = - K R \hspace{2cm} \xi R' = 1 - K^2 + R
\label{equ768}
\eeq
where the constants $\alpha$, $\beta$ and $\gamma$ are related to $\theta$
through the relations:
\beq
\alpha \sin \theta + \beta \cos \theta =1 \hspace{2cm}
\gamma = \alpha \cos \theta - \beta \sin \theta
\label{rel96}
\eeq 
that imply $ \alpha^2 + \beta^2 - \gamma^2 =1$ and determine $\alpha$, 
$\beta$ and $\gamma$ as functions of $\theta$. 

In order to solve eqs.(\ref{linequa}) we introduce the new functions 
$h$ and $k$:
\beq
{\hat{H}} (\xi)  = -1 - \xi h(\xi) \hspace{2cm} K(\xi)  = \xi k (\xi)
\label{newfu}
\eeq
In terms of these new functions  eqs.(\ref{linequa}) become:
\beq
k' = hk \hspace{2cm} h' = k^2
\label{newequa2}
\eeq
They imply
\beq
\frac{d}{d \xi} \left( k^2 - h^2 \right)=0 \hspace{1cm} \Rightarrow 
\hspace{1cm} k^2 - h^2 = \alpha
\label{eq31}
\eeq
where $\alpha$ is a constant that is determined by imposing the boundary 
conditions:
\beq
\lim_{\xi \to \infty} k( \xi ) = 0 \hspace{2cm} \lim_{\xi \to \infty} 
h (\xi) = -1
\label{bound2}
\eeq
that are obtained from eqs.(\ref{bouncon}). Those boundary conditions 
require  $\alpha = -1$ and then from eq.(\ref{eq31}) we get
\beq
h^2 - k^2 = 1
\label{const2}
\eeq
Inserting it in the second eq. of (\ref{newequa2}) we get 
\beq
h' = h^2 -1
\label{equa3}
\eeq
whose solution is
\beq
h (\xi) = - \coth (\xi + \beta)
\label{sol1}
\eeq
where $\beta$ is a constant to be determined.
Inserting $h(\xi)$ given in eq.(\ref{sol1}) in the first eq. of 
(\ref{newequa2}) we get
\beq
k' = - k \left[ \coth( \xi + \beta) \right]
\label{equa4}
\eeq
whose solution is:
\beq
k(\xi) = \frac{\gamma}{\sinh (\xi + \beta)} 
\label{sol5}
\eeq
The finiteness of the energy in eq. (\ref{ene2}) requires that 
$\lim_{ \xi \to 0} K^2 = 1$. This limit is satisfied only if $\beta =0$.
Then eq.(\ref{const2}) implies $\gamma^2 =1$. Choosing $\gamma=1$ we
arrive at the solution:
\beq
h ( \xi) = - \coth \xi \hspace{2cm} k(\xi) = \frac{1}{\sinh \xi}
\label{sol8}
\eeq
that, through the relations in eq.(\ref{newfu}), correspond to the 
expressions in eqs.(\ref{dyonsol1}) and (\ref{dyonsol2}) and for 
$\theta=0$ to those in eqs.(\ref{bpslim}).

\sect{Appendix B}
\label{B}

In this appendix we start by introducing the Weyl spinor, the $N=1$ 
supersymmetry transformations and  we describe in some detail the chiral and
vector superfields. We give also the expansion of the various terms of a
$N=1$ supersymmetric Lagrangian in terms of component fields. We follow the
notation of the book by Wess and Bagger~\cite{WESS}.

The generators of the Poincar{\'e} algebra are the translational generators
$P_{\mu}$ and the generators of the Lorentz transformations $M_{\mu \nu}$.
Under a transformation of the Poincar{\'e} group the coordinate $x^{\mu}$ is 
transformed as
\beq
\label{A.1}
x^{\mu} \rightarrow \Lambda^{\mu}_{\;\;\nu} x^{\nu} + a^{\mu}
\eeq
A particular representation of the Lorentz group is given by the Dirac spinors
defined in terms of the fourdimensional $\gg$-matrices
\eq
\{\gg^{\mu} ,\gg^{\nu} \} = 2 g^{\mu \nu}
\label{A.4}
\en
The quantity
\eq
\frac{1}{2} \Sigma^{\mu \nu} = \frac{i}{4} ( \gg^{\mu}\gg^{\nu} -
\gg^{\nu} \gg^{\mu} )
\label{A.5}
\en
satisfies the Lorentz algebra. The Poincar{\'e generators acting on the Dirac 
spinors are
\eq
\label{A.6}
P^{\mu} = i \partial^{\mu} \hspace{2cm}
M^{\mu \nu} = x^{\mu} P^{\nu} - x^{\nu} P^{\mu} + \frac{1}{2} \Sigma^{\mu \nu}
\en
It is convenient to use the Weyl representation of the Dirac spinors:
\eq
\label{A.7}
\gg^{\mu} = \mat{0}{\sigma^{\mu}}{{\bar{\sigma}}^{\mu}}{0}
\en
where
\eq
\sigma^{\mu} = ( \sigma^{0}, \sigma^i ) \hspace{2cm}
{\bar{\sigma}}^{\mu} = ( \sigma^{0}, -\sigma^i ) 
\label{A.8}
\en
$\sigma^i$ are the Pauli matrices that satisfy the relation
\eq
\sigma^i \sigma^j = \delta^{ij} + i \epsilon^{ijk} \sigma^k
\label{A.10}
\en
Then
\eq
\gg_5 \equiv i \gg^0 \gg^1 \gg^2 \gg^3 = \mat{-1}{0}{0}{1}
\label{A.11}
\en
In this representation the upper [bottom] two components have left [right]
chirality:
\eq
\Psi = \Psi_L + \Psi_R  \hspace{.5cm};\hspace{.5cm} \Psi_L = 
\left(\frac{ 1 - \gg_5}{2} \right) \Psi \hspace{1cm} \Psi_R = 
\left(\frac{ 1 + \gg_5}{2} \right) \Psi
\label{A.12}
\en
The left and right chirality spinor fields 
\eq
\label{A.13}
\Psi_L = \psi_{\aa} \hspace{1cm} \Psi_R = {\bar{\cc}}^{\dot{\aa}}
\hspace{1cm} \Psi = \vecc{\psi_{\aa}}{ {\bar{\cc}}^{\dot{\aa}}}
\en
are called undotted and dotted Weyl spinors respectively. The generators 
of rotations and boosts are given by
\eq
\frac{1}{2} \Sigma^{ij} = \frac{1}{2} \epsilon^{ijk} 
\mat{\sigma^k}{0}{0}{\sigma^k} \hspace{1cm} \frac{1}{2} \Sigma^{0i} = 
\frac{1}{2}  \mat{-i \sigma^i}{0}{0}{i \sigma^i}   
\label{A.14}
\en
In Dirac theory the charge conjugated Dirac spinor is given by
\eq
\Psi^{c} = C {\bar{\Psi}}^{T} \hspace{2cm} C = \omega \gg^0 \gg^2
= \omega \mat{- \sigma^2}{0}{0}{\sigma^2}
\label{A.15}
\en
with $|\omega| =1$ and $ \bar{\Psi} \equiv \Psi^{\dagger} \gg^0$. Choosing for 
convenience $\omega = -i $ we get
\eq
{\bar{\Psi}}^T = \vecc{\cc^{\aa}}{{\bar{\psi}}_{\dot{\aa}}} \hspace{1cm}
\Psi^{c} = \vecc{i \sigma^2 \cc}{-i \sigma^2 {\bar{\psi}}} =  
\vecc{\epsilon_{\aa \bb} \cc^{\bb}}{ \epsilon^{\dot{\aa} \dot{\bb}} 
{\bar{\psi}}_{\dot{\bb}}}
\equiv \vecc{\cc_{\aa}}{{\bar{\psi}}^{\dot{\aa}}} 
\label{A.16}
\en
where we have introduced the antisymmetric matrices:
\eq
\epsilon_{\aa \bb} = \mat{0}{1}{-1}{0} \hspace{2cm}
\epsilon^{\dot{\aa} \dot{\bb}} = \mat{0}{-1}{1}{0} 
\label{A.17}
\en
In conclusion we have
\eq
\Psi = \vecc{ \psi_{\aa}}{{\bar{\cc}}^{\dot{\aa}}} 
\hspace{2cm}
\Psi^{c} = \vecc{ \cc_{\aa}}{{\bar{\psi}}^{\dot{\aa}}}
\label{A.18}
\en
From eq.(\ref{A.16}) one gets how to raise undotted spinors and lower dotted 
ones:
\eq
\cc^{\aa} = \ee^{\aa \bb} \cc_{\bb} \hspace{2cm}
{\bar{\psi}}_{\dot{\aa}} = \ee_{\dot{\aa} \dot{\bb} } {\bar{\psi}}^{\dot{\bb}}
\label{A.19}
\en
where

\eq
\label{A.20}
\ee^{\aa \bb} = \mat{0}{-1}{1}{0} \hspace{2cm}
\ee_{\dot{\aa} \dot{\bb}} = \mat{0}{1}{-1}{0}
\en
A Dirac spinor has four independent complex components. A Majorana spinor
satisfies the property
\eq
\Psi_{M} = \Psi_{M}^{c}
\label{A.21}
\en
and has therefore only two independent complex components.
A Dirac spinor is transformed under a Lorentz transformation by the following
matrix
\eq
S_{AB} = \left( e^{- \frac{i}{4}  \omega_{\mu \nu}  \Sigma^{\mu \nu}} 
\right)_{AB} 
\label{A.22}
\en
where
\eq
\label{A.23}
 \frac{1}{2} \Sigma^{\mu \nu} =\frac{i}{4} [\gamma^{\mu} , \gamma^{\nu} ] = 
\mat{ i \sigma^{\mu \nu}}{0}{0}{i {\bar{\sigma}}^{\mu \nu}}
\en
with
\eq
\label{A.24}
\left( \sigma^{\mu \nu} \right)_{\aa}^{\;\;\bb} \equiv \frac{1}{4} 
\left( \sigma^{\mu} {\bar{\sigma}}^{\nu} - \sigma^{\nu} {\bar{\sigma}}^{\mu}
\right)_{\aa}^{\;\;\bb} \hspace{1cm}
\left( {\bar{\sigma}}^{\mu \nu} \right)^{\dot{\aa}}_{\;\;\dot{\bb}} \equiv 
\frac{1}{4} \left( {\bar{\sigma}}^{\mu} \sigma^{\nu} - {\bar{\sigma}}^{\nu} 
\sigma^{\mu} \right)^{\dot{\aa}}_{\;\;\dot{\bb}}
\en
that are consistent with the following index structure 
\eq
( \sigma^{\mu} )_{\aa \dot{\aa}}  \hspace{2cm}
( {\bar{\sigma}}^{\mu} )^{\dot{\aa} \aa}
\label{A.25}
\en
We see that the left and right component of a Dirac spinor transform 
independently under a Lorentz transformation. A Dirac spinor is a reducible 
representation of the Lorentz group.

Under a Lorentz transformation an undotted spinor transforms as 
\eq
\psi_{\aa} \rightarrow ( S )_{\aa}^{\;\;\bb} \psi_{\bb} \equiv
\left( e^{\frac{1}{2} \omega_{\mu \nu} \sigma^{\mu \nu}} 
\right)_{\aa}^{\;\;\bb} \psi_{\bb}
\label{A.26}
\en
while the Lorentz transformation of a dotted spinor is given by
\eq
{\bar{\cc}}^{\dot{\aa}} \rightarrow 
( [S^{\dagger}]^{-1} )^{\dot{\aa}}_{\;\;\dot{\bb}} 
{\bar{\chi}}^{\dot{\beta}} \equiv
\left( e^{\frac{1}{2} \omega_{\mu \nu} {\bar{\sigma}}^{\mu \nu}} 
\right)^{\dot{\aa}}_{\;\;\dot{\bb}} \bar{\cc}^{\dot{\bb}}
\label{A.27}
\en
They are obtained from one another through the identity 
$[( \sigma^{\mu \nu})_{\alpha}^{\,\, \beta} ]^{\dagger} = - 
( {\bar{\sigma}}^{\mu \nu})^{\dot{\beta}}_{\,\, {\dot{\alpha}}}$. $S$ is a 
matrix of $SL(2,C)$. 
The Lorentz transformation of the undotted and dotted spinors obtained from
the previous one by lowering or raising the index are given by:
\beq
\psi^{\aa} \rightarrow ( [S^{-1}]^T )_{\;\;\bb}^{\aa} \psi^{\bb} 
\hspace{2cm}
{\bar{\cc}}_{\dot{\aa}} \rightarrow [( S^{\dagger})^{T}
]_{\dot{\aa}}^{\;\;\dot{\bb}} {\bar{\cc}}_{\dot{\bb}}
\label{A.30}
\en
The superscript $T$ means the transposed matrix.
Using the previous transformation rules it is easy to show that $ 
\psi^{\aa} \cc_{\aa}$, ${\bar{\psi}}_{\dot{\aa}} {\bar{\cc}}^{\dot{\aa}}$ and 
$ \psi^{\aa} (\sigma^{\mu})_{\aa \dot{\aa}} \partial_{\mu} 
{\bar{\cc}}^{\dot{\aa}} $transform as scalars under Lorentz transformations.
The following relation can also be easily shown
\eq
( {\bar{\sigma}}^{\mu} )^{\dot{\aa} \aa} = \ee^{\dot{\aa} \dot{\bb}} 
\ee^{\aa \bb}
\sigma^{\mu}_{\bb \dot{\bb}}
\label{A.31}
\en
We list here a number of the useful identities
\eq
\label{A.32}
( \bar{\psi} \bar{\chi} )^{+} = \cc \psi \equiv \cc^{\aa} \psi_{\aa} = 
-\cc_{\aa} \psi^{\aa} = \psi^{\aa} \cc_{\aa} =  \psi \cc
\en

\eq
(\psi \chi )^{+}=
{\bar{\cc}} {\bar{\psi}} \equiv {\bar{\cc}}_{\dot{\aa}} 
{\bar{\psi}}^{\dot{\aa}} = - {\bar{\cc}}^{\dot{\aa}} {\bar{\psi}}_{\dot{\aa}}
= {\bar{\psi}}_{\dot{\aa}} {\bar{\cc}}^{\dot{\aa}} = {\bar{\psi}} {\bar{\cc}}
\label{A.33}
\en


\eq
\label{A.34}
( \chi \sigma^{\mu} {\bar{\psi}} )^{+} = -
{\bar{\cc}}_{\dot{\aa}} ( {\bar{\sigma}}^{\mu})^{\dot{\aa} \aa} \psi_{\aa}  
= \psi^{\aa} ( \sigma^{\mu} )_{\aa \dot{\aa} }{\bar{\cc}}^{\dot{\aa}}
= - ( \bar{\psi} {\bar{\sigma}}^{\mu} \chi )^{+}
\en

\eq
\label{A.35}
\cc^{\aa} ( \sigma^{\mu \nu} )_{\aa}^{\;\;\bb} \psi_{\bb} =
- \psi^{\aa} ( \sigma^{\mu \nu} )_{\aa}^{\;\;\bb} \cc_{\bb} 
\en

\eq
{\bar{\cc}}_{\dot{\aa}} ( \bar{\sigma}^{\mu \nu} )^{\dot{\aa}}_{\;\;\dot{\bb}}
{\bar{\psi}}^{\dot{\bb}} = - {\bar{\psi}}_{\dot{\aa}} 
( {\bar{\sigma}}^{\mu \nu})^{\dot{\aa}}_{\dot{\;\;\bb}}{\bar{\cc}}^{\dot{\bb}}
\label{A.36}
\en
where we have used $ \{\psi , \cc \} = \{ \bar{\psi} , \bar{\cc} \} =\{ \psi ,
\bar{\cc} \} =0 $ and the definition $ ( \chi^{\aa} \psi_{\aa} )^+ \equiv 
{\bar{\psi}}_{\dot{\aa}} {\bar{\chi}}^{\dot{\aa}} $.
Other useful formulas are
\eq
\label{A.37}
\th^{\aa} \th^{\bb} = - \frac{1}{2} \ee^{\aa \bb} \th \th \hspace{2cm}
\th_{\aa} \th_{\bb} =  \frac{1}{2} \ee_{\aa \bb} \th \th 
\en

\eq
\label{A.38}
\tb^{\dot{\aa}} \tb^{\dot{\bb}} =  \frac{1}{2} \ee^{\dot{\aa} \dot{\bb}}
\tb \tb \hspace{2cm}
\tb_{\dot{\aa}} \tb_{\dot{\bb}} = - \frac{1}{2} \ee_{\dot{\aa} \dot{\bb}}
\tb \tb 
\en

\eq
\label{A.39}
\th^{\aa} \tb^{\dot{\aa}} = \frac{1}{2} \left( {\bar{\sigma}}^{\mu}
\right)^{\dot{\aa} \aa} \th \sigma_{\mu} \tb
\en

\eq
\th \sigma^{\mu} \tb \th \sigma^{\nu} \tb = \frac{1}{2} \th \th \tb \tb 
g^{\mu \nu}
\label{A.40}
\en

\eq
\th \psi \th \cc = - \frac{1}{2} \psi \cc \th \th \hspace{2cm}
\tb {\bar{\psi}}  \tb {\bar{\cc}} = - \frac{1}{2} \tb \tb {\bar{\psi}}   
{\bar{\cc}}
\label{A.41}
\en

\eq
\label{A.42}
\ee^{\aa \bb} \frac{\part}{\part \th^{\beta}} = - \frac{\part}{\part \th_{\aa}} 
\en

\eq
\label{A.43}
Tr ( \sigma^{\mu} {\bar{\sigma}}^{\nu} ) = 2 g^{\mu \nu}
\en

\eq
\label{A.44}
Tr ( \sigma^\mu {\bar{\sigma}}^\nu  \sigma^\rho {\bar{\sigma}}^\sigma ) = 
2( g^{\mu \nu}  g^{\rho \sigma} + g^{\mu \sigma}  g^{\nu \rho} - 
g^{\mu \rho}  g^{\nu \sigma} + i \ee^{\mu \nu \rho \sigma} ) 
\en
Finally we list a set of relations that allow to go from a formulation
in terms of Weyl spinors into a formulation in terms of Dirac spinors:
\eq
\label{A.50}
{\bar{\Psi}}_1 \Psi_2 = \cc_1 \psi_2 + {\bar{\psi}}_1  {\bar{\cc}}_2 
\en

\eq
\label{A.51}
{\bar{\Psi}}_1 \gg_5 \Psi_2 = - \cc_1 \psi_2 + {\bar{\psi}}_1  {\bar{\cc}}_2 
\en

\eq
\label{A.52}
{\bar{\Psi}}_1  \gg^\mu \Psi_2 =  {\bar{\psi}}_1  {\bar{\sigma}}^\mu \psi_2 -  
{\bar{\cc}}_2  {\bar{\sigma}}^\mu \cc_1 
\en

\eq
\label{A.53}
{\bar{\Psi}}_1  \gg^\mu \gg_5 \Psi_2 = - {\bar{\psi}}_1  
{\bar{\sigma}}^\mu \psi_2 -  {\bar{\cc}}_2  {\bar{\sigma}}^\mu \cc_1 
\en

\eq
\label{A.54}
{\bar{\Psi}}_1   \Sigma^{\mu \nu} \Psi_2 = i \chi_{1} \sigma^{\mu \nu} \psi_2 
+ i {\bar{\psi}}_1  {\bar{\sigma}}^{\mu \nu} {\bar{\chi}}_{2}
\en
where we have used the following representation for the Dirac spinors in terms
of Weyl spinors
\eq
\label{A.55}
\Psi_{1} = \vecc{ \psi_{1\aa}}{{\bar{\cc}}^{\dot{\aa}}_{1}} 
\hspace{2cm}
\Psi_{2} = \vecc{ \psi_{2\aa}}{{\bar{\cc}}^{\dot{\aa}}_{2}}
\en
Having established the formalism of Weyl spinors we introduce now the
supersymmetry transformations and their action on the superfields. The
supersymmetry algebra is an extension of the Poincar{\'{e}} algebra to 
include the supersymmetry generators $Q_{\alpha}$ and 
${{\bar{Q}}}_{{\dot{\alpha}}}$. They satisfy the following (anti)commutation
relations with themselves and with the generators of the Poincar{\'{e}} 
group:
\[
[ P^{\mu}, {\bar{Q}}_{\dot{\aa}}] = [P^{\mu}, Q_\aa ] = 0
\]

\[
[ M^{\mu \nu}, Q_{\aa}] = -i (\sigma^{\mu \nu})_{\aa}^{\;\;\bb} Q_{\bb}
\]

\[
[ M^{\mu \nu}, {\bar{Q}}^{\dot{\aa}}] = -i 
({\bar{\sigma}}^{\mu \nu})^{\dot{\aa}}_{\;\;\dot{\bb}} {\bar{Q}}^{\dot{\bb}}
\]

\[
\{ Q_{\aa} , Q_{\bb} \} = \{ {\bar{Q}}_{\dot{\aa}}, {\bar{Q}}_{\dot{\bb}} \}=0
\]

\eq
\{ Q_{\aa}, {\bar{Q}}_{\dot{\bb}} \} = 2 \sigma^{\mu}_{\aa \dot{\bb}}P_{\mu}
\label{V.1}
\en
The supersymmetry algebra can be viewed as a Lie algebra with 
anticommuting parameters. This observation motivates to define the 
corresponding group element:
\eq
G( x , \theta, \tb ) = e^{ \left[ i x_{\mu} P^{\mu} + \th Q + \tb 
\bar{Q} \right] }
\label{R.1}
\en
Using the Hausdorff's relation
\eq
e^A e^B = e^{A+B + 1/2 [A,B] + \dots}
\label{R.2}
\en
where higher order terms are vanishing, we get 
\eq
\label{R.3}
G(a, \xi, {\bar{\xi}} ) G ( x^{\mu} , \th , \tb ) = G( x^{\mu} + a^{\mu} + i 
\th \sigma^{\mu} {\bar{\xi}} - i \xi \sigma^{\mu} \tb , \th + \xi, \tb + 
{\bar{\xi}} )
\en
As usual, a multiplication of two elements induces a change in the 
parameter space:
\[
x^{\mu} \rightarrow x^{\mu} + i \th \sigma^{\mu} {\bar{\xi}} - i \xi 
\sigma^{\mu} \tb + a^{\mu}
\]
\eq
\th \rightarrow \th + \xi \hspace{2cm} 
\tb \rightarrow \tb + {\bar{\xi}}
\label{R.4}
\en
This transformation for $a^{\mu} =0$ is generated by the following operator:
\eq
\label{R.5}
\xi Q + {\bar{\xi}} {\bar{Q}}
\en
where
\[
Q_{\aa} = \frac{\part}{ \part \th^{\aa} } - i \sigma^{\mu}_{\aa 
\dot{\aa}} \tb^{\dot{\aa}} \part_{\mu} \hspace{1cm}
Q^{\aa}= - \frac{\part}{ \part \th_{\aa} } + i \tb_{\dot{\aa}} 
({\bar{\sigma}}^{\mu})^{{\dot{\aa}}\aa} \part_{\mu}
\hspace{1cm} 
\]
\eq
{\bar{Q}}^{\dot{\aa}} = \frac{\part}{ \part {\bar{\th}}_{\dot{\aa}} } - i 
({\bar{\sigma}}^{\mu})^{\dot{\aa} \aa} \th_{\aa} \part_{\mu} \hspace{1cm}
{\bar{Q}}_{\dot{\aa}} = -\frac{\part}{ \part {\bar{\th}}^{\dot{\aa}} } + i 
\th^{\aa} \sigma^{\mu}_{\aa \dot{\aa}} \part_{\mu} 
\label{R.6}
\en
They satisfy the following algebra
\[
\{ Q_{\aa}, {\bar{Q}}_{\dot{\aa}} \} = 2 i \sigma^{\mu}_{\aa \dot{\aa}} 
\part_{\mu}
\]
\eq
\{ Q_{\aa}, Q_{\bb} \} = \{ {\bar{Q}}_{\aa}, {\bar{Q}}_{\dot{\bb}} \} =0
\label{R.7}
\en
A superfield $\Phi ( x , \th , \tb )$ is a function of the space-time 
variable $x_{\mu}$ and of the two Weyl spinors $ \th_{\aa}$ and 
$\tb_{\dot{\aa}}$. Under a supersymmetry transformation with parameters 
$ \xi^{\aa}$ and $ {\bar{\xi}}_{\dot{\aa}}$ a superfield transforms as 
follows
\eq
\delta \Phi ( x, \th , \tb ) = \left[ (\xi Q ) + ({\bar{\xi}} {\bar{Q}}) 
\right] \Phi (x , \th , \tb )
\label{R.8}
\en
It is useful to define the supersymmetric covariant derivative
\[
D_{\aa} = \frac{\part}{ \part \th^{\aa} } + i \sigma^{\mu}_{\aa 
\dot{\aa}} \tb^{\dot{\aa}} \part_{\mu} \hspace{1cm}
{\bar{D}}_{\dot{\aa}} = - \frac{\part}{ \part {\bar{\th}}^{\dot{\aa}} } - i 
\th^{\aa} \sigma^{\mu}_{\aa \dot{\aa}} \part_{\mu}
\]
\eq
D^{\aa} = -\frac{\part}{ \part \th_{\aa} } - i \tb_{\dot{\aa}}
(\bar{\sigma}^{\mu})^{\dot{\aa} \aa}  \part_{\mu} \hspace{1cm}
{\bar{D}}^{\dot{\aa}} =  \frac{\part}{ \part {\bar{\th}}_{\dot{\aa}} } + i 
(\bar{\sigma}^{\mu})^{ \dot{\aa} \aa} \th_{\aa} \part_{\mu}
\label{R.9}
\en
They anticommute with the supersymmetry generators given in eq.(\ref{R.6}).

An arbitrary superfield can be expanded in terms of normal fields as 
follows
\[
F (x , \th , \tb ) = f (x) + (\th \varphi (x) ) + ( \tb \bar{\chi}(x) ) +
(\th \th ) m (x) + ( \tb \tb ) n (x) + 
\]
\eq
+ \th \sigma^{\mu} \tb v_{\mu} (x)
+ ( \th \th ) ( \tb \bar{\lambda} (x) ) + ( \tb \tb ) ( \th \psi (x)) +
( \th \th ) ( \tb \tb ) d (x)
\label{R.11}
\en
All higher powers of $\th$ and $ \tb$ vanish.

From eq.(\ref{R.5}) one can construct the transformations of the 
ordinary fields under supersymmetry. In particular it can be seen that the 
last component of a superfield transforms as a total derivative under 
supersymmetry:
\eq     
\dd d(x) =\frac{i}{2} \part_{\mu} O^{\mu}
\label{R.12}
\en
where $O^{\mu}= \xi \sigma^{\mu} \bar{\lambda} + {\bar{\xi}} 
{\bar{\sigma}}^{\mu} \psi  $ . This observation will be very useful for
constructing supersymmetric Lagrangians using the superfield formalism.
The superfield introduced in eq.(\ref{R.11}) is not reducible in general.
In four dimensions the irreducible superfields are the chiral and the
vector ones.

A chiral superfield is characterized by the condition
\eq
\label{T.1}
{\bar{D}}_{\dot{\aa}} \Phi =0
\en
The above constraint is easily solved in terms of the two 
quantities
\eq
y^{\mu}_{+} = x^{\mu} + i \th \sigma^{\mu} \tb \hspace{1cm}; \hspace{1cm} 
\th^{\alpha}
\label{T.2}
\en
that satisfy the conditions
\eq
{\bar{D}}_{\dot{\aa}} y^{\mu}_{+} = {\bar{D}}_{\dot{\aa}} \th =0
\label{T.3}
\en
Any function of these two variables will satisfy the condition
in eq.(\ref{T.1}) and therefore a chiral superfield can be written as 
follows
\[
\Phi = A ( y_{+} ) + \sqrt{2} ( \th \psi (y_{+})) + ( \th \th ) F( y_{+}) =
\]
\[
= A ( x) + i ( \th \sigma^{\mu} \tb ) \part_{\mu} A (x) - \frac{1}{4} ( \th 
\th ) ( \tb \tb ) \part_{\mu} \part^{\mu} A ( x) +
\]
\eq
+ \sqrt{2} ( \th \pp ( x) )  - \frac{i}{\sqrt{2}} ( \th \th ) ( 
\part_{\mu} \pp (x) \sigma^{\mu} \tb ) + ( \th \th ) F (x)
\label{T.4}
\en
The supersymmetry transformations of a chiral superfield in terms of the 
component fields are given by
\[
\dd A = \sqrt{2} \xi \psi
\]
\[
\dd \psi_{\aa} = \sqrt{2} \xi_{\aa} F + i \sqrt{2} \sigma^{\mu}_{\aa \dot{\aa}}
{\bar{\xi}}^{\dot{\aa}}\part_{\mu} A
\]
\eq
\dd F = i \sqrt{2} {\bar{\xi}} {\bar{\sigma}}^{\mu} \part_{\mu}\psi
\label{T.5}
\en

The superfield $ \bar{\Phi}$ will instead satisfy the constraint
\eq
D_{\aa} \bar{\Phi} =0
\label{T.6}
\en
It can be conveniently expressed in terms of the two variables
\eq
y^{\mu}_{-} = x^{\mu} - i \th \sigma^{\mu} \tb \hspace{1cm}; \hspace{1cm} 
\tb^{{\dot{\alpha}}}
\label{T.7}
\en
It is given by
\[
{\bar{\Phi}} = \bar{A} ( y_{-} ) + \sqrt{2} ( \tb {\bar{\psi}}(y_{-})) + 
( \tb \tb ) \bar{F} ( y_{-}) =
\]
\[
= \bar{A} ( x) - i ( \th \sigma^{\mu} \tb ) \part_{\mu} \bar{A} (x) - 
\frac{1}{4} ( \th \th ) ( \tb \tb ) \part_{\mu} \part^{\mu} \bar{A} ( x) +
\]
\eq
+ \sqrt{2} ( \tb {\bar{\psi}} ( x) )  + \frac{i}{\sqrt{2}} ( \tb \tb ) 
( \th \sigma^{\mu} \part_{\mu} {\bar{\psi}} (x) ) + ( \tb \tb ) \bar{F} (x)
\label{T.8}
\en
It is also useful to give the transformations of the component fields 
belonging to the antichiral superfield
\[
\dd \bar{A} = \sqrt{2} {\bar{\xi}} \bar{\psi}
\]
\[
\dd {\bar{\psi}}^{\dot{\aa}} = \sqrt{2} {\bar{\xi}}^{\dot{\aa}} \bar{F}
+ i \sqrt{2} ( {\bar{\sigma}}^{\mu} )^{\dot{\aa} \aa} \xi_{\aa} \part_{\mu} 
\bar{A}
\]
\eq
\dd \bar{F} = i \sqrt{2} \xi \sigma^{\mu} \part_{\mu} {\bar{\psi}} 
\label{T.9}
\en
In order to write the kinetic term of the Lagrangian of a chiral superfield
in superfield notations it is useful to have 
the following product of superfields in terms of component fields
\[
\bar{\Phi}_{i} \Phi_{j} = \bar{A}_i A_j + \sqrt{2} \th \psi_j {\bar{A}}_i
+ \sqrt{2} \tb {\bar{\psi}}_i A_j + \th^2 {\bar{A}}_i F_j + \tb^2 {\bar{F}}_i
z_j +
\]

\[
+\th \sigma^{\mu} \tb\left[ i \left({\bar{A}}_i \partial_{\mu} A_j -
A_j \partial_{\mu} {\bar{A}}_i  \right) - {\bar{\psi}}{\bar{\sigma}}_{\mu} 
\psi \right] + \sqrt{2} \tb^2 \th \psi_j {\bar{F}}_i + \sqrt{2} \th^2 \tb 
{\bar{\psi}}_i F_j +
\]

\[
+ \frac{i}{\sqrt{2}} \th^2 \left[ \tb {\bar{\sigma}}^{\mu} \part_{\mu} 
\psi_{i} {\bar{A}}_j - \tb \sigma^{\mu} \psi_j \part_{\mu} {\bar{A}}_i
\right] +
\] 

\[
+ \frac{i}{\sqrt{2}} \tb^2 \left[ \th \sigma^{\mu} \part_{\mu} 
{\bar{\psi}}_{j} A_i - \th \sigma^{\mu} {\bar{\psi}}_i \part_{\mu} A_j
\right] +
\] 

\[
+ (\th \th ) ( \tb \tb ) \left[ 
{\bar{F}}_{i} F_{j} - \frac{1}{4} {\bar{A}}_{i} \part_{\mu} \part^{\mu} 
A_{j} - \frac{1}{4} \part_{\mu} \part^{\mu} {\bar{A}}_{i}  A_{j} + 
\frac{1}{2} \part_{\mu} {\bar{A}}_{i} \part^{\mu} A_{j} \right. +
\]
\eq
\left.+ \frac{i}{2} \part_{\mu} {\bar{\pp}}_{i} {\bar{\sigma}}^{\mu} \pp_{j} -
\frac{i}{2}  {\bar{\pp}}_{i} {\bar{\sigma}}^{\mu} \part_{\mu} \pp_{j} \right] 
\label{T.10}
\en
We give also the following formulas for the term quadratic in the fields 
\eq
\Phi_i (x, \th ) \Phi_j (x, \th ) = \dots + ( \th \th )\left[ A_i F_j + F_i A_j 
- \psi_i \psi_j \right] 
\label{T.11}
\en
and for the term cubic in the fields

\[
\Phi_i (x, \th ) \Phi_j (x, \th ) \Phi_k (x, \th ) = \dots + 
\]
\eq
\label{T.12}
+ ( \th \th )
\left[ F_i A_j A_k + A_i F_j A_k  + A_i  A_j F_k - \psi_i \psi_j A_k -
A_i \psi_j \psi_k - \psi_i  A_j \psi_k \right]
\en
The most general renormalizable and supersimmetric action containing scalar 
and spinor fields is given by the sum of a kinetic term and a potential term:
\eq
S = \int d^4 x \left\{ \int \;\;d^2 \th \;\; d^2 \tb \;\;{\bar{\Phi}}_{i} 
\Phi_{i} + 
\left[\left( \int d^2 \th  W ( \Phi) \right) + h.c.  \right] \right\} 
\label{susyact}
\en
This action is automatically supersymmetric as follows from the observation 
(see eq.(\ref{R.12})) that
the last component of a superfield transforms as a total derivative and from
the fact that the integral in $d^2 \theta d^2 {\bar{\theta}}$ in the first term 
in eq.(\ref{susyact}) selects just the last component of the real superfield in 
eq.(\ref{T.10}), while the integral in $d^2 \theta$ selects the last term of a 
chiral  superfield. Renormalizability implies then that $W$ must contain at most
a cubic power of the chiral superfields. By performing the integral over the
Grassmann variables $\theta$ and ${\bar{\theta}}$ one gets:
\eq
\int d^4 x \left\{ \left[ \bar{F}_{i} F_{i} +
\part_{\mu} {\bar{A}}_{i} \part^{\mu} A_{i} - i
{\bar{\pp}}_{i} {\bar{\sigma}}^{\mu} \part_{\mu} \pp_{i} \right] + 
\left[F_i \frac{\part W}{\part A_i} 
- \frac{1}{2} \psi_i \psi_j 
\frac{\part^2 W}{\part A_i \part A_j} + h.c. \right] \right\} 
\label{T.15}
\en
The fields $F_i$ are non dynamical fields that can be eliminated by using
their classical equation of motion:
\eq
\label{T.16}
{\bar{F}}_i = - \frac{\part W}{\part A_i} 
\en
One gets finally the following Lagrangian
\eq
\label{T.17}
L= \part_{\mu} {\bar{A}}_{i} \part^{\mu} A_{i} - i
{\bar{\pp}}_{i} {\bar{\sigma}}^{\mu} \part_{\mu} \pp_{i}  
-| \frac{\part W}{\part A_i} |^2  - \frac{1}{2} \psi_i \psi_j 
\frac{\part^2 W}{\part A_i \part A_j}  
- \frac{1}{2} {\bar{\psi}}_i {\bar{\psi}}_j 
\frac{\part^2 \bar{W}}{\part {\bar{A}}_i \part {\bar{A}}_j}  
\en
The vector superfield is a real superfield
\eq
\label{S.1}
V = {\bar{V}} 
\en
If we expand it in component fields we get
\[
V ( x , \th , \tb ) = f (x) + \th \psi + \tb {\bar{\psi}} + \th \th m 
(x) + \tb \tb \bar{m} - \th \sigma^{\mu} \tb V_{\mu}
\]
\eq
\label{S.2}
+ (\th \th) ( \tb {\bar{\lambda}}) + ( \tb \tb ) ( \th \lambda ) + (\th 
\th ) ( \tb \tb ) D(x) 
\en
Under a gauge transformation the chiral superfields transform as follows
\eq
\label{S.3}
\Phi \rightarrow e^{-2i g \Lambda } \Phi \hspace{2cm}
{\bar{\Phi}} \rightarrow  {\bar{\Phi}} e^{2i g {\bar{\Lambda}} }
\en
where $ \Lambda$ is a chiral superfield. The matrix $ \Lambda$ can be 
written as
\eq
\label{S.4}
\Lambda_{AB} = T^{a}_{AB} \Lambda^a \hspace{2cm} [ T^a , T^b ] = i 
f^{abc}T^c
\en
where $T^a$ is the generator of the gauge group in the representation 
defined by the scalar superfield $\Phi$.

Therefore the kinetic term for the matter can be made supersymmetric and 
locally gauge invariant if we make the following substitution
\eq
\int d^4 \th {\bar{\Phi}}_{A} \Phi_A \Longrightarrow \int d^4 \th 
{\bar{\Phi}}_{A} 
\left( e^{2g V} \right)_{AB} \Phi_B
\label{S.5}
\en
if the vector superfield transforms as follows under a gauge 
transformation

\eq
e^{2gV} \rightarrow e^{-2ig {\bar{\Lambda}}} e^{2gV} e^{2ig \Lambda}
\label{S.6}
\en
The previous gauge transformation for $V^a$ is independent on the 
particular representation that one uses. In fact in computing the 
product of exponentials we encounter via the Hausdorff relation only  
commutators of generators. For an infinitesimal gauge transformation it 
can be shown that

\eq
\delta ( 2g V ) = 2ig ( \Lambda - {\bar{\Lambda}} ) + 2 i g^2 [ V, \Lambda + 
{\bar{\Lambda}} ] + \frac{2ig^3}{3} [ V, [ V, \Lambda - {\bar{\Lambda}}]] + 
O(V^3 )
\label{S.7}
\en
The existence of the inhomogenous term implies that we can choose the 
so-called Wess-Zumino gauge where we can gauge away many of the 
component fields present in $V$. Since
\[
i ( \Lambda - {\bar{\Lambda}} ) = i \left\{ A - \bar{A} + i \th \sigma^{\mu} 
\tb \part_{\mu} ( A + \bar{A} ) - \frac{1}{4} \th^2 \tb^2 \part_{\mu} 
\part^{\mu}( A - \bar{A} ) \right. +
\]
\eq
\label{S.8}  
\left.
+ \sqrt{2} \th \psi - \sqrt{2} \tb {\bar{\psi}} + \frac{i}{\sqrt{2}} 
\th^2 \tb {\bar{\sigma}}^{\mu} \part_{\mu} \psi - \frac{i}{\sqrt{2}} 
\tb^2 \th \sigma^{\mu} \part_{\mu} {\bar{\psi}} + \th^2 F - \tb^2 \bar{F}
\right\}
\en

By means of a gauge transformation we can choose $ A - \bar{A}, \psi , 
{\bar{\psi}}, F $ and $\bar{F}$ in order to gauge away some of the 
component fields appearing in $V$.   
In the Wess-Zumino gauge $V$ can be written as follows
\eq
\label{S.9}
V ( x , \th , \tb )  = - \th \sigma^{\mu} \tb A_{\mu}(x) + 
i (\th \th ) ( \tb \lb (x)) -
i (\tb \tb ) ( \th \lambda (x)) + \frac{1}{2} (\th \th ) (\tb \tb ) D(x)
\en
Remember also that
\eq
\label{S.10}
V^2 = \frac{1}{2} \th^2 \tb^2 A^{\mu} A_{\mu}
\en
The superfield field strenght is given by
\eq
W_{\aa} = - \frac{1}{8g} ( \bar{D} \bar{D} ) e^{-2gV} D_{\aa} e^{2gV}
\hspace{1cm}
{\bar{W}}_{\dot{\aa}} = - \frac{1}{8g} ({D} {D} ) e^{-2gV} 
{\bar{D}}_{\dot{\aa}} e^{2gV}
\label{S.11}
\en
They are chiral superfields
\eq
\label{S.12}
{\bar{D}}_{\dot{\bb}} W_{\aa} = D_{\bb} {\bar{W}}_{\dot{\aa}}=0
\en
In terms of component fields we get

\eq
W_{\aa}^{a} = - i \lambda_{\aa}^{a} + \left[ \delta_{\aa}^{\;\;\bb} D^a - 
i ( \sigma^{\mu \nu})_{\aa}^{\;\;\bb} F_{\mu \nu}^{a} \right] \th_{\bb} +
(\th \th) (\sigma^{\mu})_{\aa \dot{\aa} } (D_{\mu} \lb^{\dot{\aa}} )^a 
\label{S.13}
\en
and

\eq
\label{S.14}
({\bar{W}}^{\dot{\aa}})^a =  i (\lb^{\dot{\aa}})^{a} + \left[ 
\delta^{\dot{\aa}}_{\,\,\dot{\\bb}} D^{a} + i ( 
{\bar{\sigma}}^{\mu \nu} )^{\dot{\aa}}_{\;\;\dot{\bb}} F_{\mu \nu}^{a} 
\right] 
\tb^{\dot{\bb}} - (\tb \tb ) ({\bar{\sigma}}^{\mu})^{\dot{\aa} \aa } 
(D_{\mu} \lambda_{\aa})^a 
\en
where
\eq
(D_{\mu} \lambda^{\aa} )^a  = \partial_{\mu} (\lambda^{\aa})^a - g f^{abc} 
A_{\mu}^{b} (\lambda^{\aa})^c
\label{S.15}
\en
For constructing a supersymmetric action the following formula is very 
useful
\[
W^{\aa} W_{\aa} = - \lambda^{\aa} \lambda_{\aa} - 2i \lambda^{\aa} \left[ 
\delta_{\aa}^{\;\;\bb} D - i (\sigma^{\mu \nu})_{\aa}^{\;\;\bb} 
F_{\mu \nu}\right] \th_{\bb} + 
\]
\eq
+ ( \th \th ) \left[ D^2 -2i \lambda^{\aa} (\sigma^{\mu})_{\aa \dot{\aa}} 
D_{\mu} {\lb}^{\dot{\aa}} - \frac{1}{2} ( F_{\mu \nu} F^{\mu \nu} - i F_{\mu 
\nu} {}^{*}{F}^{\mu \nu}) \right]
\label{S.16}
\en
and
\[
{\bar{W}}_{\dot{\aa}} {\bar{W}}^{\dot{\aa}} = - \lb_{\dot{\aa}} 
\lb^{\dot{\aa}} + 2i \lb_{\dot{\aa}} 
\left[ \delta^{\dot{\aa}}_{\;\;\dot{\bb}} D + i ({\bar{\sigma}}^{\mu 
\nu})^{\dot{\aa}}_{\;\;\dot{\bb}} F_{\mu \nu}\right] \tb^{\dot{\bb}} + 
\]
\eq
+ ( \tb \tb ) \left[ D^2 -2i \lambda^{\aa} (\sigma^{\mu})_{\aa \dot{\aa}} 
D_{\mu} {\lb}^{\dot{\aa}} - \frac{1}{2} ( F_{\mu \nu} F^{\mu \nu} + iF_{\mu 
\nu} {}^{*}F^{\mu \nu}) \right]
\label{S.17}
\en 
where
\eq
\label{S.18}
{}^{*}F_{\mu \nu} = \frac{i}{2} \epsilon_{\mu \nu \rho \sigma} 
F^{\rho \sigma}
\en
The supersymmetric extension of Yang-Mills theory is given by
\eq
\label{S.19}
\int d^2 \th  \frac{1}{4} \left[ W^{\aa} W_{\aa} + h.c.\right] =
-\frac{1}{4} F_{\mu \nu}^{a} F_{\mu \nu}^{a} - i {\bar{\lambda}}^a 
{\bar{\sigma}}^{\mu} D_{\mu} \lambda^a + \frac{1}{2} D^2
\en
while the supersymmetric extension of matter interacting with Yang-Mills 
theory is given by
\[
\int d^4 \th {\bar{\Phi}} e^{2gV} \Phi = ( D_{\mu}A )^{+} ( D_{\mu} A )
- i {\bar{\psi}} {\bar{\sigma}}^{\mu} D_{\mu} \psi + \bar{F} F +
\]
\eq
+ g \bar{A} T^a D^a A
+ \sqrt{2} ig \bar{A} T^a \lambda^a \psi - i \sqrt{2} g {\bar{\psi}} T^a
{\bar{\lambda}}^a A
\label{S.20}
\en 
Finally by introducing a $\theta$ term in eq.(\ref{S.19}) and gauge and gaugino
fields normalized in such a way to include the gauge coupling as in 
eq.(\ref{nefi}) we can rewrite the Lagrangian of pure Yang-Mills theory as 
follows
\beq
L= - \frac{i}{16 \pi} \int d^2 \theta \tau \, W^2 + h.c. = \frac{1}{8 \pi} 
Im \left\{ \int d^2  \theta  \tau \, W^2 \right\}
\label{thym}
\eeq
where $\tau$ is given in eq.(\ref{tau}). 

At the end of this appendix we give the one-loop formula for the 
$\beta$-function in a gauge theory containing together with the gluon also
an arbitrary number of fermions and scalars. It is equal to:
\beq
\beta (e) = \frac{e^3}{(4 \pi)^2} \left[ - \frac{11}{3} c_G + \frac{1}{6} N_S 
c_S + \frac{4}{3} N_F c_F \right] 
\label{bfunct}
\eeq
where $N_S$ is the number of real scalars, $N_F$ is the number of Dirac fermions
and the constant $c$ depends on the representation of the various fields:
\beq
Tr ( T^a T^b ) = c \delta^{ab}
\label{trace9}
\eeq
The generators $T$ are normalized in such a way to have $c=1/2$ for the 
fundamental and $c=N$ for the adjoint of $SU(N)$.

From eq.(\ref{bfunct}) we obtain the $\beta$-function of $N=2$ super Yang-Mills,
given in eq.(\ref{betafu}), if we insert $c = N_c$ for all fields and $N_S =2$ 
and $N_F =1$ since we have two real scalar fields and one Dirac fermion.

Finally inserting again $c = N_c$ and $N_S = 6$ together with $N_F =2$ we 
obtain the $\beta$-function of $N=4$ super Yang-Mills that is equal to zero.
 
The chiral anomaly for a system of $M$ Majorana fermions is given by:
\beq
\partial_{\mu} J_{5}^{\mu} = 2 M c_F q_F  q(x)
\label{chiano}
\eeq
where $q(x)$ is the topological charge density defined in eq.(\ref{u1rano}),
$c_F$ is defined in eq.(\ref{trace9}) and is related to the fermion
representation and $q_F$ is the chiral weight of the fermions. In the case of 
$N=2$ super Yang-Mills we have two Majorana
fermions ($M=2$) in the adjoint representation of $SU(N_c )$ ($c_F = N_c ) $ 
with chiral weight equal to $1$ obtaining eq.(\ref{u1rano}). 
In the case of $N=4$ super Yang-Mills it is easy to see
that the Lagrangian in eq.(\ref{n=4}) is invariant if the superfields 
$W$ and $\Phi_i$  transform under  $U(1)_R$ as the fields $W$ 
and $\Phi$ in eq.(\ref{u1r}) with weight $1$ and $2/3$ (instead of $2$ as in
the second eq.(\ref{u1r})) respectively.
As a consequence the fermionic components of the superfields $W$ and $\Phi_i$
transform with weight $+1$ and $-1/3$ respectively and all according to the
adjoint representation of the gauge group. Adding the contributions of the
four fermionic fields one gets zero and therefore $N=4$ super Yang-Mills 
does not have a $U(1)_R$ anomaly.  

\sect{Appendix C}
\label{C}

In this Appendix we compute the central charge $Z$ of  $N=2$ super Yang-Mills
given in eq.(\ref{cecha3}).
In the case of $N=2$ super Yang-Mills the supercharge in four dimensions
can be obtained from the dimensional reduction of the supercurrent in 
eq.(\ref{sucurre}). After rescaling the fields as in eq.(\ref{nefi}) one gets:
\beq
Q_A = \frac{1}{e^2} \int d^3 x \left\{ \left[ 
\sigma_{\mu \nu} F^{\mu \nu}_{a} -i \gamma_{\rho} (D^{\rho} A_{4})_a 
\gamma_5  + \gamma_{\rho} (D^{\rho} A_5 )^a 
+ i f^{abc} A_{4}^{b} A_{5}^{c} \right]_{AB} \left( \gamma^0 \chi^a
\right)_{B} \right\}
\label{sucha}
\eeq
By saturating the supercharge in the previous equation with the supersymmetry
parameter ${\bar{\alpha}}_A$, introducing the Weyl spinors through the eqs.:
\beq
\chi = \left( \begin{array}{c} \psi_{\alpha} \\
		       {\bar{\lambda}}^{{\dot{\alpha}}} \end{array}\right) 
\hspace{2cm}
{\bar{\chi}} = \left( \begin{array}{cc} \lambda^{\alpha} &
		       {\bar{\psi}}_{{\dot{\alpha}}} \end{array}\right) 
\label{weylspi8}
\eeq
\beq
\alpha = \left( \begin{array}{c} \epsilon_{\alpha} \\
		       {\bar{\beta}}^{{\dot{\alpha}}} \end{array}\right) 
\hspace{2cm}
{\bar{\alpha}} = \left( \begin{array}{cc} \beta^{\alpha} &
		       {\bar{\epsilon}}_{{\dot{\alpha}}} \end{array}\right) 
\label{weylspi9}
\eeq
and
\beq
Q = \left( \begin{array}{c} Q^{(1)}_{\alpha} \\
		       {\bar{Q}}^{(2) {\dot{\alpha}}} \end{array}\right) 
\hspace{2cm}
{\bar{Q}} = \left( \begin{array}{cc} Q^{(2) \alpha} &
		       {\bar{Q}}^{(1)}_{{\dot{\alpha}}} \end{array}\right) 
\label{weylspi7}
\eeq
and remembering eq.(\ref{A.7}) for $\gamma^{\mu}$ and eq.(\ref{A.11}) for 
$\gamma_5$ together with
\beq
\sigma^{\mu \nu} = \frac{1}{4} [ \gamma^{\mu} , \gamma^{\nu} ]= 
\left( \begin{array}{cc} (\sigma^{\mu \nu} )_{\alpha}^{\,\beta} & 0 \\
  0 &    ({\bar{\sigma}}^{\mu \nu})^{\,{\dot{\beta}}}_{{\dot{\alpha}}} 
\end{array}\right) 
\label{sigma45}
\eeq
we get
\[
{\bar{\alpha}}_A Q_A = \beta^{\alpha} Q_{\alpha}^{(1)} + 
{\bar{\epsilon}}_{{\dot{\alpha}}}{\bar{Q}}^{(2) {\dot{\alpha}}} =
\]
\[
= \frac{1}{e^2}\int d^3 x 
\beta^{\alpha} \left[ \left(\sigma^{\mu \nu} \sigma^0 
\right)_{\alpha {\dot{\beta}}} F_{\mu \nu}^{a} 
{\bar{\lambda}}_{a}^{{\dot{\beta}}} + \left( (D_{\mu} A_5 )^a - i 
(D_{\mu} A_4 )^a  \right) ( \sigma^{\mu} {\bar{\sigma}}^0 )_{\alpha}^{\,\, 
\beta} \psi^{a}_{\beta} \right]+
\]
\beq
+ \frac{1}{e^2}\int d^3 x  
{\bar{\epsilon}}_{{\dot{\alpha}}} \left[ \left({\bar{\sigma}}^{\mu \nu} 
{\bar{\sigma}}^0 \right)^{{\dot{\alpha}} \beta} F_{\mu \nu}^{a} 
\psi^{a}_{\beta} + \left( (D_{\mu} A_5 )^a + i 
(D_{\mu} A_4 )^a  \right) ( {\bar{\sigma}}^{\mu} 
\sigma^0 )^{{\dot{\alpha}}}_{\,\, {\dot{\beta}}}
{\bar{\lambda}}_{a}^{{\dot{\beta}}} \right]
\label{q1q2}
\eeq
where we have omitted the last term in eq.(\ref{sucha}) because it is 
inessential
in the calculation of the central charge of the supersymmetry algebra.
From this eq. we extract
\beq
Q_{\alpha}^{(1)} = \frac{1}{e^2} \int d^3 x \left\{ \left( \sigma^{\mu \nu} 
\sigma^0 \right)_{\alpha {\dot{\beta}}} F^{a}_{\mu \nu} 
{\bar{\lambda}}_{a}^{{\dot{\beta}}} + \sqrt{2} \left( \sigma^{\mu} 
{\bar{\sigma}}^0 \right)_{\alpha}^{\,\,\, \beta} \psi^{a}_{\beta} 
({\overline{D_{\mu} \phi}} )^a \right\}
\label{qa1}
\eeq
and
\beq
{\bar{Q}}^{(2) {\dot{\alpha}}} = \frac{1}{e^2} \int d^3 x \left\{ \left( 
{\bar{\sigma}}^{\mu \nu} {\bar{\sigma}}^0 
\right)^{{\dot{\alpha}} \beta} F^{a}_{\mu \nu} 
\psi^{a}_{\beta} + \sqrt{2} \left({\bar{\sigma}}^{\mu} 
\sigma^0 \right)^{{\dot{\alpha}}}_{\,\, {\dot{\beta}}} 
{\bar{\lambda}}_{a}^{{\dot{\beta}}} 
(D_{\mu} \phi )^a \right\}
\label{qbara2}
\eeq
where $\phi$ is given in eq.(\ref{identi5}).
On the other hand, using that $(\gamma^{\mu})^{\dagger} = \gamma^0 \gamma^{\mu}
\gamma^0$, we get
\[
( {\bar{\alpha}}_A Q_A )^{*} = {\bar{Q}}_A \alpha_A = 
\]
\beq
=\frac{1}{e^2}  \int d^3 x \, 
{\bar{\chi}}^a  \left\{ -  F_{a}^{\mu \nu} \gamma^0 \sigma_{\mu \nu}  +
 (D_{\mu} A_5 )^a \gamma^0 \gamma^{\mu} -i 
 (D_{\mu} A_4 )^a \gamma^0 \gamma_5 \gamma^{\mu} \right\} \alpha
\label{sucha45}
\eeq
Rewriting it in Weyl notation we get
\[
{\bar{Q}}_A \alpha_A = Q^{(2) \alpha} \epsilon_{\alpha} +
{\bar{Q}}^{(1)}_{{\dot{\alpha}}} {\bar{\beta}}^{{\dot{\alpha}}}=
\]
\[
= \frac{1}{e^2} \int d^3 x \left\{  \left[- \lambda_{a}^{\alpha} 
F_{\mu \nu}^{a} 
( \sigma^0 {\bar{\sigma}}^{\mu \nu})_{\alpha {\dot{\beta}}} + 
(D_{\mu} \phi )^a {\bar{\psi}}^{a}_{{\dot{\alpha}}} ( {\bar{\sigma}}^0 
\sigma^{\mu} )^{{\dot{\alpha}}}_{\,\,{\dot{\beta}}} \right]
{\bar{\beta}}^{{\dot{\beta}}} + \right.
\]
\beq
\left.+ \left[- {\bar{\psi}}^{a}_{{\dot{\alpha}}} F^{a}_{\mu \nu} 
({\bar{\sigma}}^0 \sigma^{\mu \nu} )^{{\dot{\alpha}} \beta} + 
({\overline{D_{\mu} \phi}})^a \lambda^{\alpha}_{a} ( \sigma^0 {\bar{\sigma}}^{\mu}
)_{\alpha}^{\,\beta} \right]\alpha_{\beta} \right\}
\label{dipa4}
\eeq
From it we get

\beq
Q^{(2)\alpha } = \frac{1}{e^2} \int d^3 x \left\{- 
{\bar{\psi}}_{{\dot{\alpha}}}^{a}
F_{\mu \nu}^{a} ( {\bar{\sigma}}^0 \sigma^{\mu \nu} )^{{\dot{\alpha}} \alpha}
+ \sqrt{2} ({\overline{D_{\mu} \phi}})^a \lambda_{a}^{\beta} ( \sigma^0 
{\bar{\sigma}}^{\mu}
)_{\beta}^{\,\, \alpha} \right\}
\label{q2alpha3}
\eeq
and
\beq
{\bar{Q}}^{(1)}_{{\dot{\alpha}}} = \frac{1}{e^2} \int d^3 x \left\{- 
\lambda^{\alpha}_{a} F_{\mu \nu}^{a} ( \sigma^0 
{\bar{\sigma}}^{\mu \nu} )_{\alpha  {\dot{\alpha}}}
+ \sqrt{2} (D_{\mu} \phi)^a {\bar{\psi}}^{a}_{{\dot{\beta}}} 
( {\bar{\sigma}}^0 \sigma^{\mu} )^{ {\dot{\beta}}}_{\,\,{\dot{\alpha}}} 
\right\}
\label{q2alpha4}
\eeq
The canonical equal-time anticommutation relations satisfied by $\chi$
\beq
\{ \chi^{a}_{A} ( {\vec{x}}, t) , \chi_{B}^{\dagger b} ( {\vec{y}}, t) \} = 
e^2 \delta^{ab} \delta^{(3)} ( {\vec{x}} - {\vec{y}} ) \delta_{AB}
\label{antico}
\eeq
imply the following anticommutation relations for $\psi$ and $\lambda$:

\beq
\{ \psi^{a}_{\alpha} ( {\vec{x}},t) , {\bar{\psi}}^{b}_{\dot{\alpha}} 
({\bar{\sigma}}^{0} )^{ {\dot{\alpha}} \beta} 
({\vec{y}}, t) \} = e^2 \delta^{ab} \delta^{(3)} ( {\vec{x}} - {\vec{y}}) 
\delta_{\alpha}^{\,\, \beta}
\label{antico2}
\eeq
and
\beq
\{ {\bar{\lambda}}_{a}^{\dot{\alpha}} ( {\vec{x}},t) , 
\lambda_{b}^{\alpha}(\sigma^0 )_{\alpha {\dot{\beta}}} 
({\vec{y}}, t) \} = e^2 \delta^{ab} \delta^{(3)} ( {\vec{x}} - {\vec{y}}) 
\delta^{\dot{\alpha}}_{\,\, \dot{\beta}}
\label{antico7}
\eeq
while all other anticommutators are vanishing. 

Using the previous anticommutation relations one can compute
\beq
\{ Q_{\alpha}^{(1)} , Q_{\beta}^{(2)} \} =\frac{\sqrt{2}}{e^2} \int d^3 x 
F^{a}_{\rho \sigma} 
\left( {\overline{D_{\mu} \phi}} \right)^a \left\{ \left(\sigma^{\rho \sigma} 
\sigma^0
{\bar{\sigma}}^{\mu} \right)_{\alpha}^{\,\, \gamma} 
- \left( \sigma^{\mu} {\bar{\sigma}}^0 \sigma^{\rho \sigma}
\right)_{\alpha}^{\,\,\gamma} \right\} \epsilon_{\beta \gamma}
\label{antico90}
\eeq
that can be written as
\beq
\{ Q_{\alpha}^{(1)} , Q_{\beta}^{(2)} \} =\frac{1}{\sqrt{2}e^2} \int d^3 x 
F^{a}_{\rho \sigma} 
\left( {\overline{D_{\mu} \phi}} \right)^a \epsilon_{\alpha \beta } T^{\mu \rho 
\sigma}
\label{commu87}
\eeq
where
\beq
\epsilon_{\alpha \beta } T^{\mu \rho  \sigma} =
\left\{ \left( \sigma^{\rho} {\bar{\sigma}}^{\sigma}  \sigma^0
{\bar{\sigma}}^{\mu} \right)_{\alpha}^{\,\, \gamma} 
- \left( \sigma^{\mu} {\bar{\sigma}}^0 \sigma^{\rho} {\bar{\sigma}}^{\sigma} 
\right)_{\alpha}^{\,\,\gamma} \right\} \epsilon_{\beta \gamma}
\label{rel976}
\eeq
By saturating it with $\epsilon^{\beta \alpha}$ we get
\beq
2 T^{\mu \rho \sigma} = 4\left[ \eta^{\sigma \mu} \eta^{\rho 0} -
\eta^{\rho \mu} \eta^{\sigma 0} - i \epsilon^{0 \mu \rho \sigma} \right]
\label{formu98}
\eeq
where in the last step we have used eq.(\ref{A.44}). Inserting 
eq.(\ref{formu98}) in eq.(\ref{commu87}) we get
\beq
\{ Q_{\alpha}^{(1)} , Q_{\beta}^{(2)} \} = \epsilon_{\alpha \beta} {\hat{Z}}
\label{alge58}
\eeq
where
\beq
{\hat{Z}} = \frac{2 \sqrt{2}}{e^2} \int d^3 x \, \partial_{i} 
\left\{{\bar{\phi}}_a \left[ 
(F_{a} )^{0i} - i ({}^{*} F_{a} )^{0i} \right] \right\}
\label{centrac9}
\eeq
Finally remembering that 
\beq
F_{a}^{0i} = E_{ai} \hspace{2cm} {}^{*}F_{a}^{0i} = B_{ai}
\label{67}
\eeq
and using eqs.(\ref{e23}) and (\ref{m23}) we arrive at eq.(\ref{cecha3}).

\end{document}